\pdfoutput=1
\documentclass[12pt,a4paper]{article}

\usepackage{ifthen} 
\newboolean{pdflatex}
\setboolean{pdflatex}{true} 

\newboolean{articletitles}
\setboolean{articletitles}{true} 

\newboolean{uprightparticles}
\setboolean{uprightparticles}{false} 

\def\paperauthors{LHCb collaboration} 
\def\paperasciititle{Measurement of J/psi production cross-sections in pp collisions at sqrt(s)=5 TeV} 
\def\papertitle{Measurement of \jpsi production cross-sections in $pp$ collisions \\ at $\sqs=5\tev$} 
\def\paperkeywords{{High Energy Physics}, {LHCb}} 
\def\papercopyright{\the\year\ CERN for the benefit of the LHCb collaboration} 
\def\paperlicence{CC BY 4.0 licence}
\def\paperlicenceurl{https://creativecommons.org/licenses/by/4.0/}

\usepackage[top=1in, bottom=1.25in, left=1in, right=1in]{geometry}

\usepackage{microtype}
\usepackage{lineno}  
\usepackage{xspace} 
\usepackage{caption} 

\usepackage{graphicx}  
\usepackage{color}
\usepackage{colortbl}
\graphicspath{{./figs/}} 

\usepackage{amsmath} 
\usepackage{amssymb}
\usepackage{amsfonts}
\usepackage{upgreek} 

\newcommand*\patchAmsMathEnvironmentForLineno[1]{%
\expandafter\let\csname old#1\expandafter\endcsname\csname #1\endcsname
\expandafter\let\csname oldend#1\expandafter\endcsname\csname
end#1\endcsname
 \renewenvironment{#1}%
   {\linenomath\csname old#1\endcsname}%
   {\csname oldend#1\endcsname\endlinenomath}%
}
\newcommand*\patchBothAmsMathEnvironmentsForLineno[1]{%
  \patchAmsMathEnvironmentForLineno{#1}%
  \patchAmsMathEnvironmentForLineno{#1*}%
}
\AtBeginDocument{%
\patchBothAmsMathEnvironmentsForLineno{equation}%
\patchBothAmsMathEnvironmentsForLineno{align}%
\patchBothAmsMathEnvironmentsForLineno{flalign}%
\patchBothAmsMathEnvironmentsForLineno{alignat}%
\patchBothAmsMathEnvironmentsForLineno{gather}%
\patchBothAmsMathEnvironmentsForLineno{multline}%
\patchBothAmsMathEnvironmentsForLineno{eqnarray}%
}

\usepackage{hyperxmp}

\usepackage[pdftex,
            pdfauthor={\paperauthors},
            pdftitle={\paperasciititle},
            pdfkeywords={\paperkeywords},
            pdfcopyright={Copyright (C) \papercopyright},
            pdflicenseurl={\paperlicenceurl}]{hyperref}

\usepackage[colorinlistoftodos,textsize=scriptsize]{todonotes}

\usepackage[bottom,flushmargin,hang,multiple]{footmisc}

\usepackage[all]{hypcap} 

\usepackage{xspace} 
\usepackage{upgreek}

\def\lhcb   {\mbox{LHCb}\xspace}

\def\MagUp {\mbox{\em Mag\kern -0.05em Up}\xspace}

\ifthenelse{\boolean{uprightparticles}}%
{

 \def\Pmu         {\ensuremath{\upmu}\xspace}

 \def\Ppsi        {\ensuremath{\uppsi}\xspace}

 \def\PDelta      {\ensuremath{\Delta}\xspace}                 
 \def\PXi         {\ensuremath{\Xi}\xspace}                 
 \def\PLambda     {\ensuremath{\Lambda}\xspace}                 
 \def\PSigma      {\ensuremath{\Sigma}\xspace}                 
 \def\POmega      {\ensuremath{\Omega}\xspace}                 
 \def\PUpsilon    {\ensuremath{\Upsilon}\xspace}

 \def\PB      {\ensuremath{\mathrm{B}}\xspace}                 
                  
 \def\PD      {\ensuremath{\mathrm{D}}\xspace}

 \def\PJ      {\ensuremath{\mathrm{J}}\xspace}                 
 \def\PK      {\ensuremath{\mathrm{K}}\xspace}

 \def\Pb      {\ensuremath{\mathrm{b}}\xspace}                 
 \def\Pc      {\ensuremath{\mathrm{c}}\xspace}                 
                  
 \def\Pe      {\ensuremath{\mathrm{e}}\xspace}

 \def\Pi      {\ensuremath{\mathrm{i}}\xspace}

 \def\Ps      {\ensuremath{\mathrm{s}}\xspace}

 \def\thebaroffset{0.0em}
}
{

 \def\Pmu         {\ensuremath{\mu}\xspace}

 \def\Ppsi        {\ensuremath{\psi}\xspace}                 
                  
 \mathchardef\PDelta="7101
 \mathchardef\PXi="7104
 \mathchardef\PLambda="7103
 \mathchardef\PSigma="7106
 \mathchardef\POmega="710A
 \mathchardef\PUpsilon="7107
                  
 \def\PB      {\ensuremath{B}\xspace}                 
                  
 \def\PD      {\ensuremath{D}\xspace}

 \def\PJ      {\ensuremath{J}\xspace}                 
 \def\PK      {\ensuremath{K}\xspace}

 \def\Pb      {\ensuremath{b}\xspace}                 
 \def\Pc      {\ensuremath{c}\xspace}                 
                  
 \def\Pe      {\ensuremath{e}\xspace}

 \def\Pi      {\ensuremath{i}\xspace}

 \def\Ps      {\ensuremath{s}\xspace}

 \def\thebaroffset{0.18em}
}
\newcommand{\offsetoverline}[2][\thebaroffset]{\kern #1\overline{\kern -#1 #2}}%

\makeatletter
\ifcase \@ptsize \relax
  \newcommand{\miniscule}{\@setfontsize\miniscule{4}{5}}
\or
  \newcommand{\miniscule}{\@setfontsize\miniscule{5}{6}}
\or
  \newcommand{\miniscule}{\@setfontsize\miniscule{5}{6}}
\fi
\makeatother

\DeclareRobustCommand{\optbar}[1]{\shortstack{{\miniscule (\rule[.5ex]{1.25em}{.18mm})}
  \\ [-.7ex] $#1$}}

\def\epem       {{\ensuremath{\Pe^+\Pe^-}}\xspace}

\def\mumu       {{\ensuremath{\Pmu^+\Pmu^-}}\xspace}

\def\squark    {{\ensuremath{\Ps}}\xspace}

\def\cquark    {{\ensuremath{\Pc}}\xspace}

\def\bquark    {{\ensuremath{\Pb}}\xspace}

\def\KorKbar {\kern \thebaroffset\optbar{\kern -\thebaroffset \PK}{}\xspace}

\def\D       {{\ensuremath{\PD}}\xspace}

\def\DorDbar {\kern \thebaroffset\optbar{\kern -\thebaroffset \PD}\xspace}

\def\Dp      {{\ensuremath{\D^+}}\xspace}
\def\Dm      {{\ensuremath{\D^-}}\xspace}

\def\DpDm    {\ensuremath{\Dp {\kern -0.16em \Dm}}\xspace}

\def\B       {{\ensuremath{\PB}}\xspace}

\def\BorBbar {\kern \thebaroffset\optbar{\kern -\thebaroffset \PB}\xspace}

\def\Bd      {{\ensuremath{\B^0}}\xspace}

\def\BdorBdbar {\kern \thebaroffset\optbar{\kern -\thebaroffset \Bd}\xspace}

\def\Bs      {{\ensuremath{\B^0_\squark}}\xspace}

\def\BsorBsbar {\kern \thebaroffset\optbar{\kern -\thebaroffset \Bs}\xspace}

\def\jpsi     {{\ensuremath{{\PJ\mskip -3mu/\mskip -2mu\Ppsi}}}\xspace}

\def\Y#1S{\ensuremath{\PUpsilon{(#1S)}}\xspace}

\def\LorLbar     {\kern \thebaroffset\optbar{\kern -\thebaroffset \PLambda}\xspace}

\def\BF         {{\ensuremath{\mathcal{B}}}\xspace}
\def\BR         {\BF}

\def\to                 {\ensuremath{\rightarrow}\xspace}

\newcommand{\etot}{{\ensuremath{\varepsilon_{\mathrm{ tot}}}}\xspace}

\def\AT#1     {\ensuremath{A_{\mathrm{T}}^{#1}}\xspace}           

\def\C#1      {\ensuremath{\mathcal{C}_{#1}}\xspace}                       
\def\Cp#1     {\ensuremath{\mathcal{C}_{#1}^{'}}\xspace}                    
\def\Ceff#1   {\ensuremath{\mathcal{C}_{#1}^{\mathrm{(eff)}}}\xspace}        
\def\Cpeff#1  {\ensuremath{\mathcal{C}_{#1}^{'\mathrm{(eff)}}}\xspace}       
\def\Ope#1    {\ensuremath{\mathcal{O}_{#1}}\xspace}                       
\def\Opep#1   {\ensuremath{\mathcal{O}_{#1}^{'}}\xspace}                    

\newcommand{\nospaceunit}[1]{\ensuremath{\text{#1}}}       
\newcommand{\aunit}[1]{\ensuremath{\text{\,#1}}}       

\newcommand{\tev}{\aunit{Te\kern -0.1em V}\xspace}
\newcommand{\gev}{\aunit{Ge\kern -0.1em V}\xspace}
\newcommand{\mev}{\aunit{Me\kern -0.1em V}\xspace}
\newcommand{\kev}{\aunit{ke\kern -0.1em V}\xspace}
\newcommand{\ev}{\aunit{e\kern -0.1em V}\xspace}
 
\newcommand{\mevc}{\ensuremath{\aunit{Me\kern -0.1em V\!/}c}\xspace}
\newcommand{\gevc}{\ensuremath{\aunit{Ge\kern -0.1em V\!/}c}\xspace}
\newcommand{\mevcc}{\ensuremath{\aunit{Me\kern -0.1em V\!/}c^2}\xspace}
\newcommand{\gevcc}{\ensuremath{\aunit{Ge\kern -0.1em V\!/}c^2}\xspace}

\def\mum  {\ensuremath{\,\upmu\nospaceunit{m}}\xspace}

\def\mub{\ensuremath{\,\upmu\nospaceunit{b}}\xspace}

\def\pb {\aunit{pb}\xspace}
\def\invpb {\ensuremath{\pb^{-1}}\xspace}

\def\ps   {\ensuremath{\aunit{ps}}\xspace}

\newcommand{\chisqip}{\ensuremath{\chi^2_{\text{IP}}}\xspace}

\def\deriv {\ensuremath{\mathrm{d}}}

\def\gsim{{~\raise.15em\hbox{$>$}\kern-.85em
          \lower.35em\hbox{$\sim$}~}\xspace}
\def\lsim{{~\raise.15em\hbox{$<$}\kern-.85em
          \lower.35em\hbox{$\sim$}~}\xspace}

\def\sPlot{\mbox{\em sPlot}\xspace}

\def\sqs   {\ensuremath{\protect\sqrt{s}}\xspace}
\def\sqsnn {\ensuremath{\protect\sqrt{s_{\scriptscriptstyle\text{NN}}}}\xspace}
\def\pt         {\ensuremath{p_{\mathrm{T}}}\xspace}

\def\ptot       {\ensuremath{p}\xspace}

\newcommand{\lum} {\ensuremath{\mathcal{L}}\xspace}

\def\evtgen     {\mbox{\textsc{EvtGen}}\xspace}

\def\geant      {\mbox{\textsc{Geant4}}\xspace}

\def\photos     {\mbox{\textsc{Photos}}\xspace}

\def\pythia     {\mbox{\textsc{Pythia}}\xspace}

\def\tell1  {TELL1\xspace}
\def\ukl1   {UKL1\xspace}

\newcommand{\ie}{\mbox{\itshape i.e.}\xspace}

\usepackage{cite} 
\usepackage{mciteplus}

\usepackage{longtable} 
\usepackage{multirow}
\usepackage[utf8]{inputenc}

\begin{document}

\renewcommand{\thefootnote}{\fnsymbol{footnote}}
\setcounter{footnote}{1}


\begin{titlepage}
\pagenumbering{roman}

\vspace*{-1.5cm}
\centerline{\large EUROPEAN ORGANIZATION FOR NUCLEAR RESEARCH (CERN)}
\vspace*{1.5cm}
\noindent
\begin{tabular*}{\linewidth}{lc@{\extracolsep{\fill}}r@{\extracolsep{0pt}}}
\ifthenelse{\boolean{pdflatex}}
{\vspace*{-1.5cm}\mbox{\!\!\!\includegraphics[width=.14\textwidth]{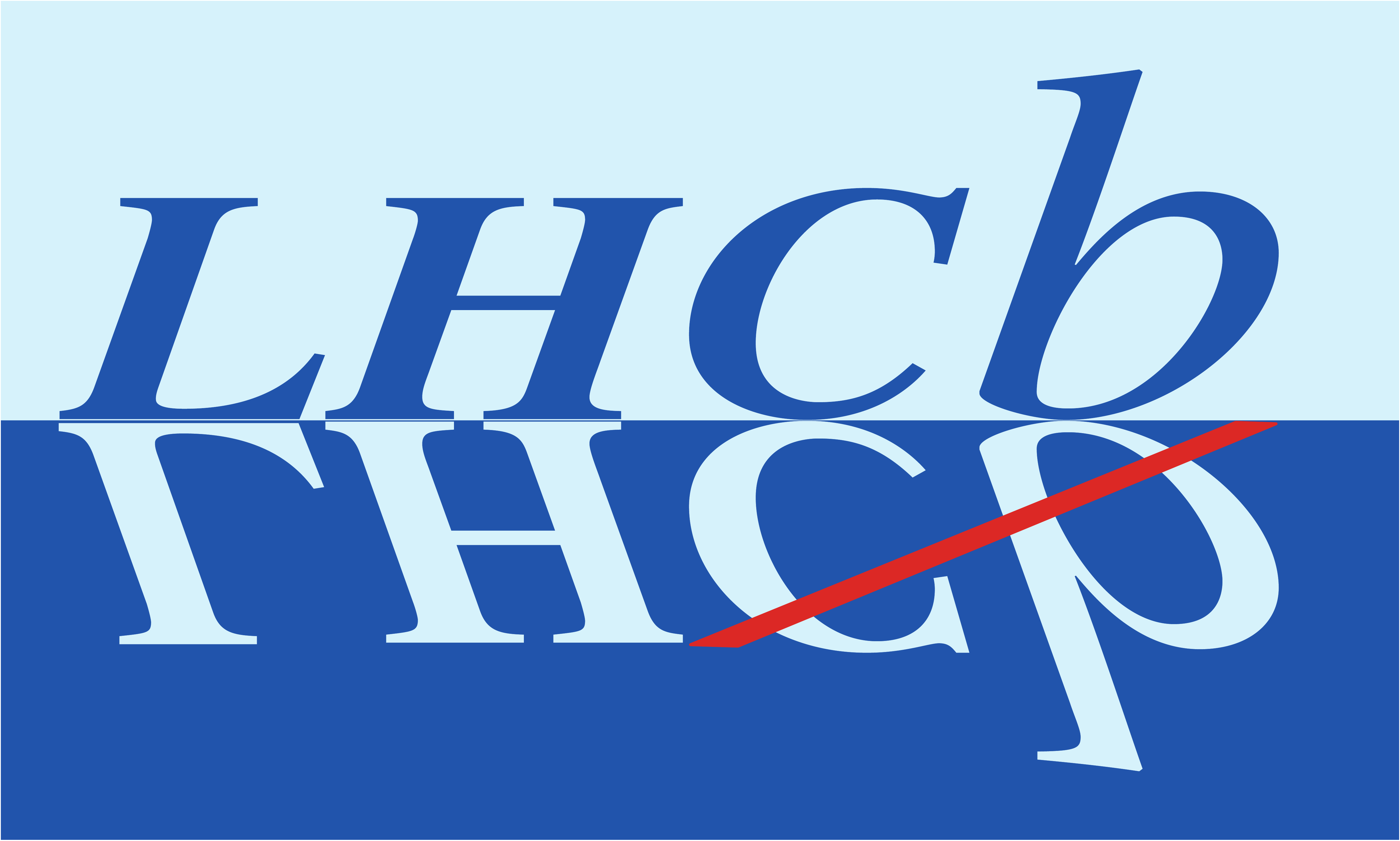}} & &}%
{\vspace*{-1.2cm}\mbox{\!\!\!\includegraphics[width=.12\textwidth]{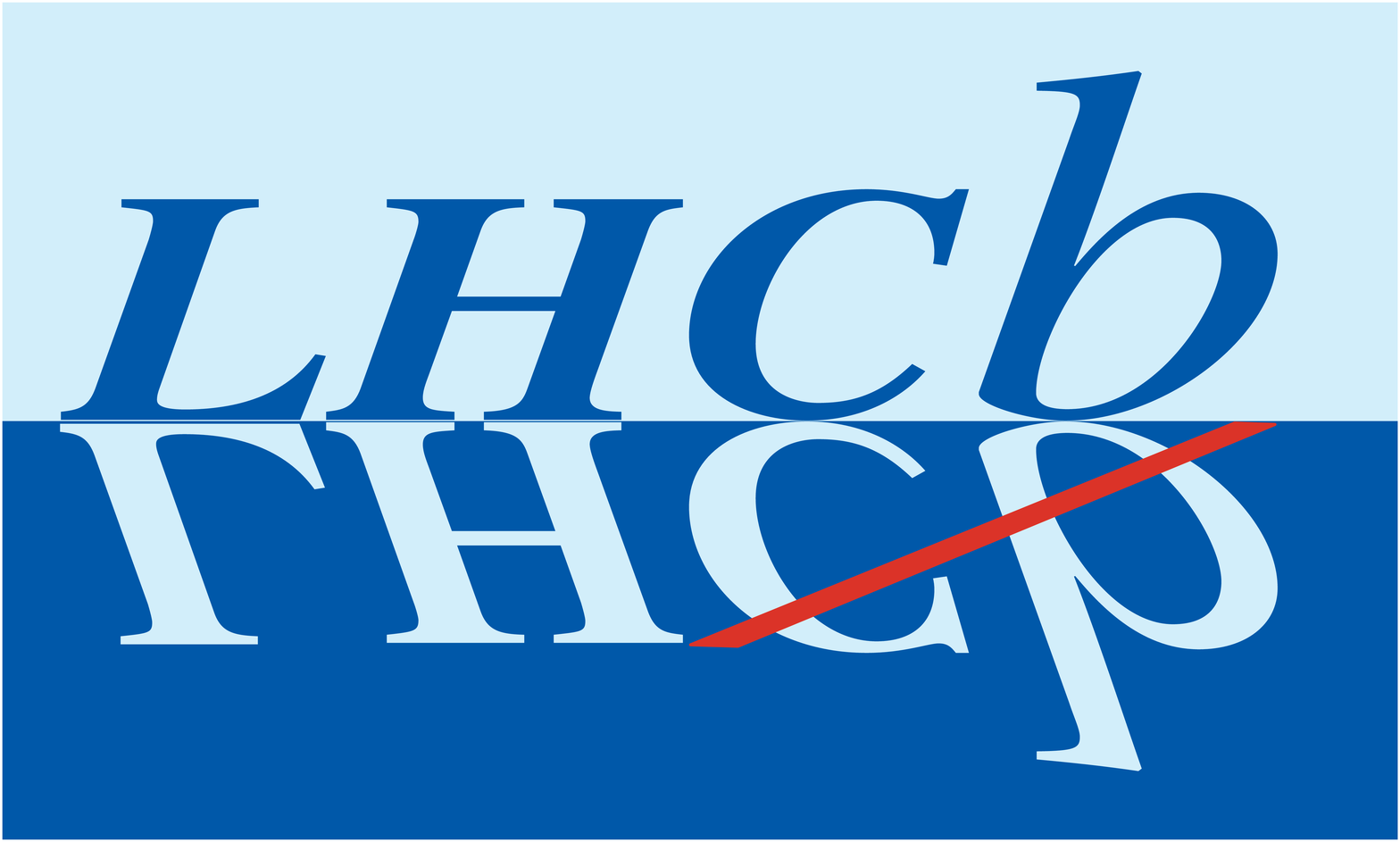}} & &}%
\\
 & & CERN-EP-2021-156 \\  
 & & LHCb-PAPER-2021-020 \\  
 & & 24 November 2021 \\ 
\end{tabular*}

\vspace*{4.0cm}

{\normalfont\bfseries\boldmath\huge
\begin{center}
  \papertitle 
\end{center}
}

\vspace*{1.0cm}

\begin{center}
\paperauthors\footnote{Authors are listed at the end of this paper.}
\end{center}

\vspace{\fill}

\begin{abstract}
  \noindent
    The production cross-sections of \jpsi mesons in proton-proton collisions
	at a centre-of-mass energy of $\sqs=5\tev$ are measured
	using a data sample corresponding to an integrated luminosity of $9.13\pm0.18\invpb$,
	collected by the \lhcb experiment.
	The cross-sections are measured differentially as a function of transverse momentum, \pt, and rapidity, $y$,
	and separately for \jpsi mesons produced promptly and
	from beauty hadron decays (nonprompt).
	With the assumption of unpolarised \jpsi mesons,
	the production cross-sections integrated over the kinematic range $0<\pt<20\gevc$ and $2.0<y<4.5$ are
\begin{equation*}
    \begin{split}
        \sigma_{\text{prompt \jpsi}} &= 8.154 \pm 0.010 \pm 0.283 \mub,\\
        \sigma_{\text{nonprompt \jpsi}} &= 0.820 \pm 0.003 \pm 0.034 \mub,
    \end{split}
\end{equation*}
where the first uncertainties are statistical and the second systematic.
	These cross-sections are compared with those at $\sqs=8\tev$ and $13\tev$, and are used to update the measurement of
	the nuclear modification factor in proton-lead collisions for \jpsi mesons at a centre-of-mass energy per nucleon pair of $\sqsnn=5\tev$. The results are compared with theoretical predictions.
  
\end{abstract}

\vspace*{1.0cm}

\begin{center}
  Published in
  JHEP 11 (2021) 181
\end{center}

\vspace{\fill}

{\footnotesize 
\centerline{\copyright~\papercopyright. \href{\paperlicenceurl}{\paperlicence}.}}
\vspace*{2mm}

\end{titlepage}


\newpage
\setcounter{page}{2}
\mbox{~}


\renewcommand{\thefootnote}{\arabic{footnote}}
\setcounter{footnote}{0}

\cleardoublepage


\pagestyle{plain} 
\setcounter{page}{1}
\pagenumbering{arabic}


\section{Introduction}
\label{sec:Introduction}
Quantum chromodynamics (QCD) is the fundamental theory that describes the strong interaction between quarks and gluons.
One of the most important properties of QCD is that the coupling constant increases at small momentum transfers.
Non-perturbative corrections, which are challenging to control theoretically, are required to describe many observables.
The study of heavy quarkonium production in proton-proton ($pp$) collisions can provide important information to improve QCD predictions in the non-perturbative regime.
The process involves the production of a $Q\overline{Q}$ system, where $Q$ denotes a beauty or charm quark, followed by its hadronisation into the heavy quarkonium state.
Predictions based on the assumption of factorisation have been found to agree well with experimental data so far.
The $Q\overline{Q}$ production step can be calculated with perturbative QCD but the hadronisation step, being of non-perturbative nature, needs to be described by models with inputs from experiments.
The colour singlet model~\cite{Carlson:1976cd,Donnachie:1976ue,Ellis:1976fj,Fritzsch:1977ay,Gluck:1977zm,Chang:1979nn,Baier:1981uk} assumes that the intermediate $Q\overline{Q}$ state is colourless and has the same spin-parity quantum numbers as the quarkonium state.
In the nonrelativistic QCD (NRQCD) approach~\cite{Bodwin:1994jh,Cho:1995vh,Cho:1995ce} intermediate $Q\overline{Q}$ states with all possible colour and spin-parity quantum numbers may evolve into a quarkonium state.
The transition probabilities are described by long-distance matrix elements (LDME) that cannot be calculated perturbatively and are therefore determined from experimental data.

In $pp$ collisions \jpsi mesons can be produced either directly from hard collisions of partons, through the feed-down of excited charmonium states, or via decays of beauty hadrons.
The \jpsi mesons from the first two sources originate from the primary $pp$ collision vertex (PV) and are called \emph{prompt} \jpsi mesons,
while those from the last source originate from decay vertices of beauty hadrons, which are typically separated from the PV, and are called \emph{nonprompt} \jpsi mesons.
The differential cross-sections for prompt and nonprompt \jpsi mesons in $pp$ collisions were measured in the rapidity range $2.0<y<4.5$
by the LHCb collaboration at centre-of-mass energies of $\sqs=2.76\tev$~\cite{LHCb-PAPER-2012-039}, 7\tev~\cite{LHCb-PAPER-2011-003}, 8\tev~\cite{LHCb-PAPER-2013-016} and 13\tev~\cite{LHCb-PAPER-2015-037}.
They were also measured by the ATLAS collaboration at $\sqs=5\tev$\cite{Aaboud:2017cif}, 7\tev~\cite{Aad:2015duc}, 8\tev~\cite{Aad:2015duc} and 13\tev~\cite{Zakareishvili:2020aki} in the region $|y|<2$,
and by the CMS collaboration at $\sqs=5\tev$~\cite{Sirunyan:2017mzd} and 7\tev~\cite{Khachatryan:2010yr,Chatrchyan:2011kc} in the region $|y|<2.4$.
Prompt \jpsi production cross-sections were measured by the CMS collaboration at $\sqs=13\tev$~\cite{Sirunyan:2017qdw} in the region $|y|<1.2$
and by the ALICE collaboration at $\sqs=7\tev$~\cite{Abelev:2012gx} in the region $|y|<0.9$.
The measurements for inclusive \jpsi mesons, which include both prompt and nonprompt contributions, were also performed by the ALICE collaboration at $\sqs=2.76\tev$~\cite{Abelev:2012kr}, 5\tev~\cite{Adam:2016rdg,Acharya:2019lkw} and 7\tev~\cite{Aamodt:2011gj} in the regions $|y|<0.9$ and $2.5<y<4.0$,
and at $\sqs=8\tev$~\cite{Adam:2015rta} and 13\tev~\cite{Acharya:2017hjh} in the region $2.5<y<4.0$.
In addition, the CDF experiment measured the prompt and nonprompt \jpsi cross-sections in proton-antiproton collisions at $\sqs=1.8\tev$~\cite{CDF:1997ykw} and 1.96\tev~\cite{CDF:2004jtw}.
The D0 experiment measured the inclusive \jpsi cross-sections in proton-antiproton collisions at $\sqs=1.8\tev$~\cite{D0:1996awi,D0:1998vai}.

This paper reports a \jpsi cross-section measurement in $pp$ collisions at $\sqs=5\tev$ using a data sample collected by the LHCb experiment, corresponding to an integrated luminosity of $9.13\pm0.18\invpb$~\cite{LHCb-PAPER-2014-047}.
This data sample was taken with special runs for cross-section measurements.
The measurement includes the production cross-sections of prompt and nonprompt \jpsi mesons with transverse momentum $\pt<20\gevc$ and rapidity $2.0<y<4.5$, assuming unpolarised \jpsi mesons,
and the cross-section ratios between 8\tev and 5\tev and between 13\tev and 5\tev.
The nuclear modification factor for \jpsi mesons in $p$Pb collisions at $\sqsnn=5\tev$, which was originally published in Ref.~\cite{LHCb-PAPER-2013-052}, is updated using the $pp$ cross-sections reported in here.

\section{Detector and simulation}
\label{sec:Detector}

The \lhcb detector~\cite{LHCb-DP-2008-001,LHCb-DP-2014-002} is a single-arm forward spectrometer covering the \mbox{pseudorapidity} range $2<\eta <5$, designed for the study of particles containing \bquark or \cquark quarks.
The detector includes a high-precision tracking system consisting of a silicon-strip vertex detector surrounding the $pp$ interaction region, a large-area silicon-strip detector located upstream of a dipole magnet with a bending power of about $4{\mathrm{\,Tm}}$, and three stations of silicon-strip detectors and straw drift tubes placed downstream of the magnet.
The tracking system provides a measurement of the momentum, \ptot, of charged particles with a relative uncertainty that varies from 0.5\% at low momentum to 1.0\% at 200\gevc.
The minimum distance of a track to a primary $pp$ collision vertex, the impact parameter (IP), is measured with a resolution of $(15+29/\pt)\mum$, where \pt is in\,\gevc.
Different types of charged hadrons are distinguished using information from two ring-imaging Cherenkov detectors. 
Photons, electrons and hadrons are identified by a calorimeter system consisting of scintillating-pad (SPD) and preshower detectors, an electromagnetic and a hadronic calorimeter.
Muons are identified by a system composed of alternating layers of iron and multiwire proportional chambers.
The online event selection is performed by a trigger, which consists of a hardware stage, based on information from the calorimeter and muon systems, followed by a software stage, which applies a full event reconstruction.

Simulated events are required to determine corrections for the detector resolution, acceptance and efficiency.
The $pp$ collisions are modelled using \pythia~\cite{Sjostrand:2007gs,Sjostrand:2006za} with a specific \lhcb configuration~\cite{LHCb-PROC-2010-056}.
In the \pythia model, \jpsi mesons are generated with zero polarisation and the leading order colour-singlet and colour-octet contributions~\cite{LHCb-PROC-2010-056,Bargiotti:2007zz} are considered in prompt \jpsi production.
Decays of unstable particles are described by \evtgen~\cite{Lange:2001uf} with QED final-state radiation handled by \photos~\cite{davidson2015photos}.
The interactions of the generated particles with the detector are modelled using the \geant toolkit~\cite{Allison:2006ve,Agostinelli:2002hh} as described in Ref.~\cite{LHCb-PROC-2011-006}. 

\section{Selection of \jpsi candidates}
\label{sec:Selection}

The \jpsi candidates are reconstructed through the \jpsi\to\mumu decay channel and are selected through two trigger stages.
The hardware trigger selects events with at least one muon candidate with $\pt>900\mevc$.
The software trigger requires two loosely identified muons, having $\pt>500\mevc$ and $p>3000\mevc$, to form a good-quality vertex.
In the offline selection the muon identification requirement is tightened and both tracks are required to have $\pt>650\mevc$ and $2.0<\eta<4.9$.
The background from fake tracks is reduced by a neural-network based algorithm~\cite{DeCian:2255039}.
The invariant mass of each \jpsi candidate, $m_{\mumu}$, is required to be within a range of $\pm 120\mevcc$ around the known \jpsi mass~\cite{PDG2020}.
All events are required to have at least one reconstructed PV.
For candidates with multiple PVs in the event, the one with the smallest \chisqip is taken as the associated PV,
where \chisqip is defined as the difference in the vertex-fit $\chi^2$ of a given PV reconstructed with and without the \jpsi candidate under consideration.

A final selection is applied to \jpsi candidates using the pseudoproper time $t_z$, defined as
\begin{equation}
\label{eq:tz}
    t_z=\frac{z_{\jpsi}-z_{\text{PV}}}{p_z} \times m_{\jpsi},
\end{equation}
where $z_{\jpsi}$ and $z_{\text{PV}}$ are positions the \jpsi decay vertex and the PV along the beam axis $z$, $p_z$ is the projection of the measured momentum of the \jpsi candidate along the $z$ axis, 
and $m_{\jpsi}$ is the known \jpsi mass~\cite{PDG2020}.
The $t_z$ uncertainty $\sigma_{t_z}$ is calculated by combining the estimated uncertainties on the $z$ position of the \jpsi decay vertex and that of the associated PV.
Candidates with $|t_z|<10\ps$ and $\sigma_{t_z}<0.3\ps$ are selected for further analysis.

\section{Cross-section determination}
\label{sec:Cross-section}
The double-differential cross-section of \jpsi production in a given (\pt,$y$) interval is defined as
\begin{equation}
\label{eq:cs}
    \frac{\deriv^2\sigma}{\deriv \pt\deriv y}=
    \frac{N(\jpsi\to\mumu)}{\lum\times\etot\times\BR\times\Delta\pt\times\Delta y},
\end{equation}
where $N(\jpsi\to\mumu)$ is the signal yield, \etot is the detection efficiency, \lum is the integrated luminosity, $\BR=(5.961\pm0.033)\%$~\cite{PDG2020} is the branching fraction of the \jpsi\to\mumu decay, and $\Delta\pt$ and $\Delta y$ are the interval widths.
Details on the interval scheme are provided in Sec.~\ref{sec:Production}.

\begin{figure}[tb]
    \centering
    \hfil
    \begin{minipage}[t]{0.49\linewidth}
        \centering
        \includegraphics[width=\linewidth]{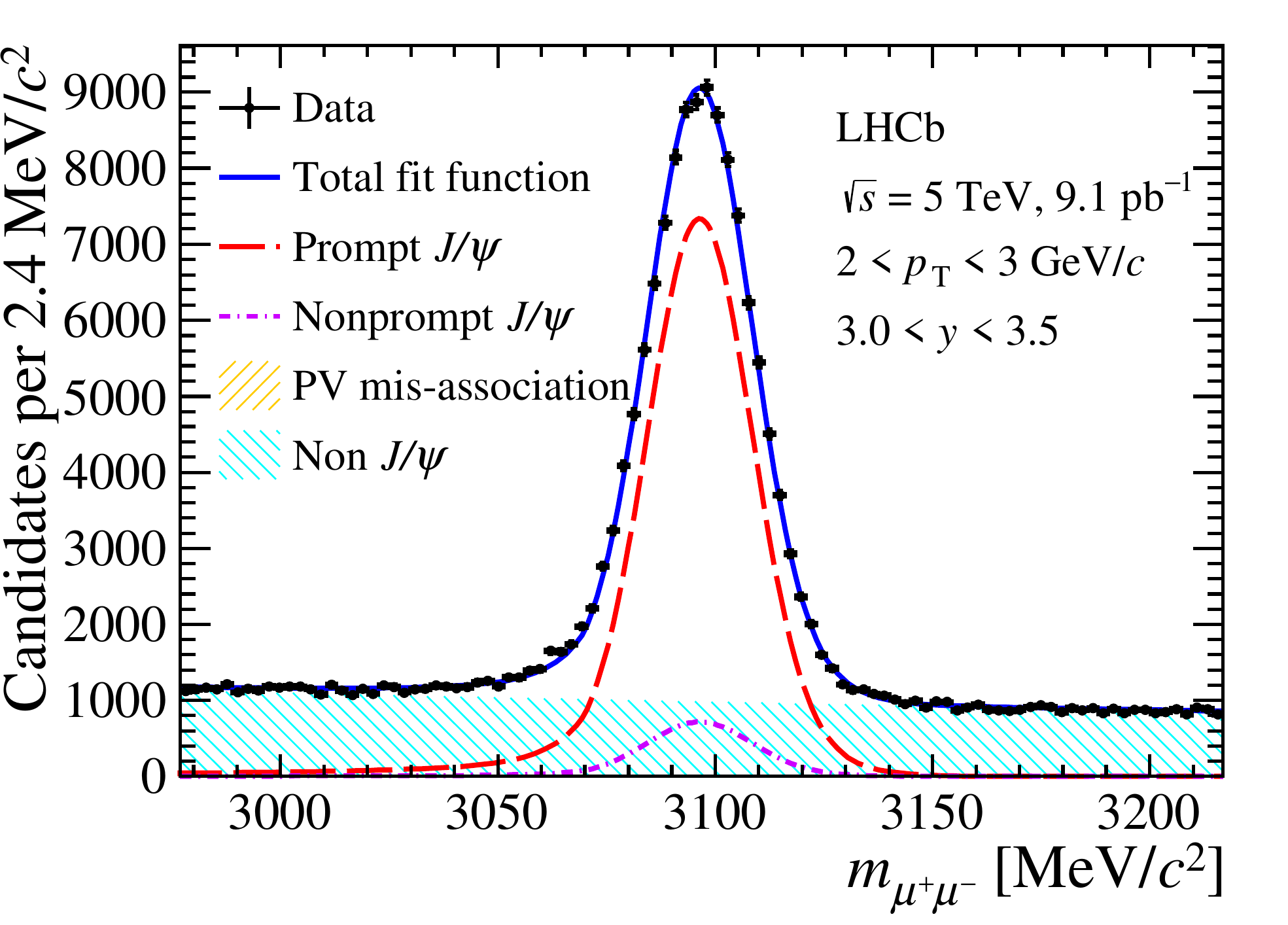}
    \end{minipage}
    \begin{minipage}[t]{0.49\linewidth}
        \centering
        \includegraphics[width=\linewidth]{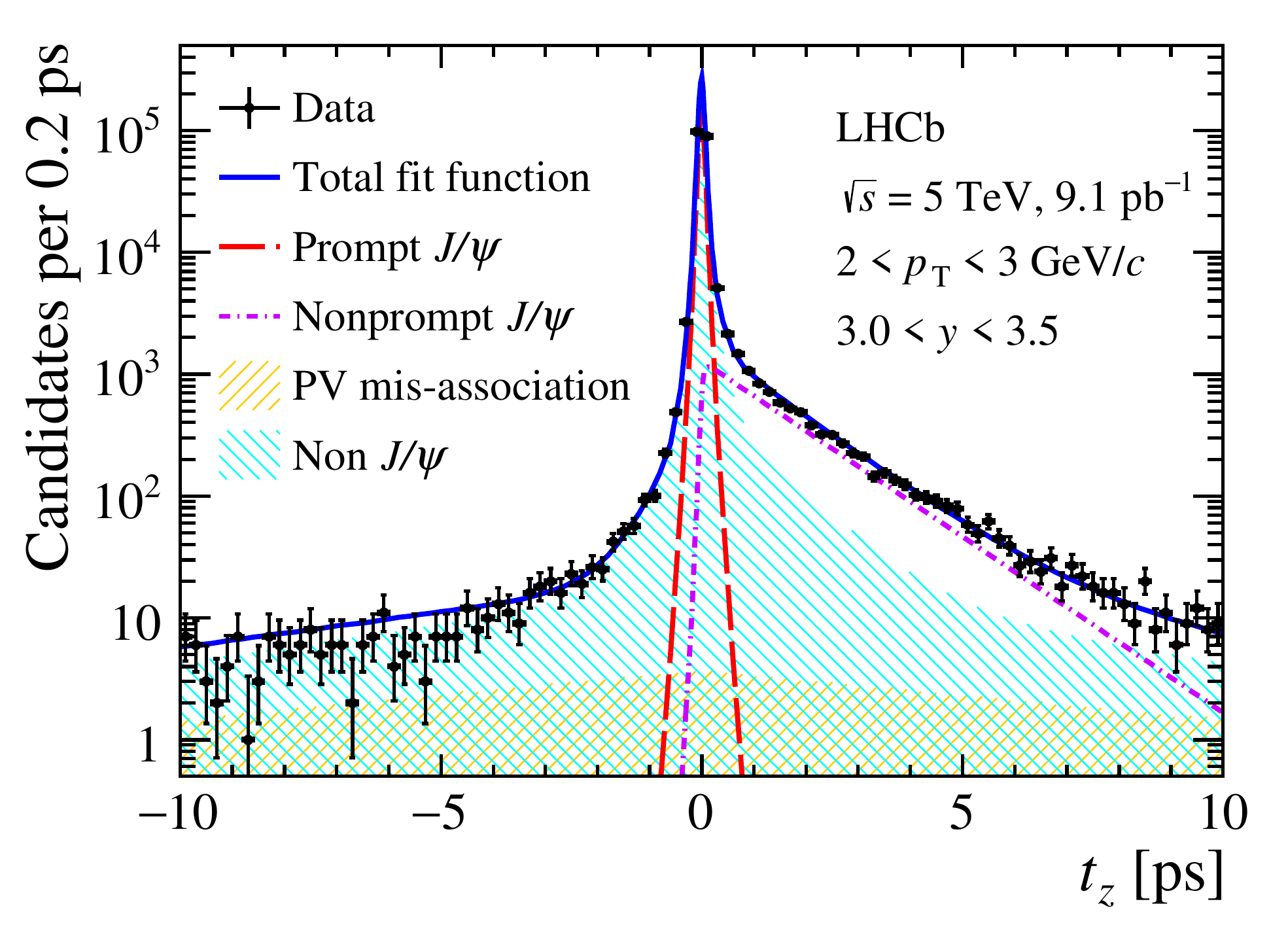}
    \end{minipage}
    \hfil
    \caption{Distributions of (left) invariant mass and (right) pseudoproper time of the \jpsi candidate for an example interval corresponding to $2<\pt<3\gevc$ and $3.0<y<3.5$. Projections of the two-dimensional fit are also shown.}
    \label{fig:2d}
\end{figure}

The yields of prompt and nonprompt \jpsi mesons are simultaneously extracted from an unbinned extended maximum-likelihood fit to the two-dimensional distribution of $m_{\mumu}$ and $t_z$ independently in each (\pt,$y$) interval.
The total \jpsi signal yield is about 1.4 (0.14) million for prompt (nonprompt) \jpsi mesons.
Figure \ref{fig:2d} shows the projections of the two-dimensional distribution, together with the fit, on $m_{\mumu}$ and $t_z$ for one (\pt,$y$) interval.
There are four components: prompt \jpsi signal, nonprompt \jpsi signal, \jpsi signal with incorrect PV association, and non-\jpsi background from random tracks.
The first three \jpsi signals have the same mass shape but their $t_z$ distributions are different.

In each interval the mass shape of \jpsi signals is described by the sum of two Crystal Ball (CB) functions~\cite{Skwarnicki:1986xj} with a common mean value and independent widths.
The simulation is used to determine the values of the two power-law tail parameters, which are shared between the two CB functions and fixed in the fit.
Only the mean and widths of the CB functions and the ratio between the two functions are left as free shape parameters in the fit.
The mass distribution of the non-\jpsi background is modelled with an exponential function.

The true $t_z$ values for prompt \jpsi mesons are assumed to be zero while those for nonprompt \jpsi mesons are assumed to follow an exponential function.
These distributions are convolved with the sum of two Gaussian functions to model the $t_z$ resolution.
The two Gaussian functions share the same mean value and their widths are proportional to the $t_z$ uncertainty $\sigma_{t_z}$.
The \jpsi signal with incorrect PV association contributes to the long tail present in the $t_z$ distribution.
This component can be modelled from data using event mixing, \ie, calculating $t_z$ with the \jpsi candidate associated to the closest PV in the next event of the sample.
The yield of this component is divided into two parts, $N_{\text{p}}^{\text{tail}}$ and $N_{\text{np}}^{\text{tail}}$, according to the ratio between prompt and nonprompt yields, and then $N_{\text{p}}^{\text{tail}}$ and $N_{\text{np}}^{\text{tail}}$ are added to the prompt and nonprompt yields respectively.
The $t_z$ distribution of the non-\jpsi background is described by an empirical function composed of a delta function and five exponential functions that are convolved with the sum of two Gaussian resolution functions sharing the same mean value.
All parameters of the empirical function are fixed to the values obtained from a fit to the $t_z$ distribution of the \jpsi mass sidebands, defined by the region $75<|m_{\mumu}-m_{\jpsi}|<150\mevcc$.

The detection efficiency is determined in each (\pt,$y$) interval using simulated samples.
The distribution of the number of SPD hits in simulation is weighted to match that in data to correct the effect of the detector occupancy in simulation.
The efficiency \etot is factorised into the product of four efficiencies:
the acceptance, $\varepsilon_{\text{acc}}$, the reconstruction-and-selection efficiency, $\varepsilon_{\text{rec\&sel}}$, the particle identification (PID) efficiency, $\varepsilon_{\text{PID}}$, and the trigger efficiency, $\varepsilon_{\text{tri}}$.
The efficiencies $\varepsilon_{\text{acc}}$ and $\varepsilon_{\text{rec\&sel}}$ are evaluated separately for prompt and nonprompt \jpsi mesons.
The efficiencies $\varepsilon_{\text{PID}}$ and $\varepsilon_{\text{tri}}$ are calculated combining the simulated samples of prompt and nonprompt \jpsi mesons, as the differences between the two production processes are observed to be negligible.
The efficiency $\varepsilon_{\text{tri}}$ is validated using data, and the efficiencies of track reconstruction and PID obtained from simulation are corrected using control channels in data, as detailed in Sec.~\ref{sec:Systematic}.

\section{Systematic uncertainties}
\label{sec:Systematic}

A summary of systematic uncertainties is presented in Table~\ref{table:sys}. Uncertainties arising from signal extraction and efficiency determination are mostly evaluated in each (\pt,$y$) interval, while those due to branching fraction and luminosity measurement are common to all intervals. The details of the evaluation are discussed in the following.

An uncertainty is attributed to the choice of the probability density function used to model the dimuon invariant-mass distribution of the signal components.
As an alternative to the sum of two CB functions, the signal invariant-mass distribution is described by a model derived from simulation using the approach of kernel density estimation~\cite{Cranmer:2000du}.
To account for the resolution difference between data and simulation, the alternative model is convolved with a Gaussian function with zero mean and width varied freely.
The default and alternative model are compared in each (\pt,$y$) interval and the relative difference, which is up to 2.0\%, is taken as a systematic uncertainty.

The exponential function describing the background is replaced by a linear function and the relative difference, varying up to 0.7\%, is taken as a systematic uncertainty.
The resulting uncertainty is considered as fully correlated between intervals.

The $t_z$ model used for the description of the non-\jpsi background is replaced by the use of the \sPlot method~\cite{Pivk:2004ty} using the $m_{\mumu}$ as the discriminating variable.
The relative difference between the two methods varies up to 1.2\% for prompt and 4.0\% for nonprompt \jpsi mesons in different intervals.

An uncertainty is attributed to the method that is used to separate prompt and nonprompt \jpsi mesons, \ie, two-dimensional fits to $m_{\mumu}$ and $t_z$ distributions.
To evaluate this uncertainty in each (\pt,$y$) interval, the same $t_z$ probability density function is used to fit the simulation.
While the relative differences between the fitted and the true yields are small for most intervals, they are significant for nonprompt \jpsi mesons in a few small-\pt intervals.
These differences, varying up to 0.8\% for prompt and 14.7\% for nonprompt \jpsi mesons, are taken as systematic uncertainties, and are assumed to be fully correlated between \pt intervals and uncorrelated between $y$ intervals as indicated by simulation.

A systematic uncertainty related to the tracking efficiency is evaluated as follows.
The efficiency correction factors are obtained from dedicated data and simulation samples of \jpsi\to\mumu decays in which one muon track is fully reconstructed and the other track is reconstructed using a subset of tracking systems~\cite{LHCb-DP-2013-002}.
These correction factors are found to depend on different event multiplicity variables.
This introduces a systematic uncertainty of 0.8\% per track.
The statistical uncertainties on these factors are propagated to the systematic uncertainties of the cross-sections, which vary up to 3.7\% depending on the (\pt,$y$) interval.

The PID efficiency is evaluated  using a dedicated sample of \mbox{\jpsi\to\mumu} candidates in which only one track is required to be identified as a muon.
The uncertainties of the muon identification efficiencies due to the finite size of the calibration data sample are propagated to the systematic uncertainties of the cross-sections, which are up to $2.2\%$ in different intervals.
Another uncertainty comes from the choice of interval scheme of the calibration sample.
The resulting uncertainties vary up to 1.5\% depending on the (\pt,$y$) interval.

The trigger efficiency in simulation is validated with data.
One muon is requested to pass the hardware-trigger requirement
such that the other muon can be regarded as an unbiased probe of the efficiency of one muon.
The hardware-trigger efficiency of the \jpsi candidate is the probability that at least one muon track passes the trigger requirement.
The relative difference between data and simulation, varying up to 1.9\% across intervals, is taken as a systematic uncertainty on the hardware-trigger efficiency.
The software-trigger efficiency is determined using a subset of events that would pass the trigger requirement if the \jpsi signals were excluded~\cite{LHCb-PUB-2014-039}.
The fraction of \jpsi candidates for which two tracks pass the software-trigger requirement is taken as the efficiency both for data and simulation.
The overall relative difference between data and simulation is 1.0\%, and is taken as a systematic uncertainty on the software-trigger efficiency common to all intervals.

The statistical uncertainties of the efficiencies due to the finite size of the simulated sample result in uncertainties on the cross-sections.
The values range up to 3.7\% for prompt and 7.7\% for nonprompt \jpsi mesons depending on the (\pt,$y$) interval.

The uncertainty on the \jpsi\to\mumu branching fraction~\cite{PDG2020} results in an uncertainty on the measured cross-sections of 0.6\%.
The luminosity is determined using methods similar to those described in Ref.~\cite{LHCb-PAPER-2014-047} and the relative uncertainty is 2.0\%.
The tail shape on the left side of the CB function is used to describe the effect of QED radiation, which leads to energy loss in some \jpsi candidates.
A small fraction of the \jpsi signal lies outside the mass range of the fit.
This signal loss is taken into account in the efficiency $\varepsilon_{\text{rec\&sel}}$ estimated with the simulated sample.
The imperfect modelling of the radiative decay is considered as a source of systematic uncertainty.
Based on a detailed comparison between the radiative tails in simulation and data a systematic uncertainty of 1.0\% is assigned.

\begin{table}[]
    \centering
    \caption{Relative systematic uncertainties on the measurement of the \jpsi production cross-section. The symbol $\oplus$ means addition in quadrature. The detailed uncertainties for each (\pt,$y$) interval are in Appendix~\ref{sec:Tables}.}
    \label{table:sys}
    \begin{tabular}{lll}
        \hline
        Source                       & Relative uncertainty      & Correlations \\
        \hline
        Signal mass model     & $<2.0\%$     & Uncorrelated \\
        Background mass model & $<0.7\%$     & Correlated between intervals \\
        \multirow{2}{*}{Background $t_z$ model} & $<1.2\%$ (prompt)  & \multirow{2}{*}{Uncorrelated} \\
                                                & $<4.0\%$ (nonprompt)\\
		\multirow{2}{*}{Signal $t_z$ model} & $<0.8\%$ (prompt)  & \multirow{2}{*}{Correlated between \pt intervals} \\
                                                       & $<14.7\%$ (nonprompt)\\
        Tracking efficiency          & $(2\times0.8\%)\oplus(<3.7\%)$ & Correlated between intervals \\
        PID efficiency               & $(<2.2\%)\oplus(<1.5\%)$ & Correlated between intervals \\
        Hardware-trigger efficiency        & $<1.9\%$     & Correlated between intervals \\
        Software-trigger efficiency      & 1.0\%              & Correlated between intervals \\
        \multirow{2}{*}{Simulation sample size}  & $<3.7\%$ (prompt) & \multirow{2}{*}{Uncorrelated} \\
                                                 & $<7.7\%$ (nonprompt) \\
        $\BR(\jpsi\to\mumu)$         & 0.6\%              & Correlated between intervals \\
        Luminosity                   & 2.0\%              & Correlated between intervals \\
        Radiative tail               & 1.0\%              & Correlated between intervals \\
        \hline
    \end{tabular}
\end{table}

\section{Production cross-sections results}
\label{sec:Production}
The measured double-differential cross-sections for prompt and nonprompt \jpsi mesons are shown in Fig.~\ref{fig:cs} and listed in Tables~\ref{table:csPrompt} and~\ref{table:csFromb}
in Appendix~\ref{sec:Tables}, for the range $0<\pt<14\gevc$ and $2.0<y<4.5$ with $\Delta\pt$ between 1 and 4\gevc and $\Delta y=0.5$.
By integrating the double-differential results over \pt or $y$,
the single-differential cross-sections $\deriv\sigma/\deriv\pt$ and $\deriv\sigma/\deriv y$ are obtained, and are listed in Tables~\ref{table:csPromptPT},~\ref{table:csFrombPT},~\ref{table:csPromptY} and~\ref{table:csFrombY} in Appendix~\ref{sec:Tables}.
The $\deriv\sigma/\deriv\pt$ results include a further \pt interval in the range $14<\pt<20\gevc$, which is not divided into $y$ intervals due to the limited size of the data sample.
The integrated cross-sections for prompt and nonprompt \jpsi mesons in the range $0<\pt<20\gevc$ and $2.0<y<4.5$ are
\begin{equation*}
    \begin{split}
        \sigma_{\text{prompt \jpsi}} &= 8.154 \pm 0.010 \pm 0.283 \mub,\\
        \sigma_{\text{nonprompt \jpsi}} &= 0.820 \pm 0.003 \pm 0.034 \mub,
    \end{split}
\end{equation*}
where the first uncertainties are statistical and the second systematic.
These results are obtained under the assumption that the polarisation of the \jpsi mesons is negligible.
The \jpsi polarisation measurement at $\sqs=7\tev$~\cite{LHCb-PAPER-2013-008} indicates that the polarisation parameters $\lambda_{\theta}$, $\lambda_{\theta\phi}$ and $\lambda_{\phi}$ are consistent with zero while the central value of $\lambda_{\theta}$ is around $-0.2$ in the helicity frame.
The polarisation affects the detection efficiency and the dependence of the cross-sections on the polarisation is reported in Appendix~\ref{sec:Polar}.
When the polarisation parameter $\lambda_{\theta}$ is assumed to be $-0.2$~\cite{LHCb-PAPER-2013-008}, the total cross-section decreases by 2.8\% (2.9\%) for prompt (nonprompt) \jpsi mesons.

\begin{figure}[tb]
    \centering
    \hfil
    \begin{minipage}[t]{0.49\linewidth}
        \centering
        \includegraphics[width=\linewidth]{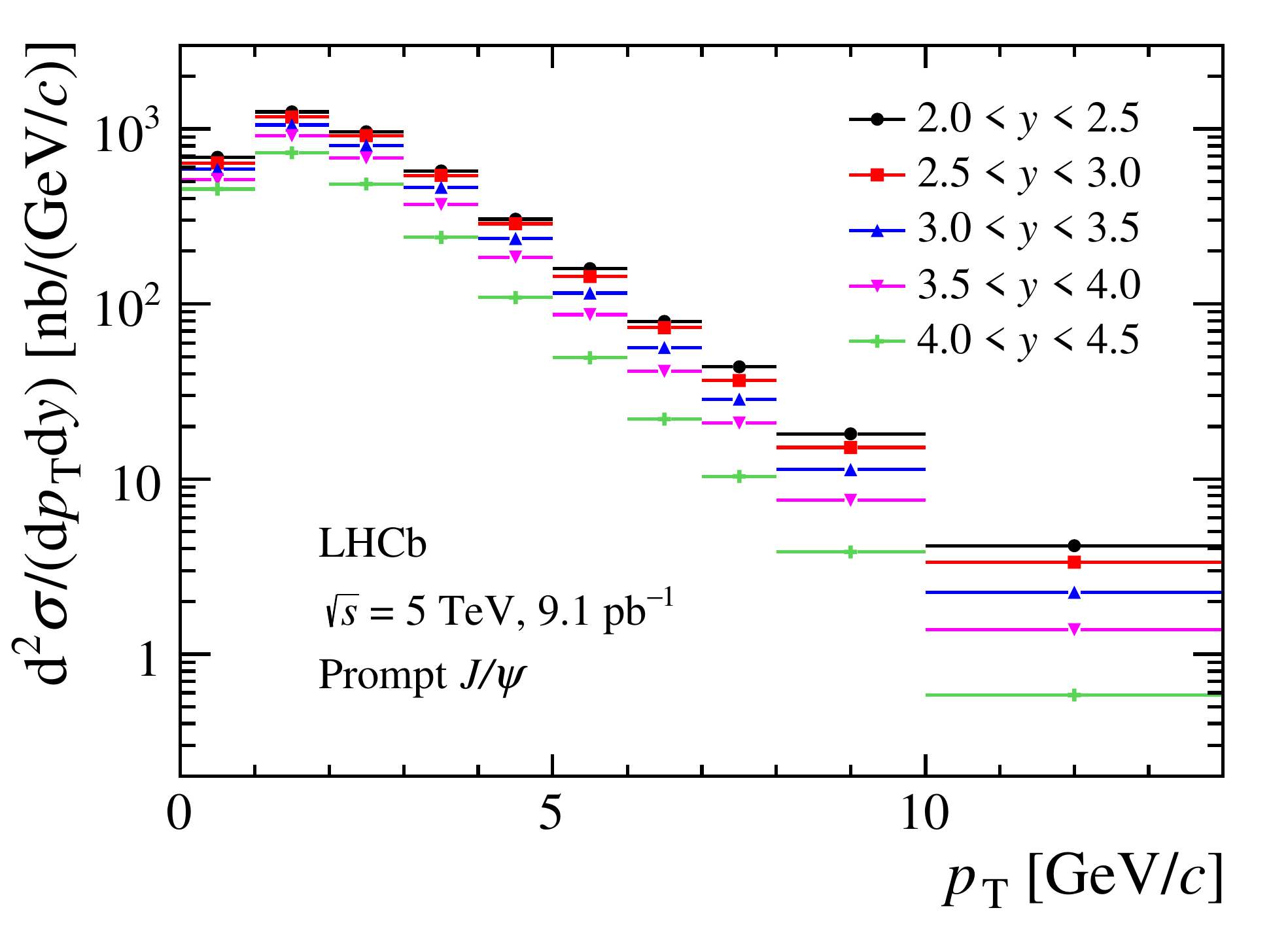}
    \end{minipage}
    \begin{minipage}[t]{0.49\linewidth}
        \centering
        \includegraphics[width=\linewidth]{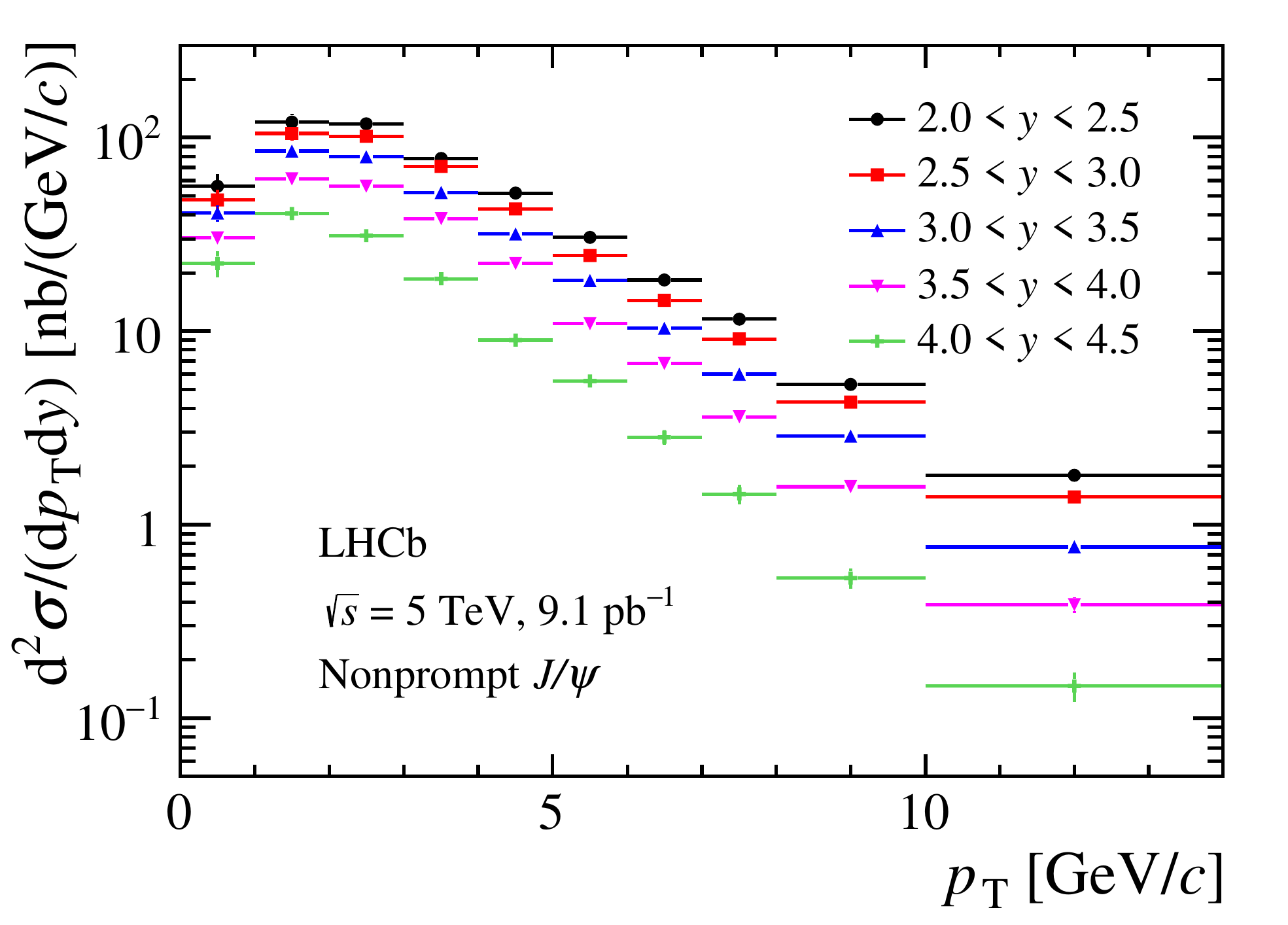}
    \end{minipage}
    \hfil
	\caption{Differential cross-sections for (left) prompt and (right) nonprompt \jpsi mesons as a function of \pt in intervals of $y$. The error bars represent the total uncertainties, which are partially correlated between intervals.}
    \label{fig:cs}
\end{figure}

\begin{figure}[tb]
    \centering
    \hfil
    \begin{minipage}[t]{0.49\linewidth}
        \centering
        \includegraphics[width=\linewidth]{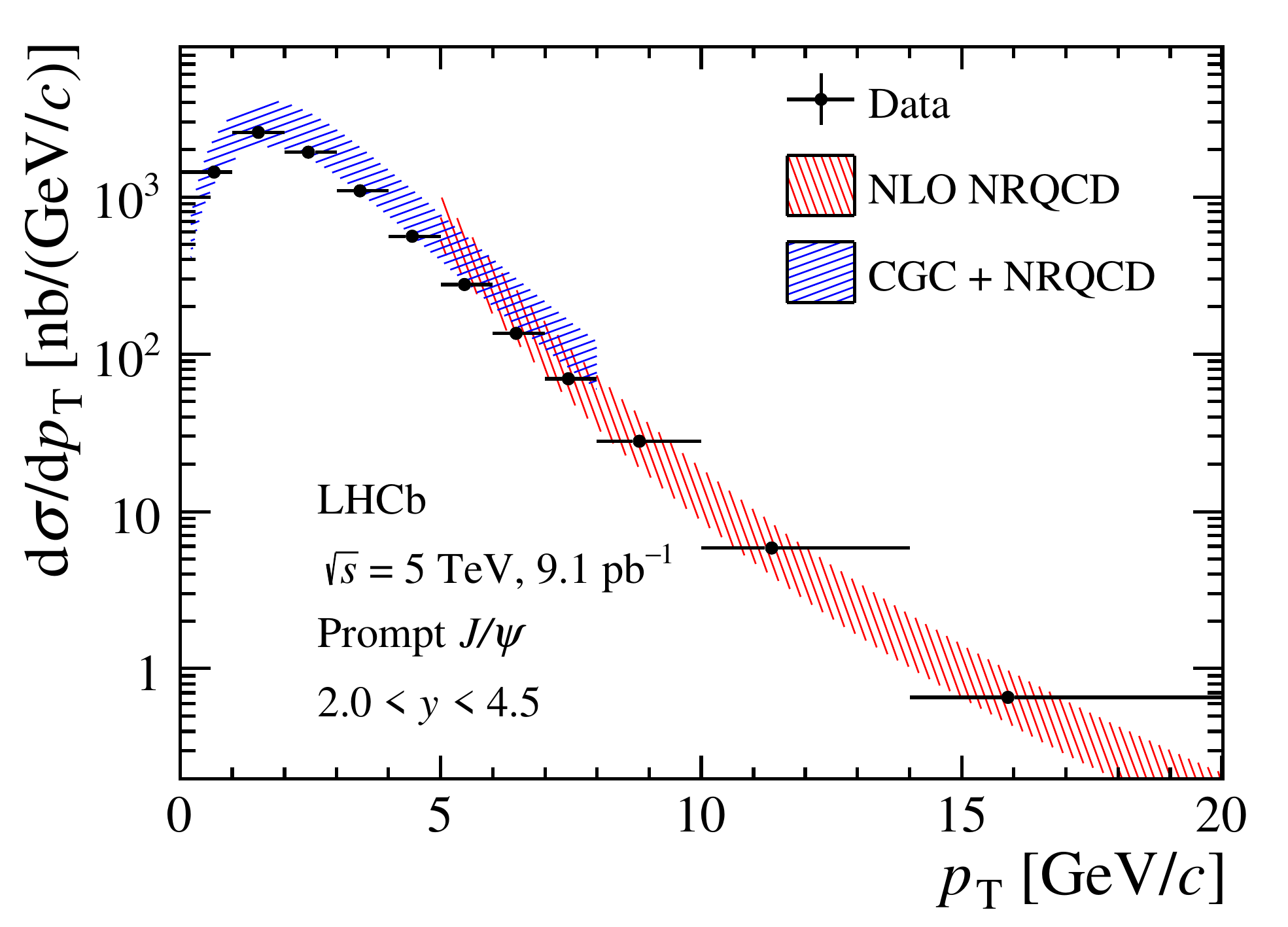}
    \end{minipage}
    \begin{minipage}[t]{0.49\linewidth}
        \centering
        \includegraphics[width=\linewidth]{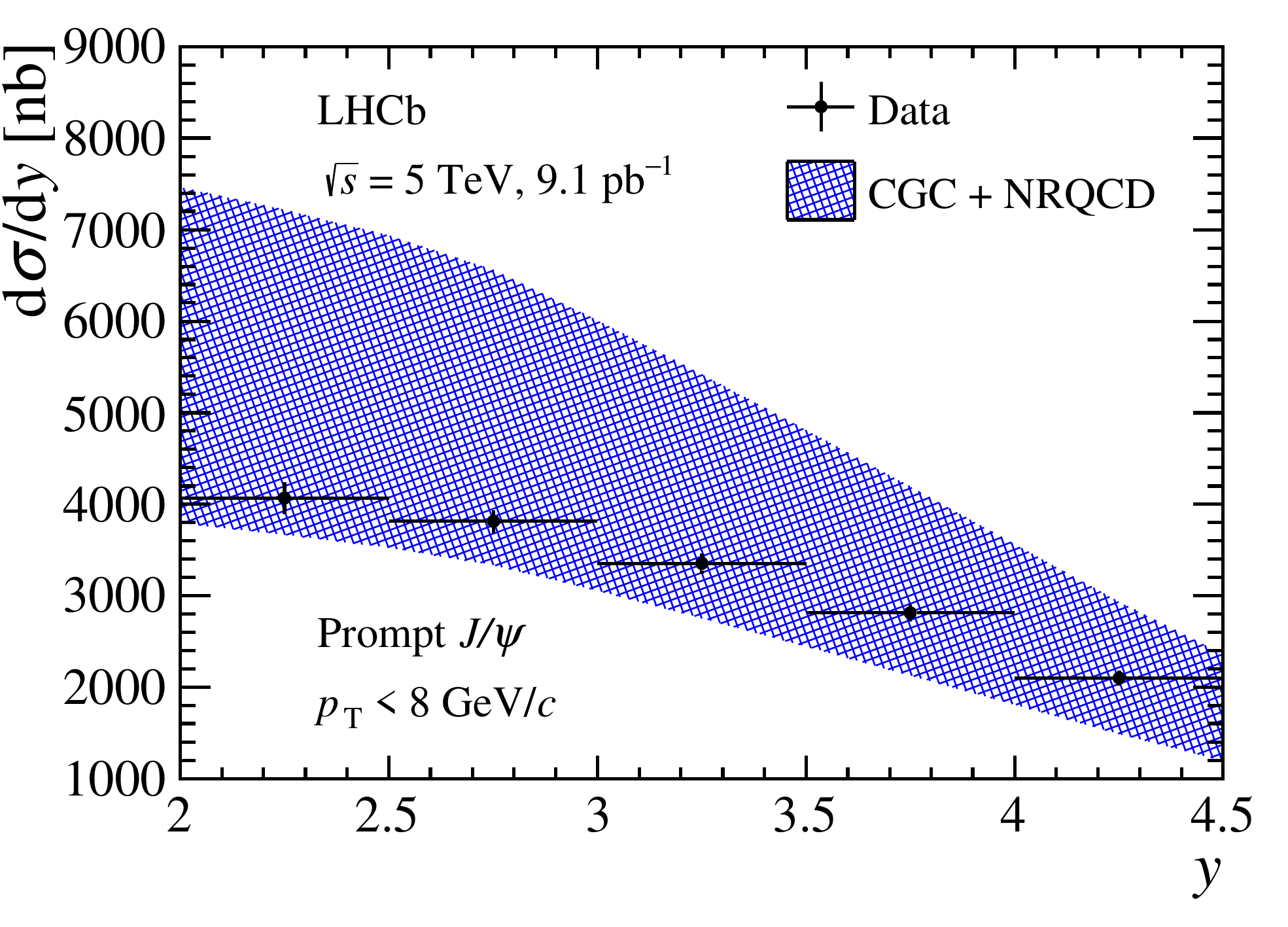}
    \end{minipage}
    \hfil
	\caption{Differential cross-sections (left) $\deriv\sigma/\deriv\pt$ and (right) $\deriv\sigma/\deriv y$ for prompt \jpsi mesons compared with NRQCD and CGC calculations~\cite{Ma:2010yw,Ma:2014mri}. Uncertainties due to LDMEs determination, renormalisation scales, and factorisation scales are included in the NRQCD and CGC predictions.}
    \label{fig:csPrompt}
\end{figure}

The single-differential cross-sections for prompt \jpsi mesons
are compared with NRQCD calculations and colour glass condensate (CGC) effective theory results, as shown in Fig.~\ref{fig:csPrompt}.
Theoretical calculations in the high \pt region are obtained from the NLO NRQCD model with LDMEs fixed from the Tevatron data~\cite{Ma:2010yw}, 
and those in the low-\pt region are obtained by combining the NRQCD model with CGC effective theory~\cite{Ma:2014mri}, in which nonperturbative parameters are fixed by fits to the Tevatron~\cite{Chao:2012iv} and HERA~\cite{Albacete:2012xq} data.
Uncertainties due to LDMEs determination, renormalisation scales, and factorisation scales are considered for the NRQCD and CGC calculations.

\begin{figure}[tb]
    \centering
    \hfil
    \begin{minipage}[t]{0.49\linewidth}
        \centering
        \includegraphics[width=\linewidth]{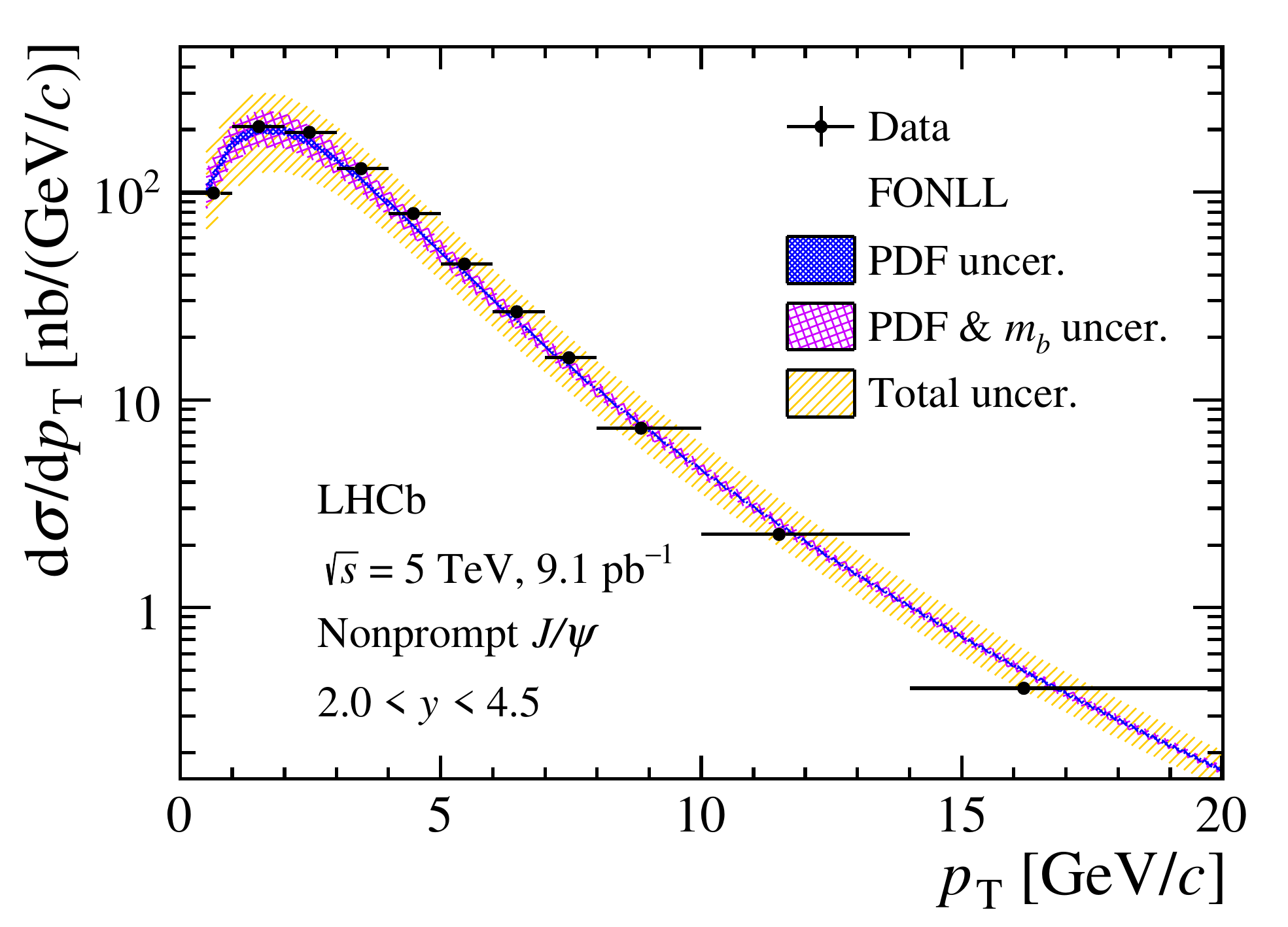}
    \end{minipage}
    \begin{minipage}[t]{0.49\linewidth}
        \centering
        \includegraphics[width=\linewidth]{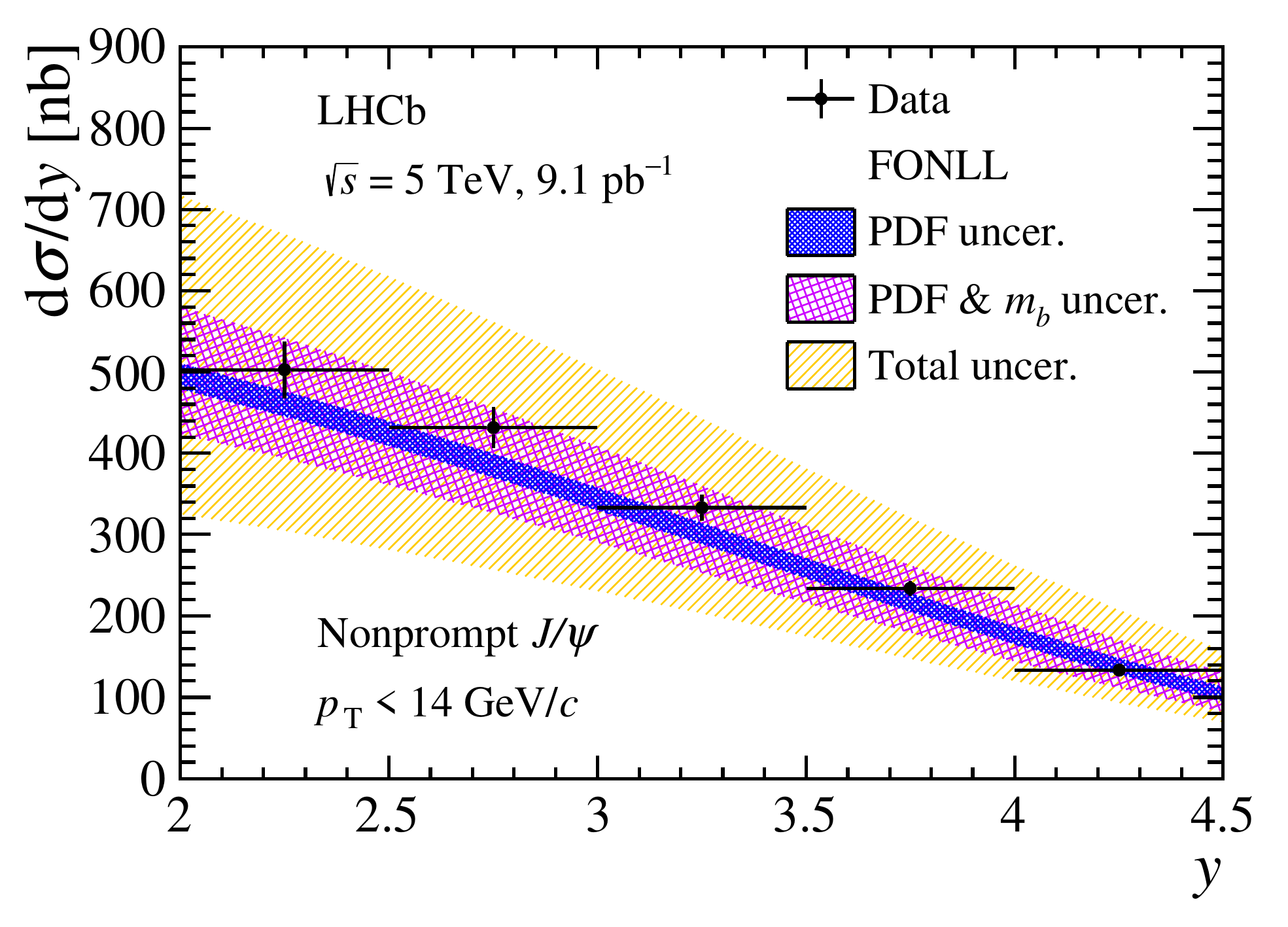}
    \end{minipage}
    \hfil
	\caption{Differential cross-sections (left) $\deriv\sigma/\deriv\pt$ and (right) $\deriv\sigma/\deriv y$ for nonprompt \jpsi mesons compared with FONLL calculations~\cite{Cacciari:2012ny,Cacciari:2015fta}. The orange band shows the total FONLL calculation uncertainty; the violet band shows the uncertainties of PDFs and that due to \bquark-quark mass added in quadrature; the blue band shows only the uncertainties on PDFs.}
    \label{fig:csFromb}
\end{figure}

A comparison between single-differential cross-sections for nonprompt \jpsi mesons and fixed order plus next-to-leading logarithms (FONLL) calculations~\cite{Cacciari:2012ny,Cacciari:2015fta} is shown in Fig.~\ref{fig:csFromb}.
The FONLL approach provides cross-sections for \bquark-quark production, and the branching fraction of the decay $\bquark\to\jpsi X$, $(1.16\pm0.10)\%$~\cite{PDG2020}, is taken from measurements performed in \epem collisions at LEP.
The FONLL calculations take into account the uncertainties of parton distribution functions (PDFs), the uncertainty due to the \bquark-quark mass, and that due to the scales of renormalisation and factorisation.
The total uncertainty of FONLL is dominated by the latter source.


The fraction of nonprompt \jpsi mesons is defined as the ratio between the nonprompt cross-section and the sum of prompt and nonprompt cross-sections, and the results in (\pt,$y$) intervals are shown in Fig.~\ref{fig:frombfrac} and Table~\ref{table:frombfrac} in Appendix~\ref{sec:Tables}.
\begin{figure}[tb]
    \centering
    \includegraphics[width=0.6\linewidth]{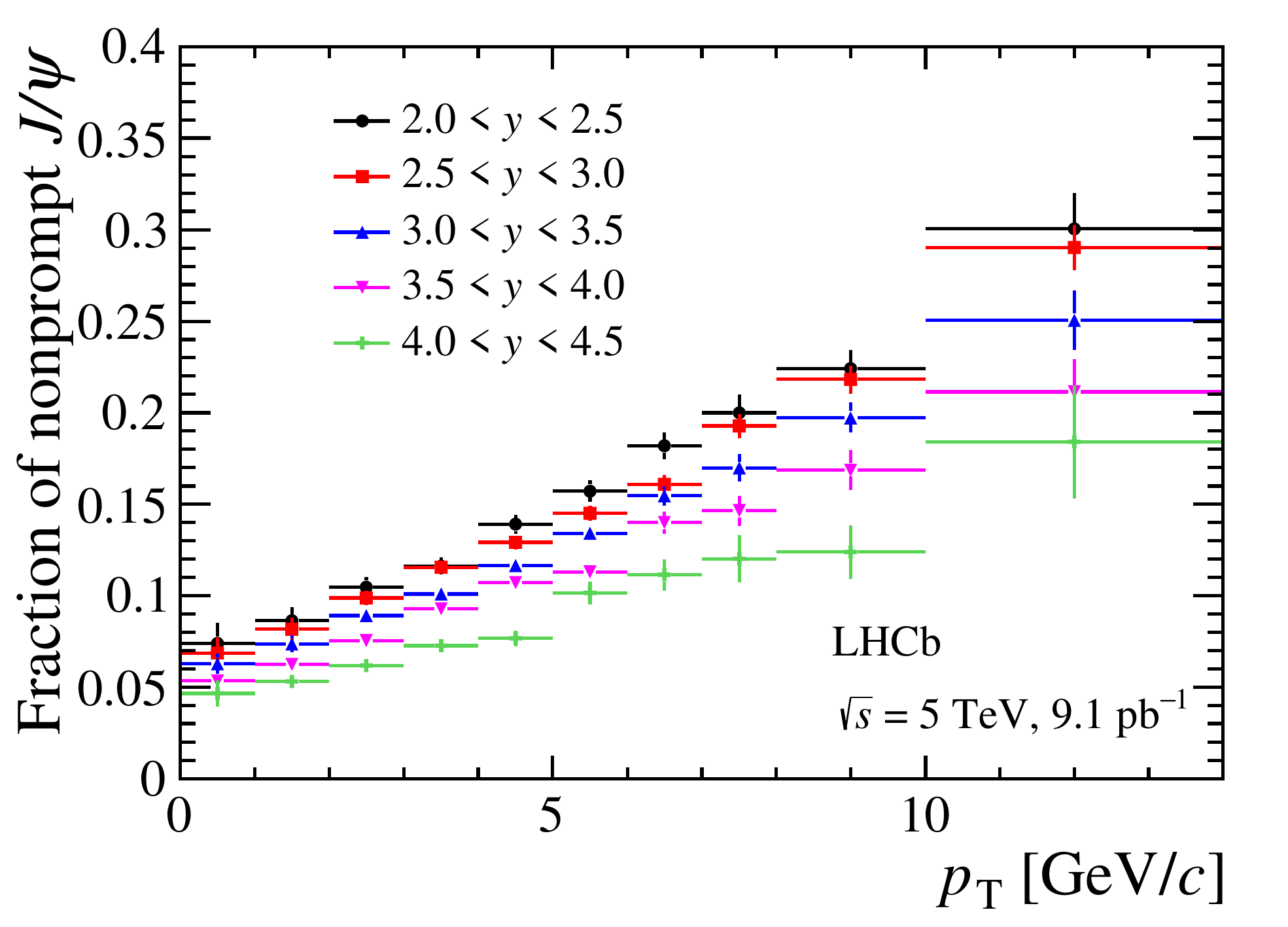}
    \caption{Fraction of nonprompt \jpsi mesons as a function of \pt in intervals of $y$. The error bars represent the total uncertainties, which are partially correlated between intervals.}
    \label{fig:frombfrac}
\end{figure}
Most systematic uncertainties cancel in the ratio.
Only the uncertainties due to the $t_z$ fit and the size of simulated sample are included.
The fraction increases as a function of \pt.
For a given \pt, the fraction decreases with increasing $y$.

\begin{figure}[tb]
    \centering
    \hfil
    \begin{minipage}[t]{0.49\linewidth}
        \centering
        \includegraphics[width=\linewidth]{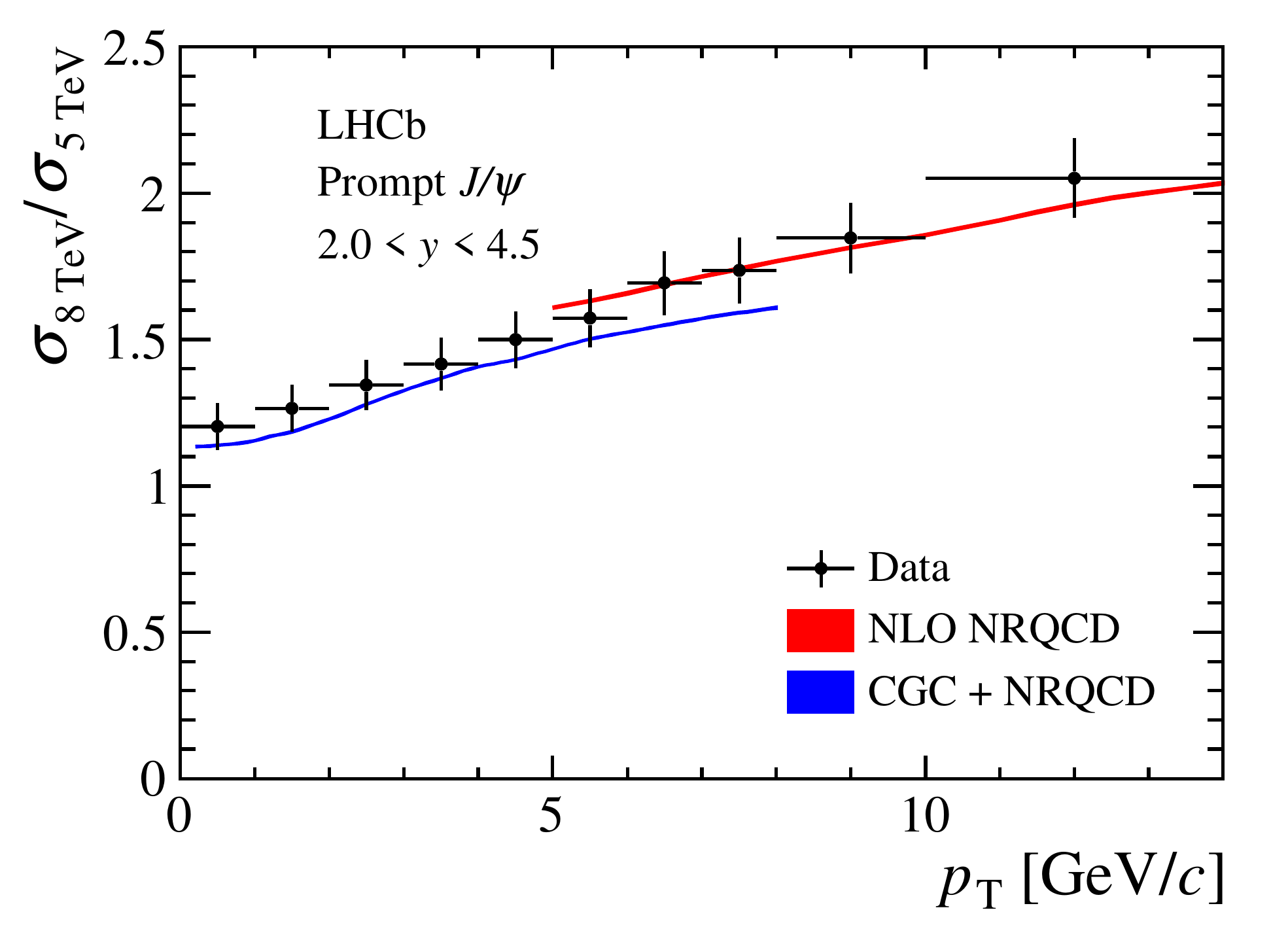}
    \end{minipage}
    \begin{minipage}[t]{0.49\linewidth}
        \centering
        \includegraphics[width=\linewidth]{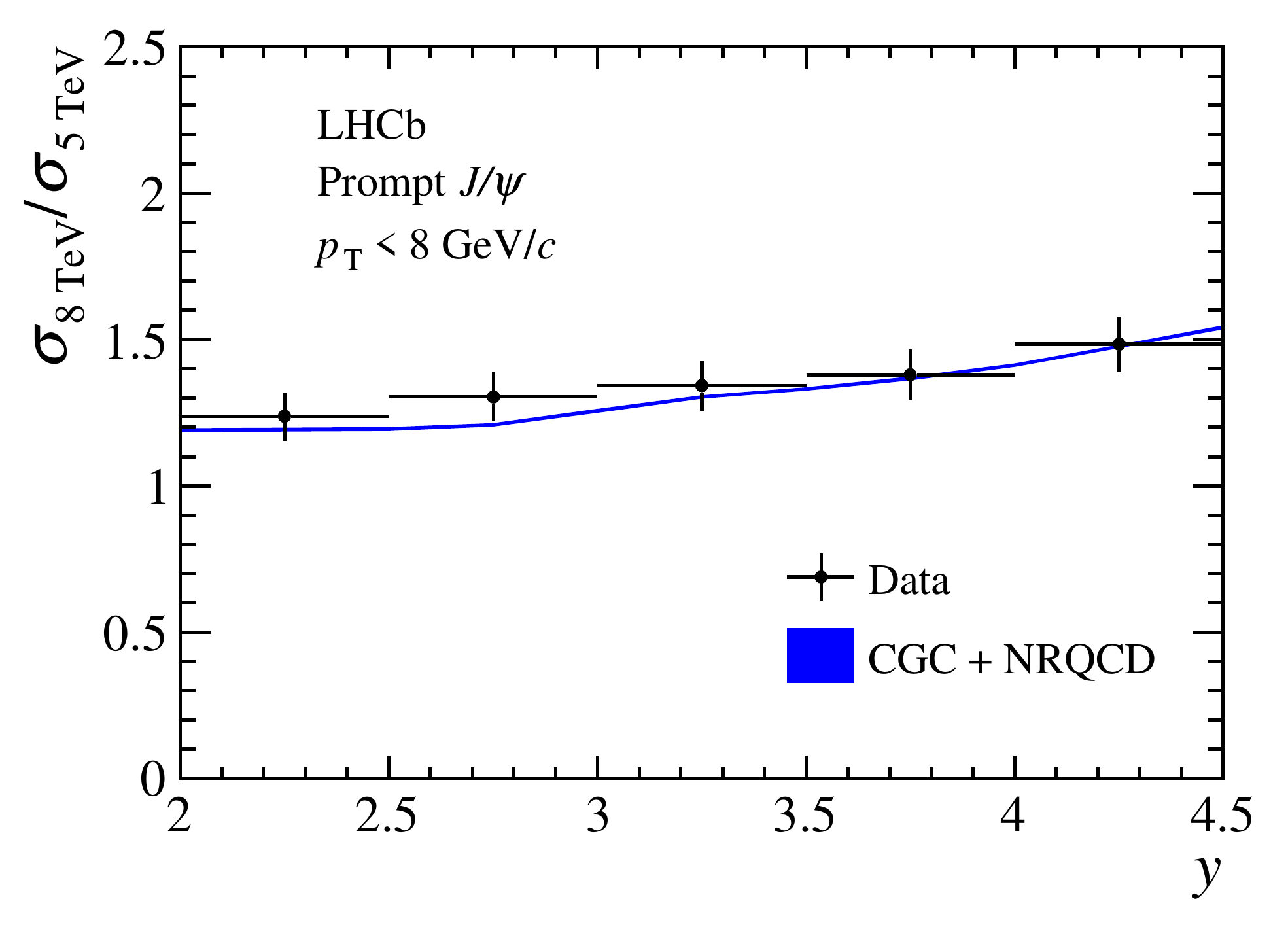}
    \end{minipage}
    \hfil
	\caption{Ratios of differential cross-sections between 8\tev and 5\tev measurements as a function of (left) \pt and (right) $y$ for prompt \jpsi mesons compared with NRQCD and CGC calculations~\cite{Ma:2010yw,Ma:2014mri}. Uncertainties due to the LDMEs determination, renormalisation scales, and factorisation scales are included in the NRQCD and CGC calculations.}
    \label{fig:R8o5Prompt}
\end{figure}

\begin{figure}[tb]
    \centering
    \hfil
    \begin{minipage}[t]{0.49\linewidth}
        \centering
        \includegraphics[width=\linewidth]{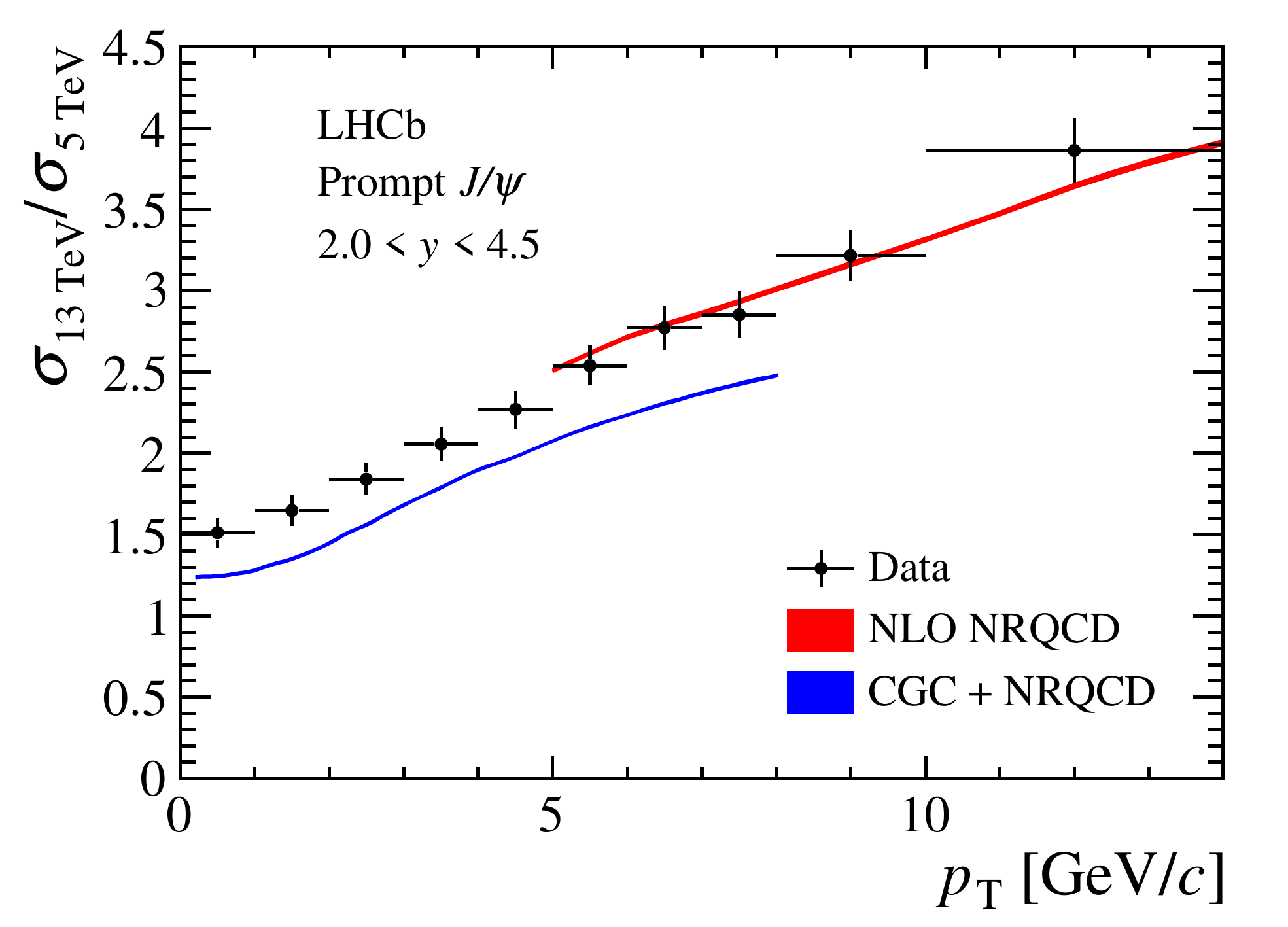}
    \end{minipage}
    \begin{minipage}[t]{0.49\linewidth}
        \centering
        \includegraphics[width=\linewidth]{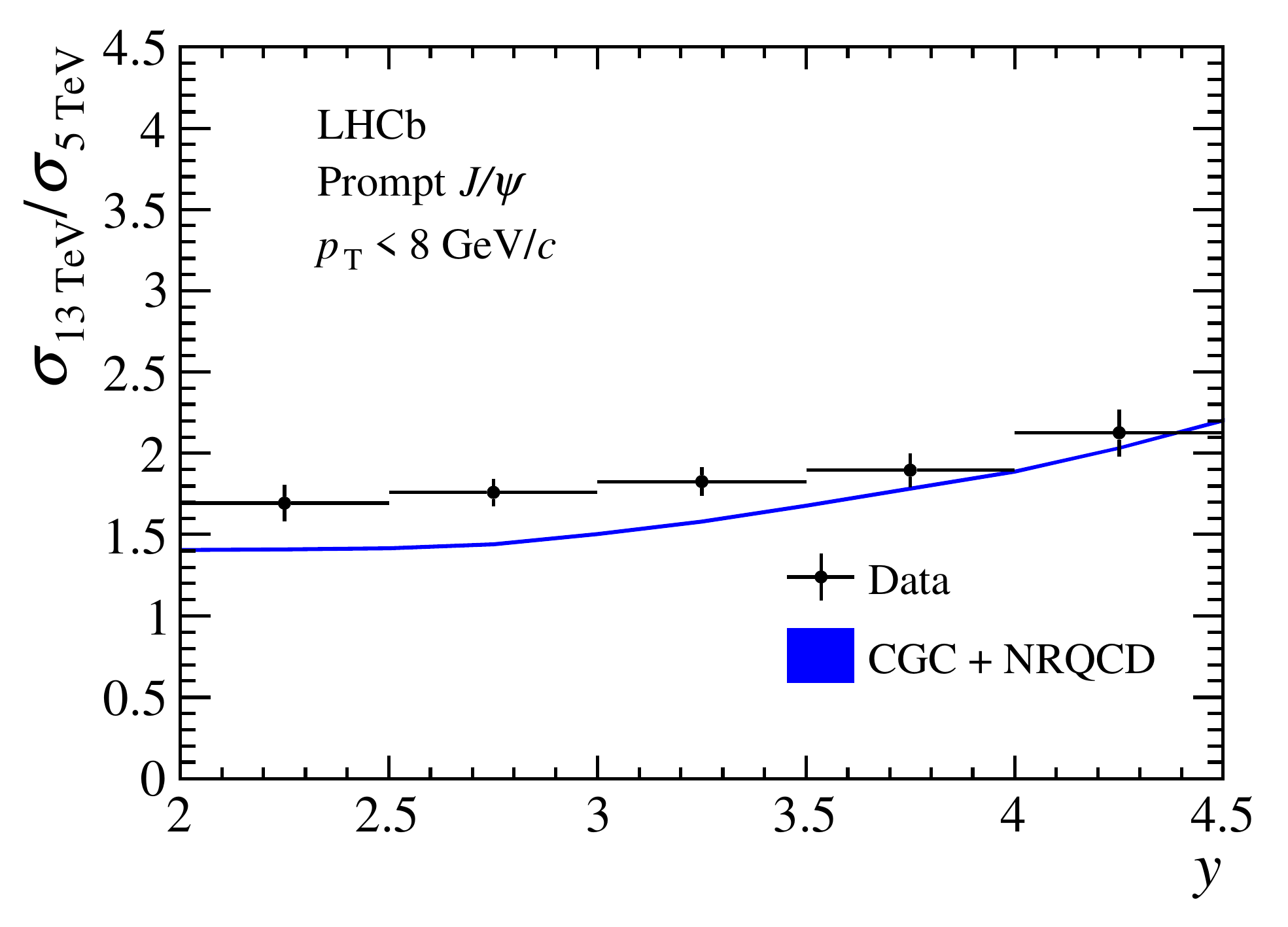}
    \end{minipage}
    \hfil
	\caption{Ratios of differential cross-sections between 13\tev and 5\tev measurements as a function of (left) \pt and (right) $y$ for prompt \jpsi mesons compared with NRQCD and CGC calculations~\cite{Ma:2010yw,Ma:2014mri}. Uncertainties due to the LDMEs determination, renormalisation scales, and factorisation scales are included in the NRQCD and CGC calculations.}
    \label{fig:R13o5Prompt}
\end{figure}

\begin{figure}[tb]
    \centering
    \hfil
    \begin{minipage}[t]{0.49\linewidth}
        \centering
        \includegraphics[width=\linewidth]{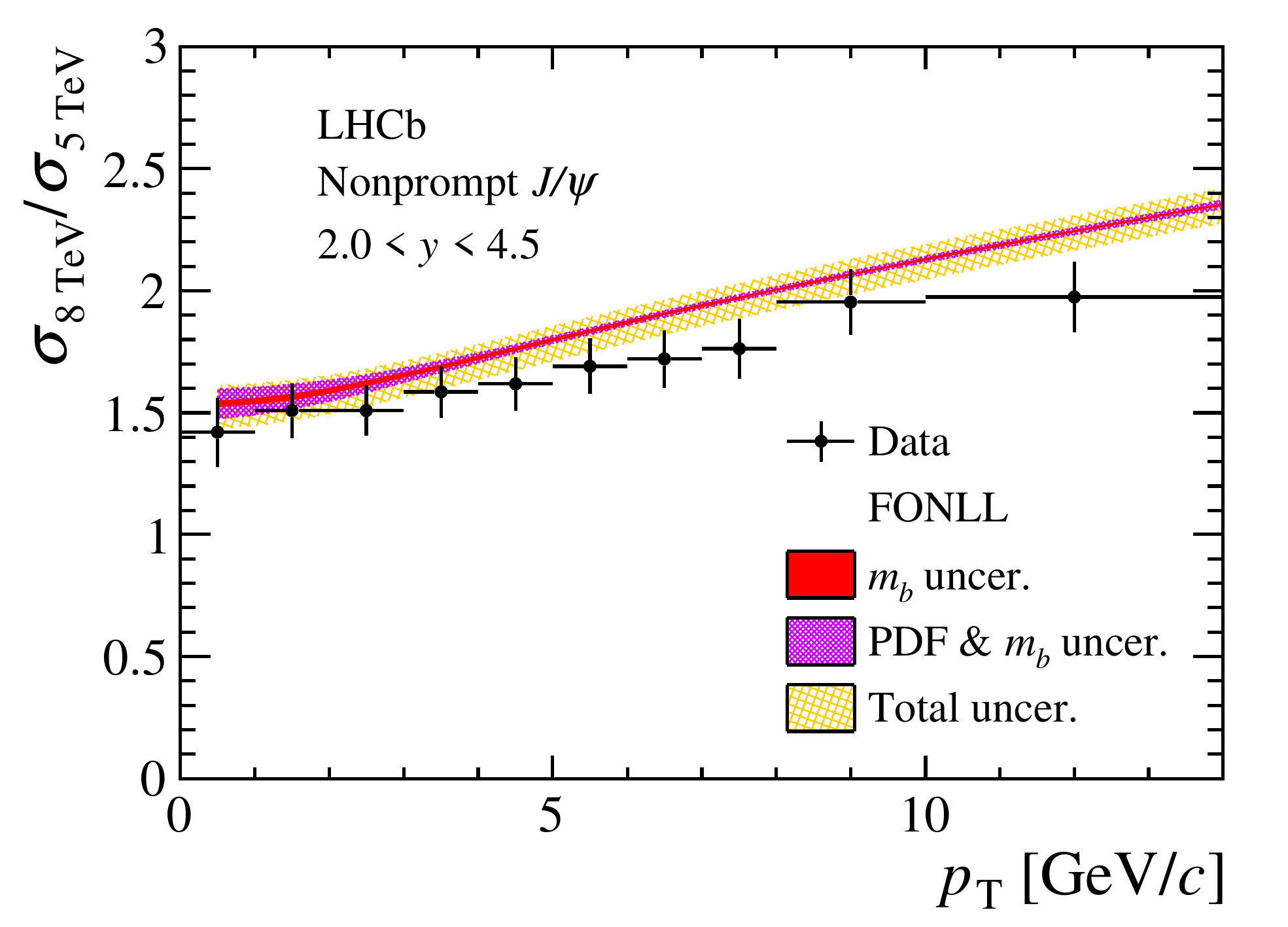}
    \end{minipage}
    \begin{minipage}[t]{0.49\linewidth}
        \centering
        \includegraphics[width=\linewidth]{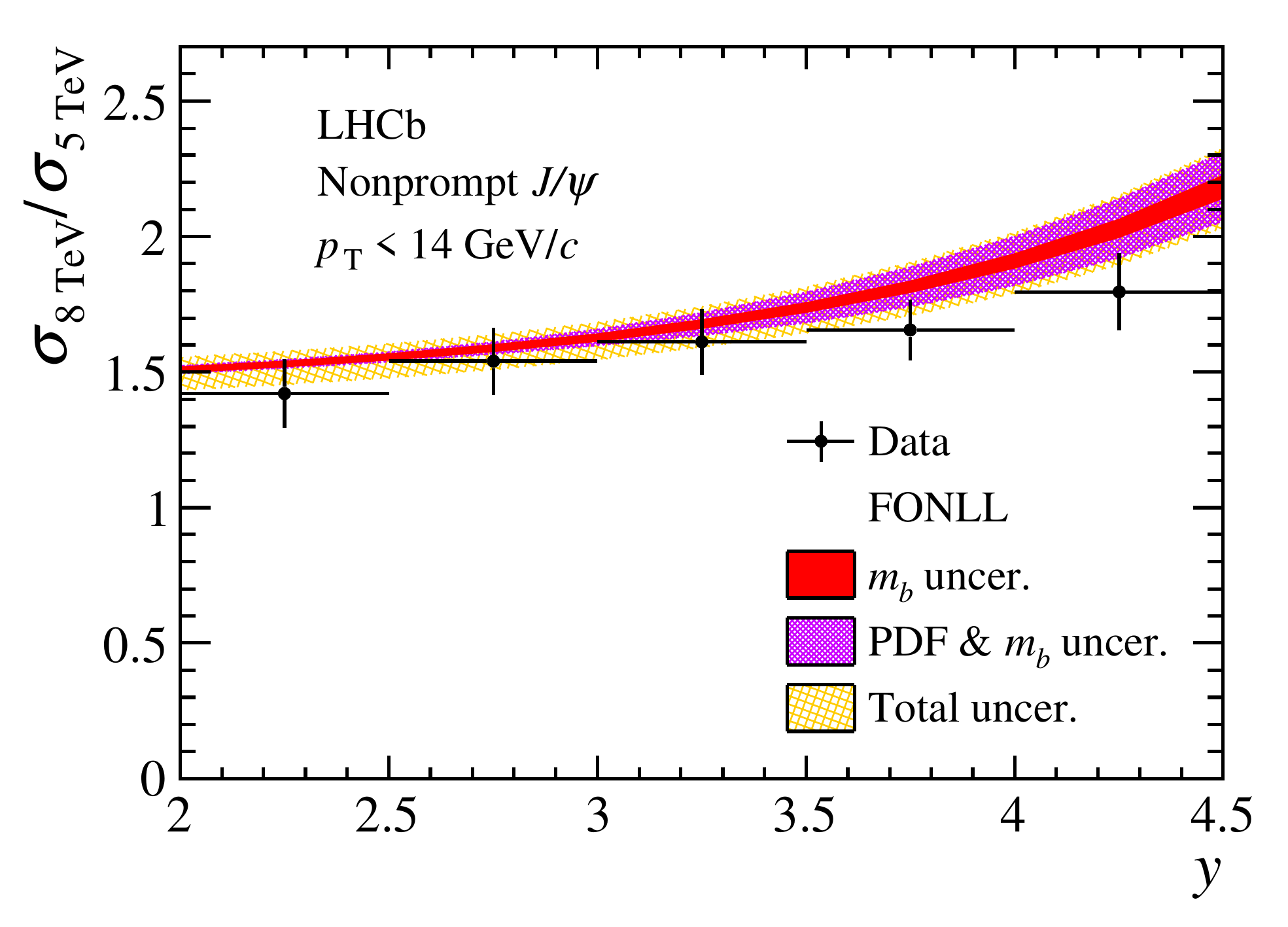}
    \end{minipage}
    \hfil
	\caption{Ratios of differential cross-sections between 8\tev and 5\tev measurements as a function of (left) \pt and (right) $y$ for nonprompt \jpsi mesons compared with FONLL calculations~\cite{Cacciari:2012ny,Cacciari:2015fta}. The orange band shows the total FONLL calculation uncertainty; the violet band shows the uncertainties on PDFs and that due to \bquark-quark mass added in quadrature; the red band shows only the uncertainty due to the \bquark-quark mass.}
    \label{fig:R8o5Fromb}
\end{figure}

\begin{figure}[tb]
    \centering
    \hfil
    \begin{minipage}[t]{0.49\linewidth}
        \centering
        \includegraphics[width=\linewidth]{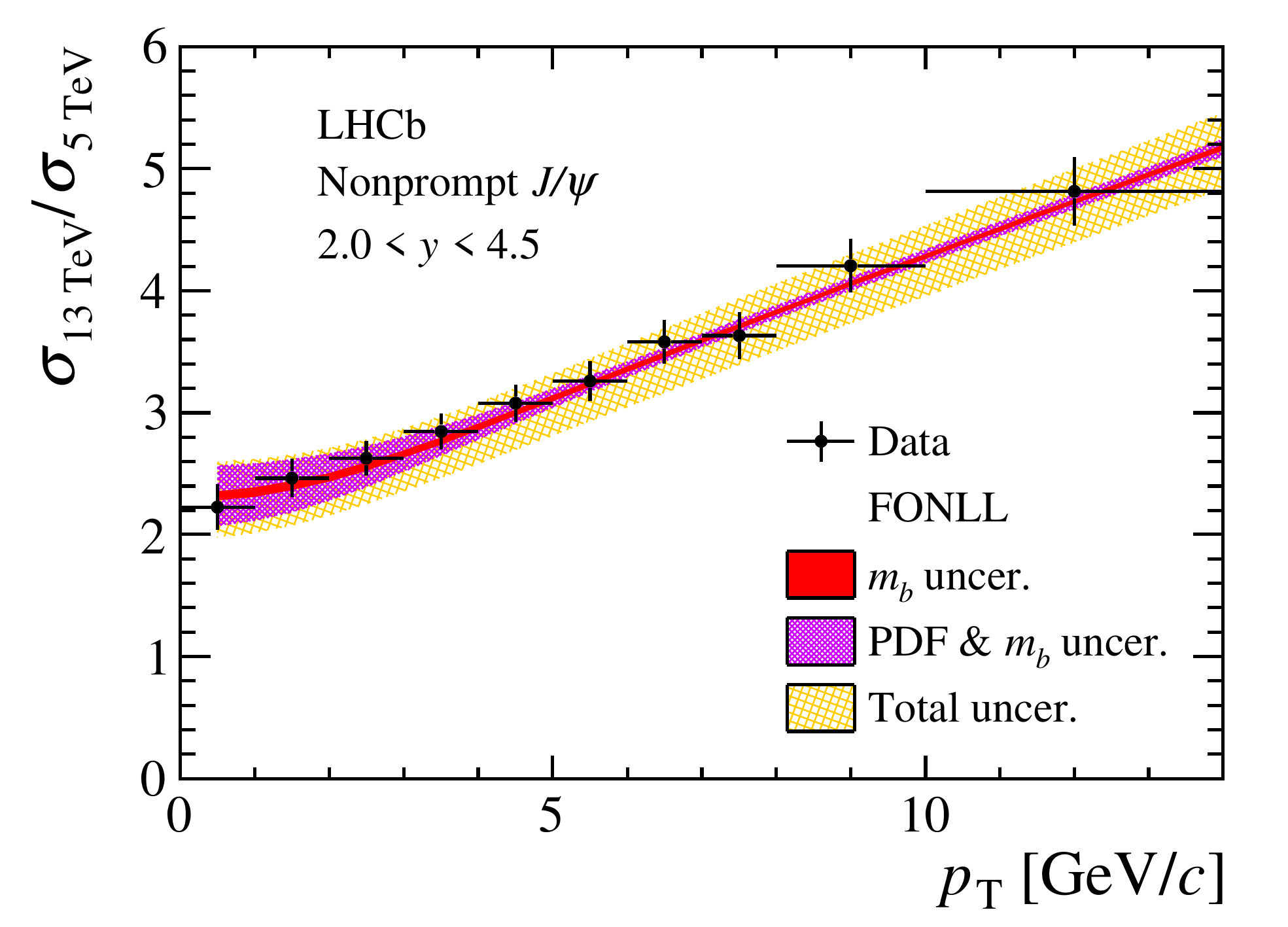}
    \end{minipage}
    \begin{minipage}[t]{0.49\linewidth}
        \centering
        \includegraphics[width=\linewidth]{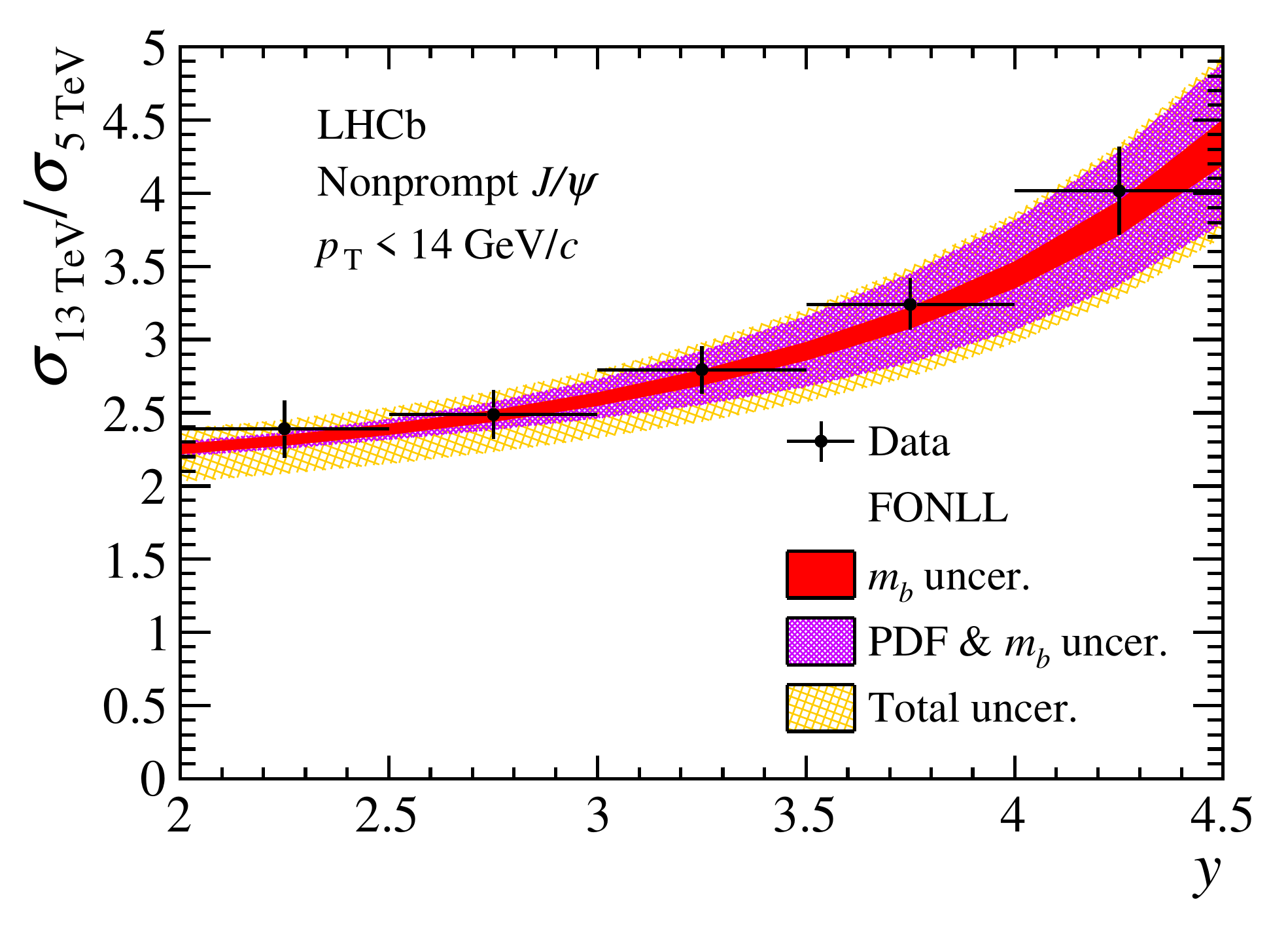}
    \end{minipage}
    \hfil
	\caption{Ratios of differential cross-sections between 13\tev and 5\tev measurements as a function of (left) \pt and (right) $y$ for nonprompt \jpsi mesons compared with FONLL calculations~\cite{Cacciari:2012ny,Cacciari:2015fta}. The orange band shows the total FONLL calculation uncertainty; the violet band shows the uncertainties on PDFs and that due to \bquark-quark mass added in quadrature; the red band shows only the uncertainty due to \bquark-quark mass.}
    \label{fig:R13o5Fromb}
\end{figure}

The production cross-sections of \jpsi mesons at 5\tev are compared with those previously measured at 8\tev~\cite{LHCb-PAPER-2013-016} and 13\tev~\cite{LHCb-PAPER-2015-037} in the range $0<\pt<14\gevc$ and $2.0<y<4.5$.
The ratios of differential cross-sections for prompt \jpsi mesons between 8\tev and 5\tev measurements are shown in Fig.~\ref{fig:R8o5Prompt}, and those between 13\tev and 5\tev in Fig.~\ref{fig:R13o5Prompt}, both compared with NRQCD and CGC calculations.
For nonprompt \jpsi mesons, the ratios of differential cross-sections between 8\tev and 5\tev measurements are shown in Fig.~\ref{fig:R8o5Fromb}, and those between 13\tev and 5\tev in Fig.~\ref{fig:R13o5Fromb}, compared with FONLL calculations.
Some of the systematic uncertainties are considered to fully cancel in the ratio, such as those due to branching fraction and the radiative tail.
The uncertainties due to the $t_z$ fit and simulation sample size are taken as uncorrelated between the two measurements, and therefore remain.
All other systematic uncertainties are assumed to cancel only partially.
For example, the systematic uncertainty due to the luminosity measurement is estimated to be correlated at 50\%.
The overall uncertainty on the measured ratio is dominated by the luminosity measurement for prompt \jpsi mesons, and by the $t_z$ fit and the luminosity measurement for nonprompt \jpsi mesons.
For the NRQCD and CGC estimates of the cross-section ratios, the uncertainties due to LDMEs determination, renormalisation scales, and factorisation scales between different energies mostly cancel.
For the FONLL calculations, the uncertainty on the ratio is dominated by the uncertainties of PDFs for the low-\pt and large-$y$ regions and by the uncertainty due to the scales of the renormalisation and factorisation for the high-\pt and small-$y$ regions.
Figures~\ref{fig:R8o5Prompt} and \ref{fig:R13o5Prompt} show good agreement between NLO NRQCD calculations and the measurement results in the high-\pt region.
The inclusion of CGC effects achieves a reasonable agreement between data and theory in the low-\pt region but a small discrepancy is still observed, which indicates that a pure fixed-order calculation may be insufficient and Sudakov resummation~\cite{Contopanagos:1996nh} may be required.
A comparison of Figs.~\ref{fig:R8o5Prompt} and~\ref{fig:R13o5Prompt} suggests that the energy dependence of the cross-sections may differ between the  theoretical calculation and the experimental measurements.
Figures~\ref{fig:R8o5Fromb} and \ref{fig:R13o5Fromb} show that the FONLL calculations agree with the experimental results for nonprompt \jpsi mesons.

\section{Nuclear modification factor}
\label{sec:Rplead}

The production cross-sections in $pp$ collisions are essential inputs for the study of nuclear effects in collisions involving heavy ions.
Nuclear effects are usually characterized by the nuclear modification factor.
In proton-lead ($p$Pb) collisions, this factor, $R_{p\text{Pb}}$, is defined as the production cross-section of a given particle per nucleon pair in $p$Pb collisions divided by that in $pp$ collisions.
The previous $R_{p\text{Pb}}$ measurement performed by the LHCb collaboration at a centre-of-mass energy per nucleon pair of $\sqsnn=5\tev$~\cite{LHCb-PAPER-2013-052} made use of \jpsi production cross-sections in $pp$ collisions at 5\tev derived from an interpolation of LHCb measurements at 2.76, 7 and $8\tev$~\cite{LHCb-PAPER-2012-039,LHCb-PAPER-2011-003,LHCb-PAPER-2013-016} using a power-law fit.
Based on the direct measurement presented in this paper, the nuclear modification factor $R_{p\text{Pb}}$ is updated.
In the data taking of $p$Pb collisions, two distinct beam configurations are used, $p$Pb and Pb$p$.
In the $p$Pb configuration particles produced in the direction of the proton beam are analysed, while in the Pb$p$ configuration particles are analysed in the Pb beam direction.
Rapidity $y$ is defined in the nucleon-nucleon rest frame, and the coverage at LHCb is $1.5<y<4.0$ ($-5.0<y<-2.5$) in the $p$Pb (Pb$p$) configuration.

\begin{figure}[tb]
    \centering
    \hfil
    \begin{minipage}[t]{0.49\linewidth}
        \centering
        \includegraphics[width=\linewidth]{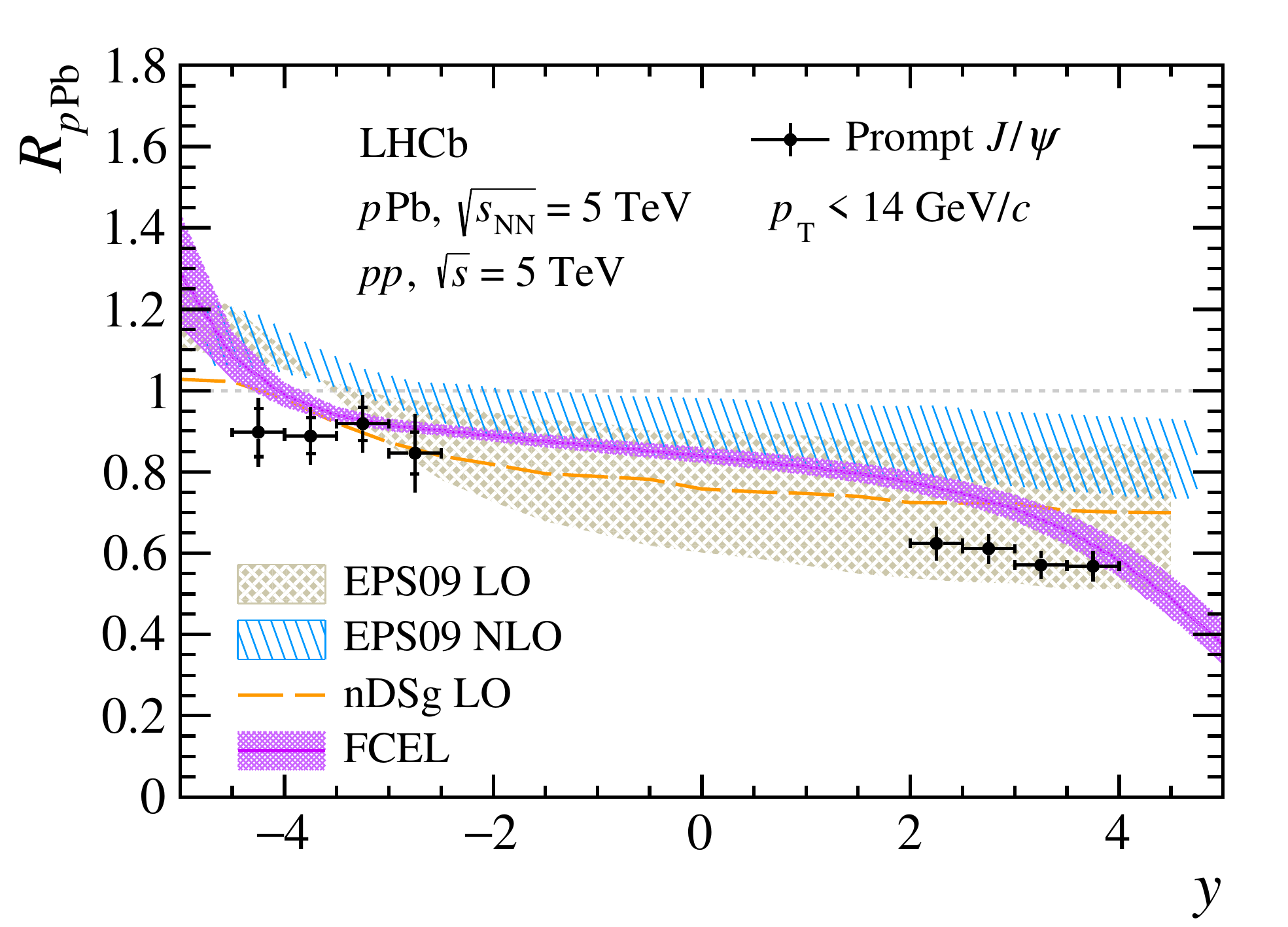}
    \end{minipage}
    \begin{minipage}[t]{0.49\linewidth}
        \centering
        \includegraphics[width=\linewidth]{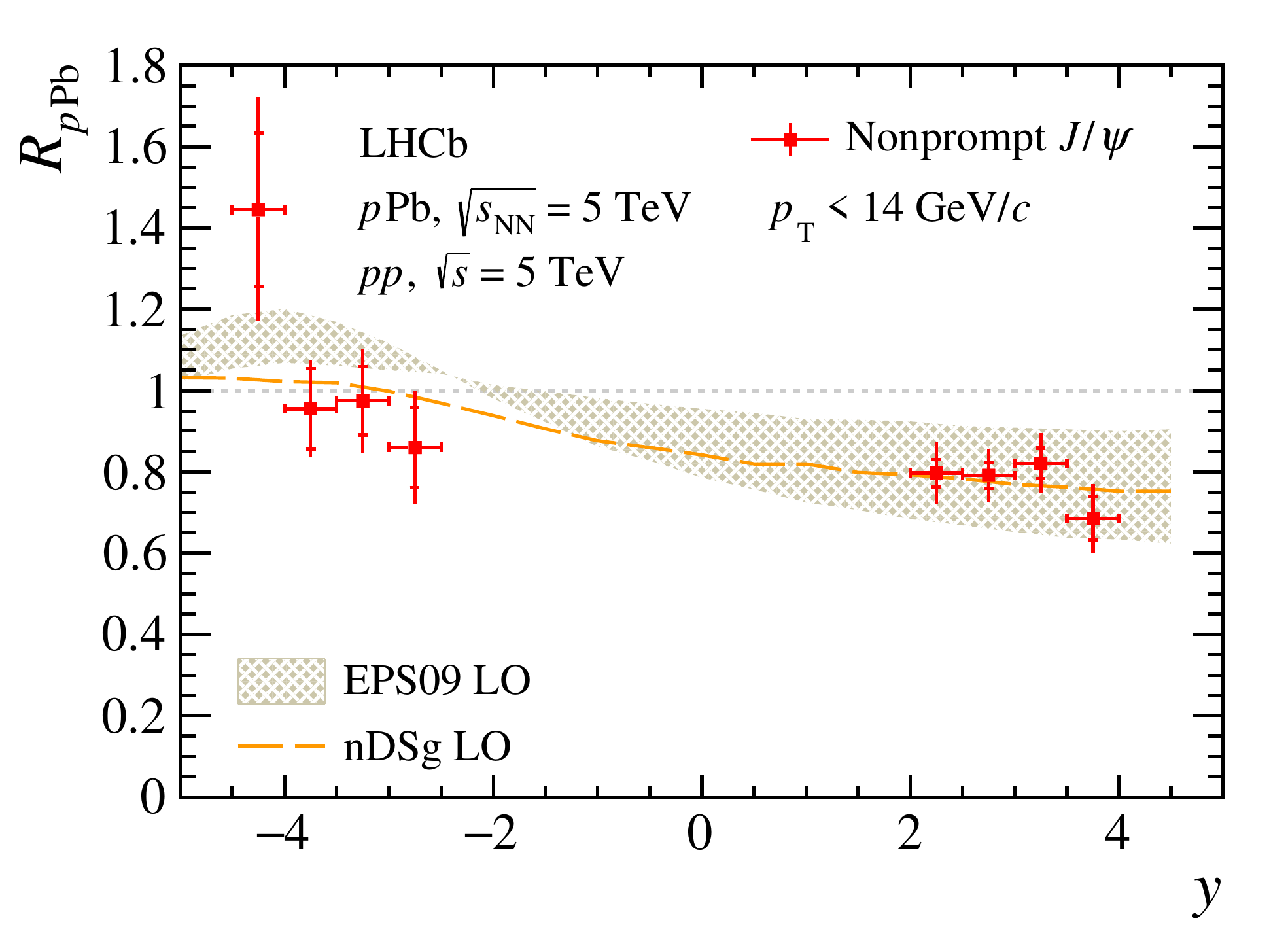}
    \end{minipage}
    \hfil
	\caption{Nuclear modification factor $R_{p\text{Pb}}$ as a function of $y$ for (left) prompt and (right) nonprompt \jpsi mesons, together with the theoretical predictions from (yellow dashed line and brown band) Ref.~\cite{Ferreiro:2013pua}, (blue band) Ref.~\cite{Albacete:2013ei}, and (violet solid line with band) Ref.~\cite{Arleo:2012rs,Arleo:2012hn}. In the data points the full error bars represent the statistical and systematic uncertainties added in quadrature, while the smaller ones represent the statistical uncertainties.}
    \label{fig:Rplead}
\end{figure}

The $R_{p\text{Pb}}$ values, as a function of $y$ in the range $\pt<14\gevc$, for prompt and nonprompt \jpsi mesons, are shown in Fig.~\ref{fig:Rplead} along with several theoretical predictions. The values are also
listed in Table~\ref{table:Rplead} in Appendix~\ref{sec:Tables}.
The predictions are obtained with the nDSg LO nuclear parton distribution function (nPDF) parameterisation~\cite{Ferreiro:2013pua}, the EPS09 LO nPDF parameterisation~\cite{Ferreiro:2013pua}, and the EPS09 NLO nPDF parameterisation~\cite{Albacete:2013ei}, and from the fully coherent energy loss (FCEL) model~\cite{Arleo:2012hn}.
Conservatively, the systematic uncertainties are assumed to be uncorrelated between the results obtained in $p$Pb and $pp$ collisions.
For prompt \jpsi mesons, the measurement agrees with most theoretical calculations, while the calculation with the EPS09 NLO nPDF parameterisation~\cite{Albacete:2013ei} gives a poorer description.
The comparison for nonprompt \jpsi mesons shows good agreement.

\section{Conclusion}
\label{sec:Conclusion}
The \jpsi production cross-sections in proton-proton collisions at a centre-of-mass energy $\sqs=5\tev$ are studied using a data sample corresponding to an integrated luminosity of $9.18\pm0.35\invpb$
collected by the \lhcb detector.
The \jpsi differential cross-sections, as a function of \pt and $y$, are measured separately for prompt and nonprompt \jpsi mesons in the range $0<\pt<20\gevc$ and $2.0<y<4.5$.
The \jpsi production cross-section ratios between 8\tev and 5\tev, and between 13\tev and 5\tev are also determined and compared with the theory models.
The measured prompt \jpsi results are in good agreement with NLO NRQCD calculations in the high-\pt region.
A small tension is observed between data for prompt \jpsi in the low-\pt region and NRQCD and CGC calculations, which may indicate the need for further corrections in the theory model.
The FONLL calculations describe the measured results for nonprompt \jpsi mesons well.
The nuclear modification factor in proton-lead collisions for \jpsi mesons at a centre-of-mass energy per nucleon pair of $\sqsnn=5\tev$ is updated and supersedes that in the previous publication~\cite{LHCb-PAPER-2013-052}.


\section*{Acknowledgements}
%
%
\noindent We express our gratitude to our colleagues in the CERN
accelerator departments for the excellent performance of the LHC. We
thank the technical and administrative staff at the LHCb
institutes.
We acknowledge support from CERN and from the national agencies:
CAPES, CNPq, FAPERJ and FINEP (Brazil); 
MOST and NSFC (China); 
CNRS/IN2P3 (France); 
BMBF, DFG and MPG (Germany); 
INFN (Italy); 
NWO (Netherlands); 
MNiSW and NCN (Poland); 
MEN/IFA (Romania); 
MSHE (Russia); 
MICINN (Spain); 
SNSF and SER (Switzerland); 
NASU (Ukraine); 
STFC (United Kingdom); 
DOE NP and NSF (USA).
We acknowledge the computing resources that are provided by CERN, IN2P3
(France), KIT and DESY (Germany), INFN (Italy), SURF (Netherlands),
PIC (Spain), GridPP (United Kingdom), RRCKI and Yandex
LLC (Russia), CSCS (Switzerland), IFIN-HH (Romania), CBPF (Brazil),
PL-GRID (Poland) and NERSC (USA).
We are indebted to the communities behind the multiple open-source
software packages on which we depend.
Individual groups or members have received support from
ARC and ARDC (Australia);
AvH Foundation (Germany);
EPLANET, Marie Sk\l{}odowska-Curie Actions and ERC (European Union);
A*MIDEX, ANR, IPhU and Labex P2IO, and R\'{e}gion Auvergne-Rh\^{o}ne-Alpes (France);
Key Research Program of Frontier Sciences of CAS, CAS PIFI, CAS CCEPP, 
Fundamental Research Funds for the Central Universities, 
and Sci. \& Tech. Program of Guangzhou (China);
RFBR, RSF and Yandex LLC (Russia);
GVA, XuntaGal and GENCAT (Spain);
the Leverhulme Trust, the Royal Society
 and UKRI (United Kingdom).

\newcommand{\xx}{\ensuremath{\kern 0.5em }}
\newcommand{\xxx}{\ensuremath{\kern 0.75em }}
\clearpage
\section*{Appendices}

\appendix

\section{Result tables}
\label{sec:Tables}

\begin{table}[h]
    \centering
    \caption{Double-differential production cross-sections $\frac{\deriv^2\sigma}{\deriv \pt\deriv y}$[nb/(\ensuremath{\text{Ge\kern -0.1em V\!/}c}) per unit rapidity] for prompt \jpsi mesons in (\pt,$y$) intervals. The first uncertainties are statistical, the second are correlated systematic uncertainties shared between intervals, the third are uncorrelated systematic uncertainties, and the last are correlated between \pt intervals and uncorrelated between $y$ intervals.}
    \label{table:csPrompt}
    \begin{tabular}{l|ll}
\hline\pt[\gevc]&$2.0<y<2.5$&$2.5<y<3.0$\\\hline
$0-1$&$\xx686.03\pm6.80\pm35.11\pm\xx9.03\pm4.92$&$\xx640.03\pm3.70\pm22.97\pm2.78\pm5.06$\\
$1-2$&$1253.23\pm8.66\pm55.59\pm15.31\pm9.82$&$1173.73\pm4.95\pm39.14\pm4.68\pm7.58$\\
$2-3$&$\xx964.82\pm6.79\pm40.41\pm12.19\pm5.23$&$\xx917.24\pm4.09\pm30.11\pm2.86\pm3.76$\\
$3-4$&$\xx575.26\pm4.74\pm21.54\pm\xx7.51\pm2.89$&$\xx540.97\pm2.82\pm17.21\pm2.52\pm1.80$\\
$4-5$&$\xx305.38\pm2.89\pm10.65\pm\xx2.93\pm1.28$&$\xx286.82\pm1.76\pm\xx8.98\pm1.74\pm1.10$\\
$5-6$&$\xx159.48\pm1.84\pm\xx5.44\pm\xx1.51\pm0.72$&$\xx143.21\pm1.12\pm\xx4.46\pm1.11\pm0.49$\\
$6-7$&$\xx\xx79.23\pm1.19\pm\xx2.63\pm\xx1.02\pm0.26$&$\xx\xx73.00\pm0.75\pm\xx2.26\pm0.66\pm0.25$\\
$7-8$&$\xx\xx43.60\pm0.82\pm\xx1.43\pm\xx0.77\pm0.01$&$\xx\xx36.67\pm0.52\pm\xx1.13\pm0.32\pm0.00$\\
$8-10$&$\xx\xx18.11\pm0.34\pm\xx0.59\pm\xx0.24\pm0.08$&$\xx\xx15.17\pm0.23\pm\xx0.47\pm0.15\pm0.04$\\
$10-14$&$\xx\xx\xx4.15\pm0.11\pm\xx0.13\pm\xx0.07\pm0.02$&$\xx\xx\xx3.34\pm0.07\pm\xx0.11\pm0.04\pm0.01$\\
\hline\pt[\gevc]&$3.0<y<3.5$&$3.5<y<4.0$\\\hline
$0-1$&$\xx589.31\pm3.22\pm19.72\pm1.82\pm2.99$&$\xx515.07\pm2.78\pm17.29\pm1.72\pm0.30$\\
$1-2$&$1056.67\pm4.29\pm34.89\pm4.63\pm4.82$&$\xx911.86\pm3.68\pm30.29\pm2.93\pm0.63$\\
$2-3$&$\xx804.99\pm3.45\pm26.10\pm6.15\pm1.60$&$\xx681.00\pm3.04\pm22.56\pm1.99\pm1.78$\\
$3-4$&$\xx461.97\pm2.24\pm14.45\pm1.89\pm0.81$&$\xx370.70\pm2.08\pm11.71\pm1.36\pm0.55$\\
$4-5$&$\xx236.68\pm1.47\pm\xx7.28\pm0.94\pm0.29$&$\xx185.30\pm1.36\pm\xx5.81\pm1.72\pm0.32$\\
$5-6$&$\xx115.63\pm0.95\pm\xx3.55\pm1.17\pm0.17$&$\xx\xx86.71\pm0.86\pm\xx2.75\pm0.78\pm0.01$\\
$6-7$&$\xx\xx56.50\pm0.64\pm\xx1.73\pm0.42\pm0.03$&$\xx\xx41.09\pm0.57\pm\xx1.34\pm0.38\pm0.10$\\
$7-8$&$\xx\xx28.48\pm0.44\pm\xx0.87\pm0.28\pm0.10$&$\xx\xx20.85\pm0.39\pm\xx0.71\pm0.28\pm0.03$\\
$8-10$&$\xx\xx11.35\pm0.19\pm\xx0.35\pm0.11\pm0.03$&$\xx\xx\xx7.59\pm0.16\pm\xx0.27\pm0.11\pm0.01$\\
$10-14$&$\xx\xx\xx2.26\pm0.06\pm\xx0.08\pm0.04\pm0.00$&$\xx\xx\xx1.38\pm0.05\pm\xx0.05\pm0.03\pm0.01$\\
\hline\pt[\gevc]&$4.0<y<4.5$&\\\hline
$0-1$&$\xx452.31\pm3.21\pm17.49\pm2.85\pm2.79$&\\
$1-2$&$\xx731.38\pm4.04\pm27.48\pm3.52\pm0.74$&\\
$2-3$&$\xx485.71\pm3.25\pm19.15\pm3.16\pm0.50$&\\
$3-4$&$\xx240.13\pm2.17\pm\xx9.05\pm2.68\pm0.21$&\\
$4-5$&$\xx108.46\pm1.33\pm\xx4.16\pm1.18\pm0.25$&\\
$5-6$&$\xx\xx49.12\pm0.84\pm\xx1.90\pm0.59\pm0.08$&\\
$6-7$&$\xx\xx22.06\pm0.52\pm\xx0.86\pm0.35\pm0.02$&\\
$7-8$&$\xx\xx10.36\pm0.34\pm\xx0.40\pm0.24\pm0.03$&\\
$8-10$&$\xx\xx\xx3.82\pm0.14\pm\xx0.15\pm0.09\pm0.01$&\\
$10-14$&$\xx\xx\xx0.58\pm0.04\pm\xx0.02\pm0.02\pm0.00$&\\
\hline
\end{tabular}

\end{table}

\begin{table}[]
    \centering
    \caption{Double-differential production cross-sections $\frac{\deriv^2\sigma}{\deriv \pt\deriv y}$[nb/(\ensuremath{\text{Ge\kern -0.1em V\!/}c}) per unit rapidity] for nonprompt \jpsi mesons in (\pt,$y$) intervals. The first uncertainties are statistical, the second are correlated systematic uncertainties shared between intervals, the third are uncorrelated systematic uncertainties, and the last are correlated between \pt intervals and uncorrelated between $y$ intervals}
    \label{table:csFromb}
    \begin{tabular}{l|ll}
\hline\pt[\gevc]&$2.0<y<2.5$&$2.5<y<3.0$\\\hline
$0-1$&$\xx56.02\pm1.88\pm2.87\pm1.35\pm7.96$&$\xx47.72\pm1.00\pm1.71\pm0.48\pm5.71$\\
$1-2$&$120.67\pm2.48\pm5.35\pm1.88\pm9.89$&$105.37\pm1.37\pm3.51\pm2.07\pm7.81$\\
$2-3$&$117.89\pm2.23\pm4.94\pm2.46\pm5.15$&$101.93\pm1.26\pm3.35\pm1.12\pm3.66$\\
$3-4$&$\xx78.32\pm1.67\pm2.93\pm1.36\pm2.57$&$\xx71.20\pm0.97\pm2.27\pm0.52\pm1.68$\\
$4-5$&$\xx51.65\pm1.24\pm1.80\pm0.89\pm1.18$&$\xx42.83\pm0.69\pm1.34\pm0.41\pm0.99$\\
$5-6$&$\xx30.55\pm0.84\pm1.04\pm0.51\pm0.64$&$\xx24.58\pm0.48\pm0.77\pm0.30\pm0.43$\\
$6-7$&$\xx18.43\pm0.60\pm0.61\pm0.39\pm0.22$&$\xx14.39\pm0.36\pm0.45\pm0.22\pm0.22$\\
$7-8$&$\xx11.57\pm0.45\pm0.38\pm0.36\pm0.01$&$\xx\xx9.13\pm0.27\pm0.28\pm0.13\pm0.00$\\
$8-10$&$\xx\xx5.29\pm0.19\pm0.17\pm0.12\pm0.07$&$\xx\xx4.31\pm0.13\pm0.13\pm0.06\pm0.03$\\
$10-14$&$\xx\xx1.80\pm0.07\pm0.06\pm0.08\pm0.02$&$\xx\xx1.39\pm0.05\pm0.04\pm0.02\pm0.01$\\
\hline\pt[\gevc]&$3.0<y<3.5$&$3.5<y<4.0$\\\hline
$0-1$&$40.73\pm0.84\pm1.36\pm0.38\pm3.55$&$30.25\pm0.76\pm1.02\pm0.40\pm0.34$\\
$1-2$&$85.36\pm1.15\pm2.82\pm1.56\pm5.16$&$61.11\pm1.00\pm2.03\pm0.60\pm0.64$\\
$2-3$&$79.84\pm1.00\pm2.59\pm1.19\pm1.59$&$55.94\pm0.89\pm1.85\pm0.79\pm1.80$\\
$3-4$&$51.96\pm0.75\pm1.62\pm0.49\pm0.75$&$38.20\pm0.69\pm1.21\pm0.42\pm0.59$\\
$4-5$&$31.75\pm0.55\pm0.98\pm0.27\pm0.28$&$22.43\pm0.49\pm0.70\pm0.31\pm0.36$\\
$5-6$&$18.22\pm0.40\pm0.56\pm0.26\pm0.17$&$10.97\pm0.33\pm0.35\pm0.16\pm0.01$\\
$6-7$&$10.36\pm0.29\pm0.32\pm0.20\pm0.03$&$\xx6.83\pm0.25\pm0.22\pm0.12\pm0.11$\\
$7-8$&$\xx6.00\pm0.22\pm0.18\pm0.09\pm0.10$&$\xx3.60\pm0.17\pm0.12\pm0.10\pm0.03$\\
$8-10$&$\xx2.87\pm0.10\pm0.09\pm0.04\pm0.02$&$\xx1.57\pm0.08\pm0.06\pm0.06\pm0.01$\\
$10-14$&$\xx0.77\pm0.04\pm0.03\pm0.03\pm0.00$&$\xx0.39\pm0.03\pm0.01\pm0.01\pm0.01$\\
\hline\pt[\gevc]&$4.0<y<4.5$&\\\hline
$0-1$&$22.40\pm0.90\pm0.87\pm0.45\pm3.29$&\\
$1-2$&$40.68\pm1.11\pm1.53\pm0.61\pm0.82$&\\
$2-3$&$31.22\pm0.91\pm1.23\pm0.71\pm0.52$&\\
$3-4$&$18.68\pm0.66\pm0.70\pm0.44\pm0.22$&\\
$4-5$&$\xx9.00\pm0.42\pm0.34\pm0.18\pm0.26$&\\
$5-6$&$\xx5.51\pm0.30\pm0.21\pm0.14\pm0.10$&\\
$6-7$&$\xx2.84\pm0.20\pm0.11\pm0.09\pm0.02$&\\
$7-8$&$\xx1.44\pm0.13\pm0.06\pm0.07\pm0.04$&\\
$8-10$&$\xx0.53\pm0.06\pm0.02\pm0.02\pm0.01$&\\
$10-14$&$\xx0.15\pm0.02\pm0.01\pm0.01\pm0.01$&\\
\hline
\end{tabular}

\end{table}

\begin{table}[]
    \centering
    \caption{Single-differential production cross-sections $\frac{\deriv\sigma}{\deriv \pt}$[nb/(\ensuremath{\text{Ge\kern -0.1em V\!/}c})] for prompt \jpsi mesons in the rapidity range $2-4.5$. The first uncertainties are statistical, the second are correlated systematic uncertainties shared between intervals, and the last are uncorrelated systematic uncertainties.}
    \label{table:csPromptPT}
    \begin{tabular}{l|l}
\hline\pt[\gevc]&$2.0<y<4.5$\\\hline
$0-1$&$1441.38\pm4.70\pm53.61\pm5.09$\\
$1-2$&$2563.43\pm6.08\pm90.61\pm8.64$\\
$2-3$&$1926.88\pm4.86\pm67.32\pm7.22$\\
$3-4$&$1094.51\pm3.34\pm36.15\pm4.34$\\
$4-5$&$\xx561.32\pm2.08\pm18.10\pm2.05$\\
$5-6$&$\xx277.07\pm1.32\pm\xx8.89\pm1.21$\\
$6-7$&$\xx135.95\pm0.86\pm\xx4.34\pm0.69$\\
$7-8$&$\xx\xx69.98\pm0.59\pm\xx2.23\pm0.48$\\
$8-10$&$\xx\xx28.02\pm0.25\pm\xx0.90\pm0.17$\\
$10-14$&$\xx\xx\xx5.85\pm0.08\pm\xx0.19\pm0.05$\\
$14-20$&$\xx\xx\xx0.66\pm0.02\pm\xx0.02\pm0.02$\\
\hline
\end{tabular}

\end{table}

\begin{table}[]
    \centering
    \caption{Single-differential production cross-sections $\frac{\deriv\sigma}{\deriv \pt}$[nb/(\ensuremath{\text{Ge\kern -0.1em V\!/}c})] for nonprompt \jpsi mesons in the rapidity range $2-4.5$. The first uncertainties are statistical, the second are correlated systematic uncertainties shared between intervals, and the last are uncorrelated systematic uncertainties.}
    \label{table:csFrombPT}
    \begin{tabular}{l|l}
\hline\pt[\gevc]&$2.0<y<4.5$\\\hline
$0-1$&$\xx98.56\pm1.29\pm\xx6.61\pm0.80$\\
$1-2$&$206.60\pm1.70\pm10.03\pm1.65$\\
$2-3$&$193.41\pm1.51\pm\xx7.58\pm1.57$\\
$3-4$&$129.17\pm1.14\pm\xx4.56\pm0.83$\\
$4-5$&$\xx78.83\pm0.83\pm\xx2.66\pm0.54$\\
$5-6$&$\xx44.91\pm0.57\pm\xx1.49\pm0.34$\\
$6-7$&$\xx26.43\pm0.41\pm\xx0.86\pm0.25$\\
$7-8$&$\xx15.88\pm0.31\pm\xx0.51\pm0.20$\\
$8-10$&$\xx\xx7.29\pm0.14\pm\xx0.24\pm0.08$\\
$10-14$&$\xx\xx2.24\pm0.05\pm\xx0.07\pm0.05$\\
$14-20$&$\xx\xx0.41\pm0.02\pm\xx0.01\pm0.01$\\
\hline
\end{tabular}

\end{table}

\begin{table}[]
    \centering
    \caption{Single-differential production cross-sections $\frac{\deriv\sigma}{\deriv y}$[nb per unit rapidity] for prompt \jpsi mesons. The first uncertainties are statistical, the second are correlated systematic uncertainties shared between intervals, and the last are uncorrelated systematic uncertainties.}
    \label{table:csPromptY}
    \begin{tabular}{l|ll}
\hline$y$&$0<\pt<14\gevc$ & $0<\pt<8\gevc$\\\hline
$2.0-2.5$&$4119.9\pm14.3\pm170.6\pm34.3$&$4067.0\pm14.3\pm169.1\pm34.1$\\
$2.5-3.0$&$3855.4\pm\xx8.3\pm126.8\pm21.3$&$3811.7\pm\xx8.2\pm125.4\pm21.2$\\
$3.0-3.5$&$3382.0\pm\xx7.0\pm109.1\pm13.7$&$3350.2\pm\xx7.0\pm108.1\pm13.6$\\
$3.5-4.0$&$2833.3\pm\xx6.2\pm\xx92.6\pm\xx5.9$&$2812.6\pm\xx6.2\pm\xx91.8\pm\xx5.9$\\
$4.0-4.5$&$2109.5\pm\xx6.7\pm\xx80.0\pm\xx7.8$&$2099.5\pm\xx6.7\pm\xx79.6\pm\xx7.8$\\
\hline
\end{tabular}
\end{table}

\begin{table}[]
    \centering
    \caption{Single-differential production cross-sections $\frac{\deriv\sigma}{\deriv y}$[nb per unit rapidity] for nonprompt \jpsi mesons. The first uncertainties are statistical, the second are correlated systematic uncertainties shared between intervals, and the last are uncorrelated systematic uncertainties.}
    \label{table:csFrombY}
    \begin{tabular}{l|l}
\hline$y$&$0<\pt<14\gevc$\\\hline
$2.0-2.5$&$502.9\pm4.5\pm20.0\pm28.1$\\
$2.5-3.0$&$431.3\pm2.5\pm14.0\pm20.8$\\
$3.0-3.5$&$333.0\pm2.1\pm10.7\pm11.9$\\
$3.5-4.0$&$234.0\pm1.8\pm\xx7.6\pm\xx4.1$\\
$4.0-4.5$&$133.4\pm1.9\pm\xx5.1\pm\xx5.4$\\
\hline
\end{tabular}

\end{table}

\begin{table}[]
    \centering
    \caption{Fraction of nonprompt \jpsi mesons (in \%) in (\pt,$y$) intervals. The first uncertainty is statistical and the second is systematic.}
    \label{table:frombfrac}
    \begin{tabular}{l|lll}
\hline\pt[\gevc]&$2.0<y<2.5$&$2.5<y<3.0$&$3.0<y<3.5$\\\hline
$0-1$&$\xx7.4\pm0.3\pm1.1$&$\xx6.8\pm0.1\pm0.8$&$\xx6.3\pm0.1\pm0.6$\\
$1-2$&$\xx8.6\pm0.2\pm0.7$&$\xx8.2\pm0.1\pm0.6$&$\xx7.4\pm0.1\pm0.5$\\
$2-3$&$10.5\pm0.2\pm0.5$&$\xx9.9\pm0.1\pm0.4$&$\xx8.9\pm0.1\pm0.2$\\
$3-4$&$11.6\pm0.3\pm0.4$&$11.5\pm0.2\pm0.3$&$10.1\pm0.2\pm0.2$\\
$4-5$&$13.9\pm0.4\pm0.4$&$12.9\pm0.2\pm0.3$&$11.6\pm0.2\pm0.1$\\
$5-6$&$15.7\pm0.5\pm0.4$&$14.5\pm0.3\pm0.3$&$13.4\pm0.3\pm0.2$\\
$6-7$&$18.2\pm0.6\pm0.4$&$16.1\pm0.4\pm0.3$&$15.5\pm0.5\pm0.3$\\
$7-8$&$20.0\pm0.8\pm0.6$&$19.3\pm0.6\pm0.2$&$17.0\pm0.7\pm0.3$\\
$8-10$&$22.4\pm0.9\pm0.5$&$21.8\pm0.7\pm0.3$&$19.7\pm0.8\pm0.3$\\
$10-14$&$30.0\pm1.4\pm1.4$&$29.0\pm1.2\pm0.5$&$25.1\pm1.3\pm0.9$\\
\hline\pt[\gevc]&$3.5<y<4$&$4<y<4.5$&\\\hline
$0-1$&$\xx5.3\pm0.1\pm0.1$&$\xx4.6\pm0.2\pm0.7$&\\
$1-2$&$\xx6.2\pm0.1\pm0.1$&$\xx5.3\pm0.1\pm0.1$&\\
$2-3$&$\xx7.5\pm0.1\pm0.3$&$\xx6.2\pm0.2\pm0.2$&\\
$3-4$&$\xx9.3\pm0.2\pm0.2$&$\xx7.3\pm0.3\pm0.2$&\\
$4-5$&$10.7\pm0.2\pm0.2$&$\xx7.7\pm0.4\pm0.2$&\\
$5-6$&$11.3\pm0.4\pm0.1$&$10.2\pm0.6\pm0.2$&\\
$6-7$&$14.0\pm0.5\pm0.3$&$11.1\pm0.8\pm0.2$&\\
$7-8$&$14.6\pm0.8\pm0.3$&$12.0\pm1.2\pm0.5$&\\
$8-10$&$16.9\pm0.9\pm0.6$&$12.4\pm1.4\pm0.5$&\\
$10-14$&$21.1\pm1.7\pm0.7$&$18.4\pm2.8\pm1.4$&\\
\hline
\end{tabular}

\end{table}

\begin{table}[]
    \centering
    \caption{Nuclear modification factor $R_{p\text{Pb}}$ as a function of $y$ with $\pt<14\gevc$. The first uncertainty is statistical and the second is systematic.}
    \label{table:Rplead}
    \begin{tabular}{l|ll}
\hline $y$  & prompt \jpsi            & nonprompt \jpsi \\ \hline
$(-4.5,-4.0)$ & $0.897\pm0.060\pm0.061$ & $1.445\pm0.189\pm0.201$ \\
$(-4.0,-3.5)$ & $0.888\pm0.044\pm0.056$ & $0.955\pm0.099\pm0.064$ \\
$(-3.5,-3.0)$ & $0.918\pm0.041\pm0.058$ & $0.974\pm0.084\pm0.097$ \\
$(-3.0,-2.5)$ & $0.846\pm0.052\pm0.082$ & $0.860\pm0.099\pm0.097$ \\
$(\xxx2.0,\xxx2.5)$ & $0.624\pm0.014\pm0.039$ & $0.797\pm0.033\pm0.068$ \\
$(\xxx2.5,\xxx3.0)$ & $0.611\pm0.012\pm0.035$ & $0.791\pm0.032\pm0.058$ \\
$(\xxx3.0,\xxx3.5)$ & $0.571\pm0.012\pm0.033$ & $0.821\pm0.038\pm0.064$ \\
$(\xxx3.5,\xxx4.0)$ & $0.568\pm0.015\pm0.035$ & $0.686\pm0.054\pm0.065$ \\
\hline
\end{tabular}

\end{table}

\begin{table}[]
    \centering
    \caption{Cross-section ratios between 8\tev and 5\tev measurements for prompt \jpsi mesons as a function of \pt with $2.0<y<4.5$. The first uncertainty is statistical and the second is systematic.}
    \label{table:Ratio8over5PT}
    \begin{tabular}{l|l}
\hline\pt[\gevc]&$2.0<y<4.5$\\\hline
$0-1$&$1.20\pm0.01\pm0.08$\\
$1-2$&$1.27\pm0.01\pm0.08$\\
$2-3$&$1.34\pm0.01\pm0.09$\\
$3-4$&$1.42\pm0.01\pm0.09$\\
$4-5$&$1.50\pm0.01\pm0.10$\\
$5-6$&$1.57\pm0.01\pm0.10$\\
$6-7$&$1.69\pm0.01\pm0.11$\\
$7-8$&$1.74\pm0.02\pm0.11$\\
$8-10$&$1.85\pm0.02\pm0.12$\\
$10-14$&$2.05\pm0.03\pm0.13$\\
\hline
\end{tabular}

\end{table}

\begin{table}[]
    \centering
    \caption{Cross-section ratios between 8\tev and 5\tev measurements for prompt \jpsi mesons as a function of $y$ with $\pt<8\gevc$. The first uncertainty is statistical and the second is systematic.}
    \label{table:Ratio8
    over5Y}
    \begin{tabular}{l|l}
\hline$y$&$0<\pt<8\gevc$\\\hline
$2.0-2.5$&$1.24\pm0.01\pm0.08$\\
$2.5-3.0$&$1.30\pm0.00\pm0.08$\\
$3.0-3.5$&$1.34\pm0.00\pm0.09$\\
$3.5-4.0$&$1.38\pm0.00\pm0.09$\\
$4.0-4.5$&$1.48\pm0.01\pm0.10$\\
\hline
\end{tabular}

\end{table}

\begin{table}[]
    \centering
    \caption{Cross-section ratios between 13\tev and 5\tev measurements for prompt \jpsi mesons as a function of \pt with $2.0<y<4.5$. The first uncertainty is statistical and the second is systematic.}
    \label{table:Ratio13over5PT}
    \begin{tabular}{l|l}
\hline\pt[\gevc]&$2.0<y<4.5$\\\hline
$0-1$&$1.51\pm0.01\pm0.08$\\
$1-2$&$1.65\pm0.01\pm0.09$\\
$2-3$&$1.84\pm0.01\pm0.09$\\
$3-4$&$2.06\pm0.01\pm0.10$\\
$4-5$&$2.27\pm0.01\pm0.11$\\
$5-6$&$2.54\pm0.02\pm0.12$\\
$6-7$&$2.77\pm0.03\pm0.13$\\
$7-8$&$2.85\pm0.03\pm0.13$\\
$8-10$&$3.22\pm0.04\pm0.15$\\
$10-14$&$3.86\pm0.07\pm0.18$\\
\hline
\end{tabular}

\end{table}

\begin{table}[]
    \centering
    \caption{Cross-section ratios between 13\tev and 5\tev measurements for prompt \jpsi mesons as a function of $y$ with $\pt<8\gevc$. The first uncertainty is statistical and the second is systematic.}
    \label{table:Ratio13over5Y}
    \begin{tabular}{l|l}
\hline$y$&$0<\pt<8\gevc$\\\hline
$2.0-2.5$&$1.70\pm0.01\pm0.10$\\
$2.5-3.0$&$1.76\pm0.01\pm0.08$\\
$3.0-3.5$&$1.83\pm0.01\pm0.08$\\
$3.5-4.0$&$1.90\pm0.01\pm0.10$\\
$4.0-4.5$&$2.13\pm0.01\pm0.13$\\
\hline
\end{tabular}

\end{table}

\begin{table}[]
    \centering
    \caption{Cross-section ratios between 8\tev and 5\tev measurements for nonprompt \jpsi mesons as a function of \pt with $2.0<y<4.5$. The first uncertainty is statistical and the second is systematic.}
    \label{table:Ratio8over5FrombPT}
    \begin{tabular}{l|l}
\hline\pt[\gevc]&$2.0<y<4.5$\\\hline
$0-1$&$1.42\pm0.03\pm0.14$\\
$1-2$&$1.51\pm0.02\pm0.11$\\
$2-3$&$1.51\pm0.02\pm0.10$\\
$3-4$&$1.59\pm0.02\pm0.10$\\
$4-5$&$1.62\pm0.02\pm0.11$\\
$5-6$&$1.69\pm0.03\pm0.11$\\
$6-7$&$1.72\pm0.03\pm0.11$\\
$7-8$&$1.76\pm0.04\pm0.12$\\
$8-10$&$1.95\pm0.04\pm0.13$\\
$10-14$&$1.97\pm0.05\pm0.13$\\
\hline
\end{tabular}

\end{table}

\begin{table}[]
    \centering
    \caption{Cross-section ratios between 8\tev and 5\tev measurements for nonprompt \jpsi mesons as a function of $y$ with $\pt<14\gevc$. The first uncertainty is statistical and the second is systematic.}
    \label{table:Ratio8over5FrombY}
    \begin{tabular}{l|l}
\hline$y$&$0<\pt<14\gevc$\\\hline
$2.0-2.5$&$1.42\pm0.02\pm0.12$\\
$2.5-3.0$&$1.54\pm0.01\pm0.13$\\
$3.0-3.5$&$1.61\pm0.01\pm0.12$\\
$3.5-4.0$&$1.65\pm0.02\pm0.11$\\
$4.0-4.5$&$1.80\pm0.03\pm0.14$\\
\hline
\end{tabular}

\end{table}

\begin{table}[]
    \centering
    \caption{Cross-section ratios between 13\tev and 5\tev measurements for nonprompt \jpsi mesons as a function of \pt with $2.0<y<4.5$. The first uncertainty is statistical and the second is systematic.}
    \label{table:Ratio13over5FrombPT}
    \begin{tabular}{l|l}
\hline\pt[\gevc]&$2.0<y<4.5$\\\hline
$0-1$&$2.23\pm0.05\pm0.17$\\
$1-2$&$2.46\pm0.03\pm0.16$\\
$2-3$&$2.62\pm0.03\pm0.14$\\
$3-4$&$2.85\pm0.04\pm0.14$\\
$4-5$&$3.08\pm0.05\pm0.15$\\
$5-6$&$3.26\pm0.06\pm0.15$\\
$6-7$&$3.58\pm0.07\pm0.17$\\
$7-8$&$3.63\pm0.09\pm0.17$\\
$8-10$&$4.21\pm0.10\pm0.20$\\
$10-14$&$4.81\pm0.14\pm0.24$\\
\hline
\end{tabular}

\end{table}

\begin{table}[]
    \centering
    \caption{Cross-section ratios between 13\tev and 5\tev measurements for nonprompt \jpsi mesons as a function of $y$ with $\pt<14\gevc$. The first uncertainty is statistical and the second is systematic.}
    \label{table:Ratio13over5FrombY}
    \begin{tabular}{l|l}
\hline$y$&$0<\pt<14\gevc$\\\hline
$2.0-2.5$&$2.39\pm0.03\pm0.19$\\
$2.5-3.0$&$2.49\pm0.02\pm0.17$\\
$3.0-3.5$&$2.79\pm0.03\pm0.16$\\
$3.5-4.0$&$3.24\pm0.04\pm0.17$\\
$4.0-4.5$&$4.02\pm0.08\pm0.29$\\
\hline
\end{tabular}

\end{table}

\clearpage

\section{Dependence of cross-sections on the polarisation}
\label{sec:Polar}
The angular distribution of the \jpsi\to\mumu decay is described by
\begin{equation}
    \frac{\deriv^2N}{\deriv\cos\theta\deriv\phi}\propto 1+\lambda_{\theta}\cos^2\theta+\lambda_{\theta\phi}\sin2\theta\cos\phi+\lambda_{\phi}\sin^2\theta\cos2\phi,
\end{equation}
where $\theta$ and $\phi$ are the polar and azimuthal angles between the direction of $\mu^+$ and the chosen polarisation axis, and $\lambda_{\theta}$, $\lambda_{\theta\phi}$ and $\lambda_{\phi}$ are polarisation parameters.
In the helicity frame, the polarisation axis coincides with the flight direction of the \jpsi meson in the centre-of-mass frame of the colliding hadrons. 
The detection efficiency of the \jpsi mesons is function of the polarisation,
especially of $\lambda_{\theta}$.
Zero polarisation is assumed in the simulation
since there is no prior knowledge of the polarisation of the \jpsi mesons
in $pp$ collisions at 5\tev, and only small longitudinal polarisations
have been found in the \jpsi polarisation analyses at the LHC
\cite{Abelev:2011md,Chatrchyan:2013cla,LHCb-PAPER-2013-008}.

To evaluate the change of results assuming a non-zero polarisation,
we reweight the angular distribution of the muon tracks in rest frame of the \jpsi mesons in simulation
and calculate the change in the total efficiency, which impacts the cross-sections.
The relative change of the cross-section for a polarisation of $\lambda_{\theta}=-0.2$~\cite{LHCb-PAPER-2013-008} in the helicity frame
compared to zero polarisation in each (\pt,$y$) interval is given in Table \ref{table:polar}.
\begin{table}[b]
    \centering
    \caption{Relative changes of cross-sections (in \%), for a polarisation of $\lambda_{\theta}=-0.2$ rather than zero, in (\pt,$y$) intervals.}
    \label{table:polar}
    \begin{tabular}{l|lllll}
\hline \pt[\ensuremath{\text{Ge\kern -0.1em V\!/}c}] & $2.0<y<2.5$ & $2.5<y<3.0$ & $3.0<y<3.5$ & $3.5<y<4.0$ & $4.0<y<4.5$\\ \hline
$0-1$ &$-5.91\pm0.83$ &$-4.47\pm0.42$ &$-2.94\pm0.37$ &$-2.39\pm0.43$ &$-1.95\pm0.77$ \\
$1-2$ &$-5.22\pm0.59$ &$-4.05\pm0.32$ &$-2.47\pm0.29$ &$-1.38\pm0.35$ &$-0.47\pm0.60$ \\
$2-3$ &$-4.38\pm0.63$ &$-3.21\pm0.36$ &$-1.62\pm0.33$ &$-0.49\pm0.41$ &$\xxx0.55\pm0.72$ \\
$3-4$ &$-4.20\pm0.75$ &$-3.09\pm0.42$ &$-1.60\pm0.40$ &$-0.30\pm0.51$ &$\xxx0.56\pm0.93$ \\
$4-5$ &$-4.14\pm0.90$ &$-3.15\pm0.50$ &$-1.80\pm0.49$ &$-0.83\pm0.63$ &$\xxx0.47\pm1.16$ \\
$5-6$ &$-4.00\pm1.06$ &$-3.00\pm0.61$ &$-1.87\pm0.62$ &$-1.10\pm0.80$ &$\xxx0.19\pm1.52$ \\
$6-7$ &$-3.77\pm1.30$ &$-2.81\pm0.76$ &$-1.89\pm0.79$ &$-1.45\pm1.05$ &$-0.41\pm1.96$ \\
$7-8$ &$-3.63\pm1.61$ &$-2.70\pm0.96$ &$-1.76\pm1.04$ &$-1.63\pm1.37$ &$-0.55\pm2.66$ \\
$8-10$ &$-3.23\pm1.52$ &$-2.32\pm0.96$ &$-1.68\pm1.08$ &$-1.78\pm1.50$ &$-1.02\pm2.92$ \\
$10-14$ &$-2.85\pm1.88$ &$-2.04\pm1.28$ &$-1.47\pm1.54$ &$-1.44\pm2.22$ &$-1.29\pm5.07$ \\
$14-20$ &\multicolumn{5}{c}{$-1.55\pm1.87~(2.0<y<4.5)$}\\
\hline
\end{tabular}

\end{table}
In addition, the relative change of the cross-section for a polarisation of $\lambda_{\theta}=-1~(+1)$ in the helicity frame, which corresponds to the fully longitudinally (transversely) polarised scenario, compared to zero polarisation in each (\pt,$y$) interval is given in Table \ref{table:polar-1} (\ref{table:polar+1}).
\begin{table}[h]
    \centering
    \caption{Relative change of cross-sections (in \%), for a polarisation of $\lambda_{\theta}=-1$ rather than zero, in (\pt,$y$) intervals.}
    \label{table:polar-1}
    \begin{tabular}{l|lllll}
\hline \pt[\ensuremath{\text{Ge\kern -0.1em V\!/}c}] & $2.0<y<2.5$ & $2.5<y<3.0$ & $3.0<y<3.5$ & $3.5<y<4.0$ & $4.0<y<4.5$\\ \hline
$0-1$ &$-30.6\pm0.6$ &$-24.6\pm0.3$ &$-17.4\pm0.3$ &$-18.0\pm0.3$ &$-24.9\pm0.6$ \\
$1-2$ &$-27.8\pm0.4$ &$-22.8\pm0.2$ &$-15.2\pm0.2$ &$-13.4\pm0.3$ &$-16.2\pm0.5$ \\
$2-3$ &$-24.3\pm0.5$ &$-18.9\pm0.3$ &$-10.5\pm0.3$ &$\xx-7.5\pm0.4$ &$\xx-8.8\pm0.7$ \\
$3-4$ &$-23.5\pm0.6$ &$-18.2\pm0.3$ &$-10.2\pm0.4$ &$\xx-5.8\pm0.5$ &$\xx-5.2\pm0.9$ \\
$4-5$ &$-23.2\pm0.7$ &$-18.5\pm0.4$ &$-11.3\pm0.4$ &$\xx-7.3\pm0.6$ &$\xx-3.9\pm1.1$ \\
$5-6$ &$-22.6\pm0.8$ &$-17.8\pm0.5$ &$-11.8\pm0.5$ &$\xx-8.8\pm0.7$ &$\xx-4.8\pm1.5$ \\
$6-7$ &$-21.5\pm1.0$ &$-16.9\pm0.6$ &$-11.8\pm0.7$ &$\xx-9.8\pm1.0$ &$\xx-5.8\pm1.9$ \\
$7-8$ &$-21.1\pm1.3$ &$-16.3\pm0.8$ &$-11.1\pm0.9$ &$-10.4\pm1.2$ &$\xx-7.2\pm2.5$ \\
$8-10$ &$-19.2\pm1.2$ &$-14.3\pm0.8$ &$-10.7\pm1.0$ &$-11.2\pm1.3$ &$\xx-8.3\pm2.8$ \\
$10-14$ &$-16.9\pm1.6$ &$-12.6\pm1.1$ &$\xx-9.3\pm1.4$ &$\xx-9.0\pm2.0$ &$-11.0\pm4.6$ \\
$14-20$ &\multicolumn{5}{c}{$-10.2\pm1.7~(2.0<y<4.5)$}\\
\hline
\end{tabular}

\end{table}
\begin{table}[h]
    \centering
    \caption{Relative changes of cross-sections (in \%), for a polarisation of $\lambda_{\theta}=+1$ rather than zero, in (\pt,$y$) intervals.}
    \label{table:polar+1}
    \begin{tabular}{l|lllll}
\hline \pt[\ensuremath{\text{Ge\kern -0.1em V\!/}c}] & $2.0<y<2.5$ & $2.5<y<3.0$ & $3.0<y<3.5$ & $3.5<y<4.0$ & $4.0<y<4.5$\\ \hline
$0-1$ &$28.2\pm1.2$ &$19.6\pm0.6$ &$11.8\pm0.5$ &$12.4\pm0.5$ &$19.8\pm1.0$ \\
$1-2$ &$23.9\pm0.8$ &$17.3\pm0.4$ &$\xx9.9\pm0.3$ &$\xx8.4\pm0.4$ &$10.7\pm0.7$ \\
$2-3$ &$19.0\pm0.8$ &$13.1\pm0.4$ &$\xx6.2\pm0.4$ &$\xx4.2\pm0.4$ &$\xx5.1\pm0.8$ \\
$3-4$ &$18.1\pm1.0$ &$12.5\pm0.5$ &$\xx6.1\pm0.4$ &$\xx3.2\pm0.5$ &$\xx2.8\pm1.0$ \\
$4-5$ &$17.7\pm1.1$ &$12.8\pm0.6$ &$\xx6.8\pm0.6$ &$\xx4.1\pm0.7$ &$\xx2.1\pm1.2$ \\
$5-6$ &$17.1\pm1.3$ &$12.2\pm0.8$ &$\xx7.2\pm0.7$ &$\xx5.1\pm0.9$ &$\xx2.6\pm1.6$ \\
$6-7$ &$15.8\pm1.6$ &$11.3\pm1.0$ &$\xx7.2\pm0.9$ &$\xx5.8\pm1.2$ &$\xx3.2\pm2.1$ \\
$7-8$ &$15.3\pm2.0$ &$10.8\pm1.2$ &$\xx6.7\pm1.2$ &$\xx6.2\pm1.6$ &$\xx4.0\pm2.9$ \\
$8-10$ &$13.4\pm1.9$ &$\xx9.2\pm1.2$ &$\xx6.3\pm1.3$ &$\xx6.6\pm1.7$ &$\xx4.7\pm3.2$ \\
$10-14$ &$11.4\pm2.3$ &$\xx7.8\pm1.6$ &$\xx5.5\pm1.8$ &$\xx5.1\pm2.5$ &$\xx6.5\pm5.7$ \\
$14-20$ &\multicolumn{5}{c}{$6.0\pm2.1~(2.0<y<4.5)$}\\
\hline
\end{tabular}

\end{table}

\clearpage

\addcontentsline{toc}{section}{References}
\bibliographystyle{LHCb}
\bibliography{main,standard,LHCb-PAPER,LHCb-CONF,LHCb-DP,LHCb-TDR}

\ifx\mcitethebibliography\mciteundefinedmacro
\PackageError{LHCb.bst}{mciteplus.sty has not been loaded}
{This bibstyle requires the use of the mciteplus package.}\fi
\providecommand{\href}[2]{#2}
\begin{mcitethebibliography}{10}
\mciteSetBstSublistMode{n}
\mciteSetBstMaxWidthForm{subitem}{\alph{mcitesubitemcount})}
\mciteSetBstSublistLabelBeginEnd{\mcitemaxwidthsubitemform\space}
{\relax}{\relax}

\bibitem{Carlson:1976cd}
C.~E. Carlson and R.~Suaya,
  \ifthenelse{\boolean{articletitles}}{\emph{{Hadronic production of $\jpsi$
  mesons}}, }{}\href{https://doi.org/10.1103/PhysRevD.14.3115}{Phys.\ Rev.\
  \textbf{D14} (1976) 3115}\relax
\mciteBstWouldAddEndPuncttrue
\mciteSetBstMidEndSepPunct{\mcitedefaultmidpunct}
{\mcitedefaultendpunct}{\mcitedefaultseppunct}\relax
\EndOfBibitem
\bibitem{Donnachie:1976ue}
A.~Donnachie and P.~V. Landshoff,
  \ifthenelse{\boolean{articletitles}}{\emph{{Production of lepton pairs,
  $\jpsi$ and charm with hadron beams}},
  }{}\href{https://doi.org/10.1016/0550-3213(76)90532-0}{Nucl.\ Phys.\
  \textbf{B112} (1976) 233}\relax
\mciteBstWouldAddEndPuncttrue
\mciteSetBstMidEndSepPunct{\mcitedefaultmidpunct}
{\mcitedefaultendpunct}{\mcitedefaultseppunct}\relax
\EndOfBibitem
\bibitem{Ellis:1976fj}
S.~D. Ellis, M.~B. Einhorn, and C.~Quigg,
  \ifthenelse{\boolean{articletitles}}{\emph{{Comment on hadronic production of
  psions}}, }{}\href{https://doi.org/10.1103/PhysRevLett.36.1263}{Phys.\ Rev.\
  Lett.\  \textbf{36} (1976) 1263}\relax
\mciteBstWouldAddEndPuncttrue
\mciteSetBstMidEndSepPunct{\mcitedefaultmidpunct}
{\mcitedefaultendpunct}{\mcitedefaultseppunct}\relax
\EndOfBibitem
\bibitem{Fritzsch:1977ay}
H.~Fritzsch, \ifthenelse{\boolean{articletitles}}{\emph{{Producing heavy quark
  flavors in hadronic collisions: A test of quantum chromodynamics}},
  }{}\href{https://doi.org/10.1016/0370-2693(77)90108-3}{Phys.\ Lett.\
  \textbf{B67} (1977) 217}\relax
\mciteBstWouldAddEndPuncttrue
\mciteSetBstMidEndSepPunct{\mcitedefaultmidpunct}
{\mcitedefaultendpunct}{\mcitedefaultseppunct}\relax
\EndOfBibitem
\bibitem{Gluck:1977zm}
M.~Gl{\"u}ck, J.~F. Owens, and E.~Reya,
  \ifthenelse{\boolean{articletitles}}{\emph{{Gluon contribution to hadronic
  $\jpsi$ production}},
  }{}\href{https://doi.org/10.1103/PhysRevD.17.2324}{Phys.\ Rev.\  \textbf{D17}
  (1978) 2324}\relax
\mciteBstWouldAddEndPuncttrue
\mciteSetBstMidEndSepPunct{\mcitedefaultmidpunct}
{\mcitedefaultendpunct}{\mcitedefaultseppunct}\relax
\EndOfBibitem
\bibitem{Chang:1979nn}
C.-H. Chang, \ifthenelse{\boolean{articletitles}}{\emph{{Hadronic production of
  $\jpsi$ associated with a gluon}},
  }{}\href{https://doi.org/10.1016/0550-3213(80)90175-3}{Nucl.\ Phys.\
  \textbf{B172} (1980) 425}\relax
\mciteBstWouldAddEndPuncttrue
\mciteSetBstMidEndSepPunct{\mcitedefaultmidpunct}
{\mcitedefaultendpunct}{\mcitedefaultseppunct}\relax
\EndOfBibitem
\bibitem{Baier:1981uk}
R.~Baier and R.~R{\"u}ckl, \ifthenelse{\boolean{articletitles}}{\emph{{Hadronic
  production of $\jpsi$ and $\PUpsilon$: Transverse momentum distributions}},
  }{}\href{https://doi.org/10.1016/0370-2693(81)90636-5}{Phys.\ Lett.\
  \textbf{B102} (1981) 364}\relax
\mciteBstWouldAddEndPuncttrue
\mciteSetBstMidEndSepPunct{\mcitedefaultmidpunct}
{\mcitedefaultendpunct}{\mcitedefaultseppunct}\relax
\EndOfBibitem
\bibitem{Bodwin:1994jh}
G.~T. Bodwin, E.~Braaten, and G.~P. Lepage,
  \ifthenelse{\boolean{articletitles}}{\emph{{Rigorous QCD analysis of
  inclusive annihilation and production of heavy quarkonium}},
  }{}\href{https://doi.org/10.1103/PhysRevD.51.1125 10.1103/PhysRevD.55.5853,
  10.1103/PhysRevD.55.5853, 10.1103/PhysRevD.51.1125}{Phys.\ Rev.\
  \textbf{D51} (1995) 1125},
  \href{http://arxiv.org/abs/hep-ph/9407339}{{\normalfont\ttfamily
  arXiv:hep-ph/9407339}}\relax
\mciteBstWouldAddEndPuncttrue
\mciteSetBstMidEndSepPunct{\mcitedefaultmidpunct}
{\mcitedefaultendpunct}{\mcitedefaultseppunct}\relax
\EndOfBibitem
\bibitem{Cho:1995vh}
P.~L. Cho and A.~K. Leibovich,
  \ifthenelse{\boolean{articletitles}}{\emph{{Color octet quarkonia
  production}}, }{}\href{https://doi.org/10.1103/PhysRevD.53.150}{Phys.\ Rev.\
  \textbf{D53} (1996) 150},
  \href{http://arxiv.org/abs/hep-ph/9505329}{{\normalfont\ttfamily
  arXiv:hep-ph/9505329}}\relax
\mciteBstWouldAddEndPuncttrue
\mciteSetBstMidEndSepPunct{\mcitedefaultmidpunct}
{\mcitedefaultendpunct}{\mcitedefaultseppunct}\relax
\EndOfBibitem
\bibitem{Cho:1995ce}
P.~L. Cho and A.~K. Leibovich,
  \ifthenelse{\boolean{articletitles}}{\emph{{Color octet quarkonia production.
  2.}}, }{}\href{https://doi.org/10.1103/PhysRevD.53.6203}{Phys.\ Rev.\
  \textbf{D53} (1996) 6203},
  \href{http://arxiv.org/abs/hep-ph/9511315}{{\normalfont\ttfamily
  arXiv:hep-ph/9511315}}\relax
\mciteBstWouldAddEndPuncttrue
\mciteSetBstMidEndSepPunct{\mcitedefaultmidpunct}
{\mcitedefaultendpunct}{\mcitedefaultseppunct}\relax
\EndOfBibitem
\bibitem{LHCb-PAPER-2012-039}
LHCb collaboration, R.~Aaij {\em et~al.},
  \ifthenelse{\boolean{articletitles}}{\emph{{Measurement of \jpsi production
  in \proton\proton collisions at $\sqs=2.76$\tev}},
  }{}\href{https://doi.org/10.1007/JHEP02(2013)041}{JHEP \textbf{02} (2013)
  041}, \href{http://arxiv.org/abs/1212.1045}{{\normalfont\ttfamily
  arXiv:1212.1045}}\relax
\mciteBstWouldAddEndPuncttrue
\mciteSetBstMidEndSepPunct{\mcitedefaultmidpunct}
{\mcitedefaultendpunct}{\mcitedefaultseppunct}\relax
\EndOfBibitem
\bibitem{LHCb-PAPER-2011-003}
LHCb collaboration, R.~Aaij {\em et~al.},
  \ifthenelse{\boolean{articletitles}}{\emph{{Measurement of \jpsi production
  in \proton\proton collisions at \mbox{$\sqs=$7\tev}}},
  }{}\href{https://doi.org/10.1140/epjc/s10052-011-1645-y}{Eur.\ Phys.\ J.\
  \textbf{C71} (2011) 1645},
  \href{http://arxiv.org/abs/1103.0423}{{\normalfont\ttfamily
  arXiv:1103.0423}}\relax
\mciteBstWouldAddEndPuncttrue
\mciteSetBstMidEndSepPunct{\mcitedefaultmidpunct}
{\mcitedefaultendpunct}{\mcitedefaultseppunct}\relax
\EndOfBibitem
\bibitem{LHCb-PAPER-2013-016}
LHCb collaboration, R.~Aaij {\em et~al.},
  \ifthenelse{\boolean{articletitles}}{\emph{{Production of \jpsi and
  \Upsilonres mesons in \proton\proton collisions at \mbox{$\sqs=$8\tev}}},
  }{}\href{https://doi.org/10.1007/JHEP06(2013)064}{JHEP \textbf{06} (2013)
  064}, \href{http://arxiv.org/abs/1304.6977}{{\normalfont\ttfamily
  arXiv:1304.6977}}\relax
\mciteBstWouldAddEndPuncttrue
\mciteSetBstMidEndSepPunct{\mcitedefaultmidpunct}
{\mcitedefaultendpunct}{\mcitedefaultseppunct}\relax
\EndOfBibitem
\bibitem{LHCb-PAPER-2015-037}
LHCb collaboration, R.~Aaij {\em et~al.},
  \ifthenelse{\boolean{articletitles}}{\emph{{Measurement of forward \jpsi
  production cross-sections in \proton\proton collisions at
  \mbox{$\sqs=$13\tev}}},
  }{}\href{https://doi.org/10.1007/JHEP10(2015)172}{JHEP \textbf{10} (2015)
  172}, Erratum \href{https://doi.org/10.1007/JHEP05(2017)063}{ibid.\
  \textbf{05} (2017) 063},
  \href{http://arxiv.org/abs/1509.00771}{{\normalfont\ttfamily
  arXiv:1509.00771}}\relax
\mciteBstWouldAddEndPuncttrue
\mciteSetBstMidEndSepPunct{\mcitedefaultmidpunct}
{\mcitedefaultendpunct}{\mcitedefaultseppunct}\relax
\EndOfBibitem
\bibitem{Aaboud:2017cif}
ATLAS collaboration, M.~Aaboud {\em et~al.},
  \ifthenelse{\boolean{articletitles}}{\emph{{Measurement of quarkonium
  production in proton–lead and proton–proton collisions at $5.02\tev$ with
  the ATLAS detector}},
  }{}\href{https://doi.org/10.1140/epjc/s10052-018-5624-4}{Eur.\ Phys.\ J.\
  \textbf{C78} (2018) 171},
  \href{http://arxiv.org/abs/1709.03089}{{\normalfont\ttfamily
  arXiv:1709.03089}}\relax
\mciteBstWouldAddEndPuncttrue
\mciteSetBstMidEndSepPunct{\mcitedefaultmidpunct}
{\mcitedefaultendpunct}{\mcitedefaultseppunct}\relax
\EndOfBibitem
\bibitem{Aad:2015duc}
ATLAS collaboration, G.~Aad {\em et~al.},
  \ifthenelse{\boolean{articletitles}}{\emph{{Measurement of the differential
  cross-sections of prompt and non-prompt production of $\jpsi$ and $\psitwos$
  in $pp$ collisions at $\sqs=7$ and 8$\tev$ with the ATLAS detector}},
  }{}\href{https://doi.org/10.1140/epjc/s10052-016-4050-8}{Eur.\ Phys.\ J.\
  \textbf{C76} (2016) 283},
  \href{http://arxiv.org/abs/1512.03657}{{\normalfont\ttfamily
  arXiv:1512.03657}}\relax
\mciteBstWouldAddEndPuncttrue
\mciteSetBstMidEndSepPunct{\mcitedefaultmidpunct}
{\mcitedefaultendpunct}{\mcitedefaultseppunct}\relax
\EndOfBibitem
\bibitem{Zakareishvili:2020aki}
ATLAS collaboration, T.~Zakareishvili,
  \ifthenelse{\boolean{articletitles}}{\emph{{Measurement of the production
  cross-section of $\jpsi$ and $\psitwos$ mesons at high transverse momentum in
  pp collisions at $\sqs$ = 13 TeV with the ATLAS detector}},
  }{}\href{https://doi.org/10.1088/1742-6596/1690/1/012160}{J.\ Phys.\ Conf.\
  Ser.\  \textbf{1690} (2020) 012160}\relax
\mciteBstWouldAddEndPuncttrue
\mciteSetBstMidEndSepPunct{\mcitedefaultmidpunct}
{\mcitedefaultendpunct}{\mcitedefaultseppunct}\relax
\EndOfBibitem
\bibitem{Sirunyan:2017mzd}
CMS collaboration, A.~M. Sirunyan {\em et~al.},
  \ifthenelse{\boolean{articletitles}}{\emph{{Measurement of prompt and
  nonprompt $\jpsi$ production in $pp$ and $pPb$ collisions at $\sqsnn
  =5.02\tev$}}, }{}\href{https://doi.org/10.1140/epjc/s10052-017-4828-3}{Eur.\
  Phys.\ J.\  \textbf{C77} (2017) 269},
  \href{http://arxiv.org/abs/1702.01462}{{\normalfont\ttfamily
  arXiv:1702.01462}}\relax
\mciteBstWouldAddEndPuncttrue
\mciteSetBstMidEndSepPunct{\mcitedefaultmidpunct}
{\mcitedefaultendpunct}{\mcitedefaultseppunct}\relax
\EndOfBibitem
\bibitem{Khachatryan:2010yr}
CMS collaboration, V.~Khachatryan {\em et~al.},
  \ifthenelse{\boolean{articletitles}}{\emph{{Prompt and non-prompt $\jpsi$
  production in $pp$ collisions at $\sqs=7\tev$}},
  }{}\href{https://doi.org/10.1140/epjc/s10052-011-1575-8}{Eur.\ Phys.\ J.\
  \textbf{C71} (2011) 1575},
  \href{http://arxiv.org/abs/1011.4193}{{\normalfont\ttfamily
  arXiv:1011.4193}}\relax
\mciteBstWouldAddEndPuncttrue
\mciteSetBstMidEndSepPunct{\mcitedefaultmidpunct}
{\mcitedefaultendpunct}{\mcitedefaultseppunct}\relax
\EndOfBibitem
\bibitem{Chatrchyan:2011kc}
CMS collaboration, S.~Chatrchyan {\em et~al.},
  \ifthenelse{\boolean{articletitles}}{\emph{{$\jpsi$ and $\psitwos$ production
  in $pp$ collisions at $\sqs=7\tev$}},
  }{}\href{https://doi.org/10.1007/JHEP02(2012)011}{JHEP \textbf{02} (2012)
  011}, \href{http://arxiv.org/abs/1111.1557}{{\normalfont\ttfamily
  arXiv:1111.1557}}\relax
\mciteBstWouldAddEndPuncttrue
\mciteSetBstMidEndSepPunct{\mcitedefaultmidpunct}
{\mcitedefaultendpunct}{\mcitedefaultseppunct}\relax
\EndOfBibitem
\bibitem{Sirunyan:2017qdw}
CMS collaboration, A.~M. Sirunyan {\em et~al.},
  \ifthenelse{\boolean{articletitles}}{\emph{{Measurement of quarkonium
  production cross sections in pp collisions at $\sqs=13\tev$}},
  }{}\href{https://doi.org/10.1016/j.physletb.2018.02.033}{Phys.\ Lett.\
  \textbf{B780} (2018) 251},
  \href{http://arxiv.org/abs/1710.11002}{{\normalfont\ttfamily
  arXiv:1710.11002}}\relax
\mciteBstWouldAddEndPuncttrue
\mciteSetBstMidEndSepPunct{\mcitedefaultmidpunct}
{\mcitedefaultendpunct}{\mcitedefaultseppunct}\relax
\EndOfBibitem
\bibitem{Abelev:2012gx}
ALICE collaboration, B.~Abelev {\em et~al.},
  \ifthenelse{\boolean{articletitles}}{\emph{{Measurement of prompt $\jpsi$ and
  beauty hadron production cross sections at mid-rapidity in $pp$ collisions at
  $\sqs=7\tev$}}, }{}\href{https://doi.org/10.1007/JHEP11(2012)065}{JHEP
  \textbf{11} (2012) 065},
  \href{http://arxiv.org/abs/1205.5880}{{\normalfont\ttfamily
  arXiv:1205.5880}}\relax
\mciteBstWouldAddEndPuncttrue
\mciteSetBstMidEndSepPunct{\mcitedefaultmidpunct}
{\mcitedefaultendpunct}{\mcitedefaultseppunct}\relax
\EndOfBibitem
\bibitem{Abelev:2012kr}
ALICE collaboration, B.~Abelev {\em et~al.},
  \ifthenelse{\boolean{articletitles}}{\emph{{Inclusive $\jpsi$ production in
  $pp$ collisions at $\sqs=2.76\tev$}},
  }{}\href{https://doi.org/10.1016/j.physletb.2012.10.078}{Phys.\ Lett.\
  \textbf{B718} (2012) 295},
  \href{http://arxiv.org/abs/1203.3641}{{\normalfont\ttfamily
  arXiv:1203.3641}}\relax
\mciteBstWouldAddEndPuncttrue
\mciteSetBstMidEndSepPunct{\mcitedefaultmidpunct}
{\mcitedefaultendpunct}{\mcitedefaultseppunct}\relax
\EndOfBibitem
\bibitem{Adam:2016rdg}
ALICE collaboration, J.~Adam {\em et~al.},
  \ifthenelse{\boolean{articletitles}}{\emph{{$\jpsi$ suppression at forward
  rapidity in Pb-Pb collisions at $\sqsnn=5.02\tev$}},
  }{}\href{https://doi.org/10.1016/j.physletb.2016.12.064}{Phys.\ Lett.\
  \textbf{B766} (2017) 212},
  \href{http://arxiv.org/abs/1606.08197}{{\normalfont\ttfamily
  arXiv:1606.08197}}\relax
\mciteBstWouldAddEndPuncttrue
\mciteSetBstMidEndSepPunct{\mcitedefaultmidpunct}
{\mcitedefaultendpunct}{\mcitedefaultseppunct}\relax
\EndOfBibitem
\bibitem{Acharya:2019lkw}
ALICE collaboration, S.~Acharya {\em et~al.},
  \ifthenelse{\boolean{articletitles}}{\emph{{Inclusive $\jpsi$ production at
  mid-rapidity in $pp$ collisions at $\sqs=5.02\tev$}},
  }{}\href{https://doi.org/10.1007/JHEP10(2019)084}{JHEP \textbf{10} (2019)
  084}, \href{http://arxiv.org/abs/1905.07211}{{\normalfont\ttfamily
  arXiv:1905.07211}}\relax
\mciteBstWouldAddEndPuncttrue
\mciteSetBstMidEndSepPunct{\mcitedefaultmidpunct}
{\mcitedefaultendpunct}{\mcitedefaultseppunct}\relax
\EndOfBibitem
\bibitem{Aamodt:2011gj}
ALICE collaboration, K.~Aamodt {\em et~al.},
  \ifthenelse{\boolean{articletitles}}{\emph{{Rapidity and transverse momentum
  dependence of inclusive $\jpsi$ production in $pp$ collisions at
  $\sqs=7\tev$}}, }{}\href{https://doi.org/10.1016/j.physletb.2011.09.054,
  10.1016/j.physletb.2012.10.060}{Phys.\ Lett.\  \textbf{B704} (2011) 442},
  \href{http://arxiv.org/abs/1105.0380}{{\normalfont\ttfamily
  arXiv:1105.0380}}\relax
\mciteBstWouldAddEndPuncttrue
\mciteSetBstMidEndSepPunct{\mcitedefaultmidpunct}
{\mcitedefaultendpunct}{\mcitedefaultseppunct}\relax
\EndOfBibitem
\bibitem{Adam:2015rta}
ALICE collaboration, J.~Adam {\em et~al.},
  \ifthenelse{\boolean{articletitles}}{\emph{{Inclusive quarkonium production
  at forward rapidity in $pp$ collisions at $\sqs=8\tev$}},
  }{}\href{https://doi.org/10.1140/epjc/s10052-016-3987-y}{Eur.\ Phys.\ J.\
  \textbf{C76} (2016) 184},
  \href{http://arxiv.org/abs/1509.08258}{{\normalfont\ttfamily
  arXiv:1509.08258}}\relax
\mciteBstWouldAddEndPuncttrue
\mciteSetBstMidEndSepPunct{\mcitedefaultmidpunct}
{\mcitedefaultendpunct}{\mcitedefaultseppunct}\relax
\EndOfBibitem
\bibitem{Acharya:2017hjh}
ALICE collaboration, S.~Acharya {\em et~al.},
  \ifthenelse{\boolean{articletitles}}{\emph{{Energy dependence of
  forward-rapidity $\jpsi$ and $\psitwos$ production in $pp$ collisions at the
  LHC}}, }{}\href{https://doi.org/10.1140/epjc/s10052-017-4940-4}{Eur.\ Phys.\
  J.\  \textbf{C77} (2017) 392},
  \href{http://arxiv.org/abs/1702.00557}{{\normalfont\ttfamily
  arXiv:1702.00557}}\relax
\mciteBstWouldAddEndPuncttrue
\mciteSetBstMidEndSepPunct{\mcitedefaultmidpunct}
{\mcitedefaultendpunct}{\mcitedefaultseppunct}\relax
\EndOfBibitem
\bibitem{CDF:1997ykw}
CDF collaboration, F.~Abe {\em et~al.},
  \ifthenelse{\boolean{articletitles}}{\emph{{$\jpsi$ and $\psitwos$ production
  in $p\bar{p}$ collisions at $\sqs=1.8\tev$}},
  }{}\href{https://doi.org/10.1103/PhysRevLett.79.572}{Phys.\ Rev.\ Lett.\
  \textbf{79} (1997) 572}\relax
\mciteBstWouldAddEndPuncttrue
\mciteSetBstMidEndSepPunct{\mcitedefaultmidpunct}
{\mcitedefaultendpunct}{\mcitedefaultseppunct}\relax
\EndOfBibitem
\bibitem{CDF:2004jtw}
CDF collaboration, D.~Acosta {\em et~al.},
  \ifthenelse{\boolean{articletitles}}{\emph{{Measurement of the $\jpsi$ meson
  and $b-$hadron production cross sections in $p\bar{p}$ collisions at
  $\sqs=1960\gev$}},
  }{}\href{https://doi.org/10.1103/PhysRevD.71.032001}{Phys.\ Rev.\
  \textbf{D71} (2005) 032001},
  \href{http://arxiv.org/abs/hep-ex/0412071}{{\normalfont\ttfamily
  arXiv:hep-ex/0412071}}\relax
\mciteBstWouldAddEndPuncttrue
\mciteSetBstMidEndSepPunct{\mcitedefaultmidpunct}
{\mcitedefaultendpunct}{\mcitedefaultseppunct}\relax
\EndOfBibitem
\bibitem{D0:1996awi}
D0 collaboration, S.~Abachi {\em et~al.},
  \ifthenelse{\boolean{articletitles}}{\emph{{$\jpsi$ production in $p\bar{p}$
  collisions at $\sqs{s}=1.8\tev$}},
  }{}\href{https://doi.org/10.1016/0370-2693(96)00067-6}{Phys.\ Lett.\
  \textbf{B370} (1996) 239}\relax
\mciteBstWouldAddEndPuncttrue
\mciteSetBstMidEndSepPunct{\mcitedefaultmidpunct}
{\mcitedefaultendpunct}{\mcitedefaultseppunct}\relax
\EndOfBibitem
\bibitem{D0:1998vai}
D0 collaboration, B.~Abbott {\em et~al.},
  \ifthenelse{\boolean{articletitles}}{\emph{{Small angle $\jpsi$ production in
  $p\bar{p}$ collisions at $\sqs=1.8\tev$}},
  }{}\href{https://doi.org/10.1103/PhysRevLett.82.35}{Phys.\ Rev.\ Lett.\
  \textbf{82} (1999) 35},
  \href{http://arxiv.org/abs/hep-ex/9807029}{{\normalfont\ttfamily
  arXiv:hep-ex/9807029}}\relax
\mciteBstWouldAddEndPuncttrue
\mciteSetBstMidEndSepPunct{\mcitedefaultmidpunct}
{\mcitedefaultendpunct}{\mcitedefaultseppunct}\relax
\EndOfBibitem
\bibitem{LHCb-PAPER-2014-047}
LHCb collaboration, R.~Aaij {\em et~al.},
  \ifthenelse{\boolean{articletitles}}{\emph{{Precision luminosity measurements
  at LHCb}}, }{}\href{https://doi.org/10.1088/1748-0221/9/12/P12005}{JINST
  \textbf{9} (2014) P12005},
  \href{http://arxiv.org/abs/1410.0149}{{\normalfont\ttfamily
  arXiv:1410.0149}}\relax
\mciteBstWouldAddEndPuncttrue
\mciteSetBstMidEndSepPunct{\mcitedefaultmidpunct}
{\mcitedefaultendpunct}{\mcitedefaultseppunct}\relax
\EndOfBibitem
\bibitem{LHCb-PAPER-2013-052}
LHCb collaboration, R.~Aaij {\em et~al.},
  \ifthenelse{\boolean{articletitles}}{\emph{{Study of \jpsi production and
  cold nuclear matter effects in $\proton$Pb collisions at $\sqsnn = $5\tev}},
  }{}\href{https://doi.org/10.1007/JHEP02(2014)072}{JHEP \textbf{02} (2014)
  072}, \href{http://arxiv.org/abs/1308.6729}{{\normalfont\ttfamily
  arXiv:1308.6729}}\relax
\mciteBstWouldAddEndPuncttrue
\mciteSetBstMidEndSepPunct{\mcitedefaultmidpunct}
{\mcitedefaultendpunct}{\mcitedefaultseppunct}\relax
\EndOfBibitem
\bibitem{LHCb-DP-2008-001}
LHCb collaboration, A.~A. Alves~Jr.\ {\em et~al.},
  \ifthenelse{\boolean{articletitles}}{\emph{{The \lhcb detector at the LHC}},
  }{}\href{https://doi.org/10.1088/1748-0221/3/08/S08005}{JINST \textbf{3}
  (2008) S08005}\relax
\mciteBstWouldAddEndPuncttrue
\mciteSetBstMidEndSepPunct{\mcitedefaultmidpunct}
{\mcitedefaultendpunct}{\mcitedefaultseppunct}\relax
\EndOfBibitem
\bibitem{LHCb-DP-2014-002}
LHCb collaboration, R.~Aaij {\em et~al.},
  \ifthenelse{\boolean{articletitles}}{\emph{{LHCb detector performance}},
  }{}\href{https://doi.org/10.1142/S0217751X15300227}{Int.\ J.\ Mod.\ Phys.\
  \textbf{A30} (2015) 1530022},
  \href{http://arxiv.org/abs/1412.6352}{{\normalfont\ttfamily
  arXiv:1412.6352}}\relax
\mciteBstWouldAddEndPuncttrue
\mciteSetBstMidEndSepPunct{\mcitedefaultmidpunct}
{\mcitedefaultendpunct}{\mcitedefaultseppunct}\relax
\EndOfBibitem
\bibitem{Sjostrand:2007gs}
T.~Sj\"{o}strand, S.~Mrenna, and P.~Skands,
  \ifthenelse{\boolean{articletitles}}{\emph{{A brief introduction to PYTHIA
  8.1}}, }{}\href{https://doi.org/10.1016/j.cpc.2008.01.036}{Comput.\ Phys.\
  Commun.\  \textbf{178} (2008) 852},
  \href{http://arxiv.org/abs/0710.3820}{{\normalfont\ttfamily
  arXiv:0710.3820}}\relax
\mciteBstWouldAddEndPuncttrue
\mciteSetBstMidEndSepPunct{\mcitedefaultmidpunct}
{\mcitedefaultendpunct}{\mcitedefaultseppunct}\relax
\EndOfBibitem
\bibitem{Sjostrand:2006za}
T.~Sj\"{o}strand, S.~Mrenna, and P.~Skands,
  \ifthenelse{\boolean{articletitles}}{\emph{{PYTHIA 6.4 physics and manual}},
  }{}\href{https://doi.org/10.1088/1126-6708/2006/05/026}{JHEP \textbf{05}
  (2006) 026}, \href{http://arxiv.org/abs/hep-ph/0603175}{{\normalfont\ttfamily
  arXiv:hep-ph/0603175}}\relax
\mciteBstWouldAddEndPuncttrue
\mciteSetBstMidEndSepPunct{\mcitedefaultmidpunct}
{\mcitedefaultendpunct}{\mcitedefaultseppunct}\relax
\EndOfBibitem
\bibitem{LHCb-PROC-2010-056}
I.~Belyaev {\em et~al.}, \ifthenelse{\boolean{articletitles}}{\emph{{Handling
  of the generation of primary events in Gauss, the LHCb simulation
  framework}}, }{}\href{https://doi.org/10.1088/1742-6596/331/3/032047}{J.\
  Phys.\ Conf.\ Ser.\  \textbf{331} (2011) 032047}\relax
\mciteBstWouldAddEndPuncttrue
\mciteSetBstMidEndSepPunct{\mcitedefaultmidpunct}
{\mcitedefaultendpunct}{\mcitedefaultseppunct}\relax
\EndOfBibitem
\bibitem{Bargiotti:2007zz}
M.~Bargiotti and V.~Vagnoni, \ifthenelse{\boolean{articletitles}}{\emph{{Heavy
  quarkonia sector in PYTHIA 6.324: Tuning, validation and perspectives at
  LHC(b)}}, }{}
  \href{http://cdsweb.cern.ch/search?p=CERN-LHCB-2007-042&f=reportnumber&action_search=Search&c=LHCb}
  {CERN-LHCB-2007-042}, 2007\relax
\mciteBstWouldAddEndPuncttrue
\mciteSetBstMidEndSepPunct{\mcitedefaultmidpunct}
{\mcitedefaultendpunct}{\mcitedefaultseppunct}\relax
\EndOfBibitem
\bibitem{Lange:2001uf}
D.~J. Lange, \ifthenelse{\boolean{articletitles}}{\emph{{The EvtGen particle
  decay simulation package}},
  }{}\href{https://doi.org/10.1016/S0168-9002(01)00089-4}{Nucl.\ Instrum.\
  Meth.\  \textbf{A462} (2001) 152}\relax
\mciteBstWouldAddEndPuncttrue
\mciteSetBstMidEndSepPunct{\mcitedefaultmidpunct}
{\mcitedefaultendpunct}{\mcitedefaultseppunct}\relax
\EndOfBibitem
\bibitem{davidson2015photos}
N.~Davidson, T.~Przedzinski, and Z.~Was,
  \ifthenelse{\boolean{articletitles}}{\emph{{PHOTOS interface in C++:
  Technical and physics documentation}},
  }{}\href{https://doi.org/https://doi.org/10.1016/j.cpc.2015.09.013}{Comp.\
  Phys.\ Comm.\  \textbf{199} (2016) 86},
  \href{http://arxiv.org/abs/1011.0937}{{\normalfont\ttfamily
  arXiv:1011.0937}}\relax
\mciteBstWouldAddEndPuncttrue
\mciteSetBstMidEndSepPunct{\mcitedefaultmidpunct}
{\mcitedefaultendpunct}{\mcitedefaultseppunct}\relax
\EndOfBibitem
\bibitem{Allison:2006ve}
Geant4 collaboration, J.~Allison {\em et~al.},
  \ifthenelse{\boolean{articletitles}}{\emph{{Geant4 developments and
  applications}}, }{}\href{https://doi.org/10.1109/TNS.2006.869826}{IEEE
  Trans.\ Nucl.\ Sci.\  \textbf{53} (2006) 270}\relax
\mciteBstWouldAddEndPuncttrue
\mciteSetBstMidEndSepPunct{\mcitedefaultmidpunct}
{\mcitedefaultendpunct}{\mcitedefaultseppunct}\relax
\EndOfBibitem
\bibitem{Agostinelli:2002hh}
Geant4 collaboration, S.~Agostinelli {\em et~al.},
  \ifthenelse{\boolean{articletitles}}{\emph{{Geant4: A simulation toolkit}},
  }{}\href{https://doi.org/10.1016/S0168-9002(03)01368-8}{Nucl.\ Instrum.\
  Meth.\  \textbf{A506} (2003) 250}\relax
\mciteBstWouldAddEndPuncttrue
\mciteSetBstMidEndSepPunct{\mcitedefaultmidpunct}
{\mcitedefaultendpunct}{\mcitedefaultseppunct}\relax
\EndOfBibitem
\bibitem{LHCb-PROC-2011-006}
M.~Clemencic {\em et~al.}, \ifthenelse{\boolean{articletitles}}{\emph{{The
  \lhcb simulation application, Gauss: Design, evolution and experience}},
  }{}\href{https://doi.org/10.1088/1742-6596/331/3/032023}{J.\ Phys.\ Conf.\
  Ser.\  \textbf{331} (2011) 032023}\relax
\mciteBstWouldAddEndPuncttrue
\mciteSetBstMidEndSepPunct{\mcitedefaultmidpunct}
{\mcitedefaultendpunct}{\mcitedefaultseppunct}\relax
\EndOfBibitem
\bibitem{DeCian:2255039}
M.~De~Cian, S.~Farry, P.~Seyfert, and S.~Stahl,
  \ifthenelse{\boolean{articletitles}}{\emph{{Fast neural-net based fake track
  rejection in the LHCb reconstruction}}, }{}
  \href{http://cdsweb.cern.ch/search?p=LHCb-PUB-2017-011&f=reportnumber&action_search=Search&c=LHCb+Notes}
  {LHCb-PUB-2017-011}, 2017\relax
\mciteBstWouldAddEndPuncttrue
\mciteSetBstMidEndSepPunct{\mcitedefaultmidpunct}
{\mcitedefaultendpunct}{\mcitedefaultseppunct}\relax
\EndOfBibitem
\bibitem{PDG2020}
Particle Data Group, P.~A. Zyla {\em et~al.},
  \ifthenelse{\boolean{articletitles}}{\emph{{\href{http://pdg.lbl.gov/}{Review
  of particle physics}}}, }{}\href{https://doi.org/10.1093/ptep/ptaa104}{Prog.\
  Theor.\ Exp.\ Phys.\  \textbf{2020} (2020) 083C01}\relax
\mciteBstWouldAddEndPuncttrue
\mciteSetBstMidEndSepPunct{\mcitedefaultmidpunct}
{\mcitedefaultendpunct}{\mcitedefaultseppunct}\relax
\EndOfBibitem
\bibitem{Skwarnicki:1986xj}
T.~Skwarnicki, {\em {A study of the radiative cascade transitions between the
  Upsilon-prime and Upsilon resonances}}, PhD thesis, Institute of Nuclear
  Physics, Krakow, 1986,
  {\href{http://inspirehep.net/record/230779/}{DESY-F31-86-02}}\relax
\mciteBstWouldAddEndPuncttrue
\mciteSetBstMidEndSepPunct{\mcitedefaultmidpunct}
{\mcitedefaultendpunct}{\mcitedefaultseppunct}\relax
\EndOfBibitem
\bibitem{Cranmer:2000du}
K.~S. Cranmer, \ifthenelse{\boolean{articletitles}}{\emph{{Kernel estimation in
  high-energy physics}},
  }{}\href{https://doi.org/10.1016/S0010-4655(00)00243-5}{Comput.\ Phys.\
  Commun.\  \textbf{136} (2001) 198},
  \href{http://arxiv.org/abs/hep-ex/0011057}{{\normalfont\ttfamily
  arXiv:hep-ex/0011057}}\relax
\mciteBstWouldAddEndPuncttrue
\mciteSetBstMidEndSepPunct{\mcitedefaultmidpunct}
{\mcitedefaultendpunct}{\mcitedefaultseppunct}\relax
\EndOfBibitem
\bibitem{Pivk:2004ty}
M.~Pivk and F.~R. Le~Diberder,
  \ifthenelse{\boolean{articletitles}}{\emph{{sPlot: A statistical tool to
  unfold data distributions}},
  }{}\href{https://doi.org/10.1016/j.nima.2005.08.106}{Nucl.\ Instrum.\ Meth.\
  \textbf{A555} (2005) 356},
  \href{http://arxiv.org/abs/physics/0402083}{{\normalfont\ttfamily
  arXiv:physics/0402083}}\relax
\mciteBstWouldAddEndPuncttrue
\mciteSetBstMidEndSepPunct{\mcitedefaultmidpunct}
{\mcitedefaultendpunct}{\mcitedefaultseppunct}\relax
\EndOfBibitem
\bibitem{LHCb-DP-2013-002}
LHCb collaboration, R.~Aaij {\em et~al.},
  \ifthenelse{\boolean{articletitles}}{\emph{{Measurement of the track
  reconstruction efficiency at LHCb}},
  }{}\href{https://doi.org/10.1088/1748-0221/10/02/P02007}{JINST \textbf{10}
  (2015) P02007}, \href{http://arxiv.org/abs/1408.1251}{{\normalfont\ttfamily
  arXiv:1408.1251}}\relax
\mciteBstWouldAddEndPuncttrue
\mciteSetBstMidEndSepPunct{\mcitedefaultmidpunct}
{\mcitedefaultendpunct}{\mcitedefaultseppunct}\relax
\EndOfBibitem
\bibitem{LHCb-PUB-2014-039}
S.~Tolk, J.~Albrecht, F.~Dettori, and A.~Pellegrino,
  \ifthenelse{\boolean{articletitles}}{\emph{{Data driven trigger efficiency
  determination at LHCb}}, }{}
  \href{http://cdsweb.cern.ch/search?p=LHCb-PUB-2014-039&f=reportnumber&action_search=Search&c=LHCb+Notes}
  {LHCb-PUB-2014-039}, 2014\relax
\mciteBstWouldAddEndPuncttrue
\mciteSetBstMidEndSepPunct{\mcitedefaultmidpunct}
{\mcitedefaultendpunct}{\mcitedefaultseppunct}\relax
\EndOfBibitem
\bibitem{LHCb-PAPER-2013-008}
LHCb collaboration, R.~Aaij {\em et~al.},
  \ifthenelse{\boolean{articletitles}}{\emph{{Measurement of \jpsi polarization
  in \proton\proton collisions at \mbox{$\sqs=$7\tev}}},
  }{}\href{https://doi.org/10.1140/epjc/s10052-013-2631-3}{Eur.\ Phys.\ J.\
  \textbf{C73} (2013) 2631},
  \href{http://arxiv.org/abs/1307.6379}{{\normalfont\ttfamily
  arXiv:1307.6379}}\relax
\mciteBstWouldAddEndPuncttrue
\mciteSetBstMidEndSepPunct{\mcitedefaultmidpunct}
{\mcitedefaultendpunct}{\mcitedefaultseppunct}\relax
\EndOfBibitem
\bibitem{Ma:2010yw}
Y.-Q. Ma, K.~Wang, and K.-T. Chao,
  \ifthenelse{\boolean{articletitles}}{\emph{{$\jpsi (\psi^\prime)$ production
  at the Tevatron and LHC at ${\cal O}(\alpha_s^4v^4)$ in nonrelativistic
  QCD}}, }{}\href{https://doi.org/10.1103/PhysRevLett.106.042002}{Phys.\ Rev.\
  Lett.\  \textbf{106} (2011) 042002},
  \href{http://arxiv.org/abs/1009.3655}{{\normalfont\ttfamily
  arXiv:1009.3655}}\relax
\mciteBstWouldAddEndPuncttrue
\mciteSetBstMidEndSepPunct{\mcitedefaultmidpunct}
{\mcitedefaultendpunct}{\mcitedefaultseppunct}\relax
\EndOfBibitem
\bibitem{Ma:2014mri}
Y.-Q. Ma and R.~Venugopalan,
  \ifthenelse{\boolean{articletitles}}{\emph{{Comprehensive description of
  $\jpsi$ production in proton-proton collisions at collider energies}},
  }{}\href{https://doi.org/10.1103/PhysRevLett.113.192301}{Phys.\ Rev.\ Lett.\
  \textbf{113} (2014) 192301},
  \href{http://arxiv.org/abs/1408.4075}{{\normalfont\ttfamily
  arXiv:1408.4075}}\relax
\mciteBstWouldAddEndPuncttrue
\mciteSetBstMidEndSepPunct{\mcitedefaultmidpunct}
{\mcitedefaultendpunct}{\mcitedefaultseppunct}\relax
\EndOfBibitem
\bibitem{Chao:2012iv}
K.-T. Chao {\em et~al.}, \ifthenelse{\boolean{articletitles}}{\emph{{$\jpsi$
  polarization at hadron colliders in nonrelativistic QCD}},
  }{}\href{https://doi.org/10.1103/PhysRevLett.108.242004}{Phys.\ Rev.\ Lett.\
  \textbf{108} (2012) 242004},
  \href{http://arxiv.org/abs/1201.2675}{{\normalfont\ttfamily
  arXiv:1201.2675}}\relax
\mciteBstWouldAddEndPuncttrue
\mciteSetBstMidEndSepPunct{\mcitedefaultmidpunct}
{\mcitedefaultendpunct}{\mcitedefaultseppunct}\relax
\EndOfBibitem
\bibitem{Albacete:2012xq}
J.~L. Albacete, A.~Dumitru, H.~Fujii, and Y.~Nara,
  \ifthenelse{\boolean{articletitles}}{\emph{{CGC predictions for p + Pb
  collisions at the LHC}},
  }{}\href{https://doi.org/10.1016/j.nuclphysa.2012.09.012}{Nucl.\ Phys.\
  \textbf{A897} (2013) 1},
  \href{http://arxiv.org/abs/1209.2001}{{\normalfont\ttfamily
  arXiv:1209.2001}}\relax
\mciteBstWouldAddEndPuncttrue
\mciteSetBstMidEndSepPunct{\mcitedefaultmidpunct}
{\mcitedefaultendpunct}{\mcitedefaultseppunct}\relax
\EndOfBibitem
\bibitem{Cacciari:2012ny}
M.~Cacciari {\em et~al.},
  \ifthenelse{\boolean{articletitles}}{\emph{{Theoretical predictions for charm
  and bottom production at the LHC}},
  }{}\href{https://doi.org/10.1007/JHEP10(2012)137}{JHEP \textbf{10} (2012)
  137}, \href{http://arxiv.org/abs/1205.6344}{{\normalfont\ttfamily
  arXiv:1205.6344}}\relax
\mciteBstWouldAddEndPuncttrue
\mciteSetBstMidEndSepPunct{\mcitedefaultmidpunct}
{\mcitedefaultendpunct}{\mcitedefaultseppunct}\relax
\EndOfBibitem
\bibitem{Cacciari:2015fta}
M.~Cacciari, M.~L. Mangano, and P.~Nason,
  \ifthenelse{\boolean{articletitles}}{\emph{{Gluon PDF constraints from the
  ratio of forward heavy-quark production at the LHC at $\sqs=7$ and
  $13\tev$}}, }{}\href{https://doi.org/10.1140/epjc/s10052-015-3814-x}{Eur.\
  Phys.\ J.\  \textbf{C75} (2015) 610},
  \href{http://arxiv.org/abs/1507.06197}{{\normalfont\ttfamily
  arXiv:1507.06197}}\relax
\mciteBstWouldAddEndPuncttrue
\mciteSetBstMidEndSepPunct{\mcitedefaultmidpunct}
{\mcitedefaultendpunct}{\mcitedefaultseppunct}\relax
\EndOfBibitem
\bibitem{Contopanagos:1996nh}
H.~Contopanagos, E.~Laenen, and G.~F. Sterman,
  \ifthenelse{\boolean{articletitles}}{\emph{{Sudakov factorization and
  resummation}}, }{}\href{https://doi.org/10.1016/S0550-3213(96)00567-6}{Nucl.\
  Phys.\  \textbf{B484} (1997) 303},
  \href{http://arxiv.org/abs/hep-ph/9604313}{{\normalfont\ttfamily
  arXiv:hep-ph/9604313}}\relax
\mciteBstWouldAddEndPuncttrue
\mciteSetBstMidEndSepPunct{\mcitedefaultmidpunct}
{\mcitedefaultendpunct}{\mcitedefaultseppunct}\relax
\EndOfBibitem
\bibitem{Ferreiro:2013pua}
E.~G. Ferreiro, F.~Fleuret, J.~P. Lansberg, and A.~Rakotozafindrabe,
  \ifthenelse{\boolean{articletitles}}{\emph{{Impact of the nuclear
  modification of the gluon densities on $\jpsi$ production in $p$Pb collisions
  at $\sqsnn=5\tev$}},
  }{}\href{https://doi.org/10.1103/PhysRevC.88.047901}{Phys.\ Rev.\
  \textbf{C88} (2013) 047901},
  \href{http://arxiv.org/abs/1305.4569}{{\normalfont\ttfamily
  arXiv:1305.4569}}\relax
\mciteBstWouldAddEndPuncttrue
\mciteSetBstMidEndSepPunct{\mcitedefaultmidpunct}
{\mcitedefaultendpunct}{\mcitedefaultseppunct}\relax
\EndOfBibitem
\bibitem{Albacete:2013ei}
J.~L. Albacete {\em et~al.},
  \ifthenelse{\boolean{articletitles}}{\emph{{Predictions for $p+$Pb collisions
  at $\sqsnn=5\tev$}},
  }{}\href{https://doi.org/10.1142/S0218301313300075}{Int.\ J.\ Mod.\ Phys.\
  \textbf{E22} (2013) 1330007},
  \href{http://arxiv.org/abs/1301.3395}{{\normalfont\ttfamily
  arXiv:1301.3395}}\relax
\mciteBstWouldAddEndPuncttrue
\mciteSetBstMidEndSepPunct{\mcitedefaultmidpunct}
{\mcitedefaultendpunct}{\mcitedefaultseppunct}\relax
\EndOfBibitem
\bibitem{Arleo:2012hn}
F.~Arleo and S.~Peign\'e, \ifthenelse{\boolean{articletitles}}{\emph{{$\jpsi$
  suppression in p-A collisions from parton energy loss in cold QCD matter}},
  }{}\href{https://doi.org/10.1103/PhysRevLett.109.122301}{Phys.\ Rev.\ Lett.\
  \textbf{109} (2012) 122301},
  \href{http://arxiv.org/abs/1204.4609}{{\normalfont\ttfamily
  arXiv:1204.4609}}\relax
\mciteBstWouldAddEndPuncttrue
\mciteSetBstMidEndSepPunct{\mcitedefaultmidpunct}
{\mcitedefaultendpunct}{\mcitedefaultseppunct}\relax
\EndOfBibitem
\bibitem{Arleo:2012rs}
F.~Arleo and S.~Peign\'e,
  \ifthenelse{\boolean{articletitles}}{\emph{{Heavy-quarkonium suppression in
  p-A collisions from parton energy loss in cold QCD matter}},
  }{}\href{https://doi.org/10.1007/JHEP03(2013)122}{JHEP \textbf{03} (2013)
  122}, \href{http://arxiv.org/abs/1212.0434}{{\normalfont\ttfamily
  arXiv:1212.0434}}\relax
\mciteBstWouldAddEndPuncttrue
\mciteSetBstMidEndSepPunct{\mcitedefaultmidpunct}
{\mcitedefaultendpunct}{\mcitedefaultseppunct}\relax
\EndOfBibitem
\bibitem{Abelev:2011md}
ALICE collaboration, B.~Abelev {\em et~al.},
  \ifthenelse{\boolean{articletitles}}{\emph{{$\jpsi$ polarization in $pp$
  collisions at $\sqs=7\tev$}},
  }{}\href{https://doi.org/10.1103/PhysRevLett.108.082001}{Phys.\ Rev.\ Lett.\
  \textbf{108} (2012) 082001},
  \href{http://arxiv.org/abs/1111.1630}{{\normalfont\ttfamily
  arXiv:1111.1630}}\relax
\mciteBstWouldAddEndPuncttrue
\mciteSetBstMidEndSepPunct{\mcitedefaultmidpunct}
{\mcitedefaultendpunct}{\mcitedefaultseppunct}\relax
\EndOfBibitem
\bibitem{Chatrchyan:2013cla}
CMS collaboration, S.~Chatrchyan {\em et~al.},
  \ifthenelse{\boolean{articletitles}}{\emph{{Measurement of the prompt $\jpsi$
  and $\psitwos$ polarizations in $pp$ collisions at $\sqs=7\tev$}},
  }{}\href{https://doi.org/10.1016/j.physletb.2013.10.055}{Phys.\ Lett.\
  \textbf{B727} (2013) 381},
  \href{http://arxiv.org/abs/1307.6070}{{\normalfont\ttfamily
  arXiv:1307.6070}}\relax
\mciteBstWouldAddEndPuncttrue
\mciteSetBstMidEndSepPunct{\mcitedefaultmidpunct}
{\mcitedefaultendpunct}{\mcitedefaultseppunct}\relax
\EndOfBibitem
\end{mcitethebibliography}

\newpage
\centerline
{\large\bf LHCb collaboration}
\begin
{flushleft}
\small
R.~Aaij$^{32}$,
A.S.W.~Abdelmotteleb$^{56}$,
C.~Abell{\'a}n~Beteta$^{50}$,
T.~Ackernley$^{60}$,
B.~Adeva$^{46}$,
M.~Adinolfi$^{54}$,
H.~Afsharnia$^{9}$,
C.~Agapopoulou$^{13}$,
C.A.~Aidala$^{86}$,
S.~Aiola$^{25}$
% ...

{\footnotesize \it

% ...
$^{9}$Universit{\'e} Clermont Auvergne, CNRS/IN2P3, LPC, Clermont-Ferrand, France\\
% ...
$^{13}$LPNHE, Sorbonne Universit{\'e}, Paris Diderot Sorbonne Paris Cit{\'e}, CNRS/IN2P3, Paris, France\\
% ...
$^{25}$INFN Sezione di Milano, Milano, Italy\\
% ...
$^{32}$Nikhef National Institute for Subatomic Physics, Amsterdam, Netherlands\\
% ...
$^{46}$Instituto Galego de F{\'\i}sica de Altas Enerx{\'\i}as (IGFAE), Universidade de Santiago de Compostela, Santiago de Compostela, Spain\\
% ...
$^{50}$Physik-Institut, Universit{\"a}t Z{\"u}rich, Z{\"u}rich, Switzerland\\
% ...
$^{54}$H.H. Wills Physics Laboratory, University of Bristol, Bristol, United Kingdom\\
% ...
$^{56}$Department of Physics, University of Warwick, Coventry, United Kingdom\\
% ...
$^{60}$Oliver Lodge Laboratory, University of Liverpool, Liverpool, United Kingdom\\
% ...
$^{68}$Syracuse University, Syracuse, NY, United States\\
% ...
$^{86}$University of Michigan, Ann Arbor, United States, associated to $^{68}$\\
\bigskip
% ...
\medskip
}
\end{flushleft}

\end{document}